\documentclass[prx,onecolumn,english,superscriptaddress,floatfix,longbibliography,aps, preprint]{revtex4-2}

\usepackage[english]{babel}
\usepackage[utf8]{inputenc}
\usepackage[colorinlistoftodos, color=green!40, prependcaption]{todonotes}
\usepackage{soul}
\usepackage{amsthm}
\usepackage{mathtools}
\usepackage{physics}
\usepackage{xcolor}
\usepackage{graphicx}
\usepackage[left=23mm,right=13mm,top=35mm,columnsep=15pt]{geometry} 
\usepackage{adjustbox}
\usepackage{placeins}
\usepackage[T1]{fontenc}
\usepackage{lipsum}
\usepackage{csquotes}
\usepackage[pdftex, pdftitle={Article}, pdfauthor={Author}]{hyperref} 
\usepackage{algorithm}
\usepackage{algpseudocode}
\usepackage{simpler-wick}
\usepackage{listings}
\usepackage{nccmath}
\usepackage{comment}
\usepackage{tikz}
\usepackage{tikz-feynman}
\tikzfeynmanset{compat=1.0.0}
\usepackage{multirow}
\usepackage{relsize}
\usepackage{wasysym}
\newcommand*{\Scale}[2][4]{\scalebox{#1}{$#2$}}

\definecolor{codegreen}{rgb}{0,0.6,0}
\definecolor{codegray}{rgb}{0.5,0.5,0.5}
\definecolor{codepurple}{rgb}{0.58,0,0.82}
\definecolor{backcolour}{rgb}
{0.95,0.95,0.92}
\definecolor{ForestGreen}{RGB}{34,139,34}

\lstdefinestyle{mystyle}{
    backgroundcolor=\color{backcolour},   
    commentstyle=\color{codegreen},
    keywordstyle=\color{magenta},
    numberstyle=\tiny\color{codegray},
    stringstyle=\color{codepurple},
    basicstyle=\ttfamily\footnotesize,
    breakatwhitespace=false,         
    breaklines=true,                 
    captionpos=b,                    
    keepspaces=true,                 
    numbers=left,                    
    numbersep=5pt,                  
    showspaces=false,                
    showstringspaces=false,
    showtabs=false,                  
    tabsize=2
}
\lstdefinelanguage{Julia}%
  {morekeywords={abstract,break,case,catch,const,continue,do,else,elseif,%
      end,export,false,for,function,immutable,import,importall,if,in,%
      macro,module,otherwise,quote,return,switch,true,try,type,typealias,%
      using,while},%
   sensitive=true,%
   alsoother={},%
   morecomment=[l]\#,%
   morecomment=[n]{\#=}{=\#},%
   morestring=[s]{"}{"},%
   morestring=[m]{'}{'},%
}[keywords,comments,strings]%

\begin{document}
\title{Systematic time-coarse graining for driven quantum systems}
\author{Leon Bello†}
\author{Wentao Fan†}
\email[Correspondence email address: ]{wentaof@princeton.edu}
\author{Aditya Gandotra}
\author{Hakan E. Türeci}
\affiliation{Department of Electrical and Computer Engineering, Princeton University}
\thanks{These authors contributed equally.}
    
\date{\today} 

\begin{abstract}

Many recent advancements in quantum computing leverage strong drives on nonlinear systems for state preparation, signal amplification, or gate operation. However, the interplay within such strongly driven system introduces multi-scale dynamics that affects the long-time behavior of the system in non-trivial ways that are very difficult to model. Therefore, the analysis of these systems often relies on effective Hamiltonian models that introduce additional nonlinear processes which approximate the long-time dynamics so that highly oscillatory terms may be ignored. However, the removal of such high frequency transitions can only be performed rigorously within a systematic framework of time-coarse graining, which is a fundamentally irreversible operation. This implies that standard approaches with unitary effective models cannot accurately capture the long-time behavior of strongly driven nonlinear quantum systems in general. We introduce a systematic perturbation theory for obtaining the complete non-unitary effective model of the time-coarse grained (TCG) dynamics of a driven quantum system to any order in the coupling strengths. We derive a closed-form analytical formula for both unitary and non-unitary contributions, in the form of an effective Hamiltonian and non-unitary (pseudo-)dissipators. Remarkably, even though the effective theory presumes unitary time evolution at the microscopic level, the time-coarse grained dynamics is found to follow a non-unitary time evolution in general. This occurs even when there is no open heat reservoir for the system to become entangled with or dissipate into. We demonstrate the effectiveness of the new method using several typical models of driven nonlinear systems in superconducting circuits, and show that it generalizes and improves on existing methods by providing more accurate results and explaining phenomena that have not been accounted for.
\end{abstract}


\maketitle

\section{Introduction}
\label{sec:intro}

\begin{figure}[H]
    \centering
    \includegraphics[width=0.8\linewidth]{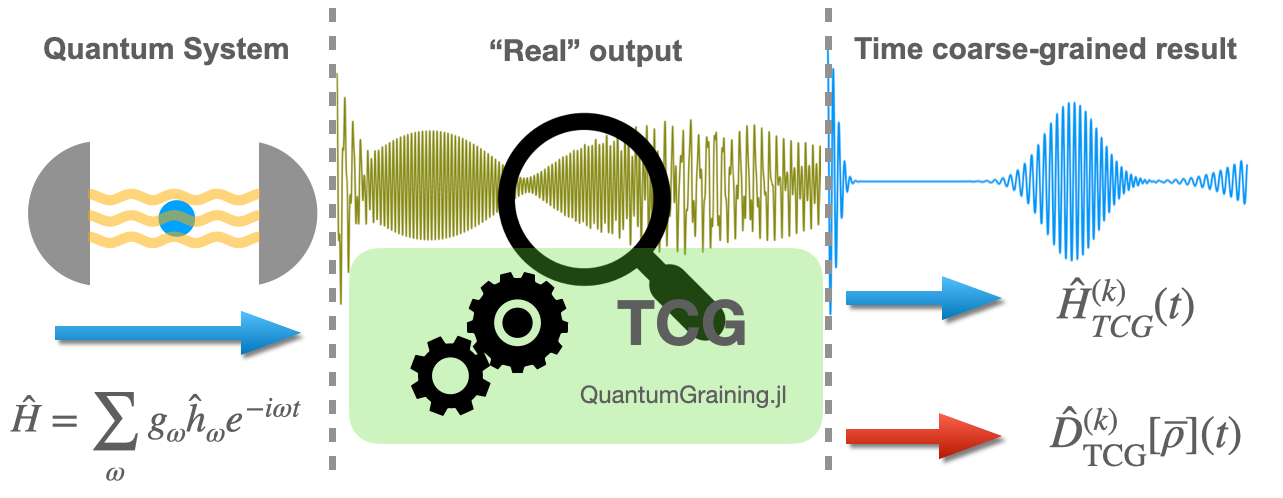}
    \caption{An illustration of the inherent time-coarse graining happening during a measurement. A quantum system produces high bandwidth signal. This signal drives a measurement apparatus which is band-limited, producing a time-coarse grained version of the real dynamics. The TCG method aims to produce a direct description of the time-coarse grained observables, which can be captured by an effective Quantum Master Equation for the coarse-grained density matrix $\overline{\rho}$ in the appropriate rotating frame determined by the central frequency of the measuring device.}
    \label{fig:tcg-scheme}
\end{figure}

When a measurement apparatus interacts with the electromagnetic fields emanating from a quantum system, its output is \emph{always} a coarse-grained function of exact physical variables at precise moments in time and/or precise locations in space. Therefore, the \textbf{observable dynamics} in any quantum system depend on the time and/or spatial resolution of the measuring apparatus. For typical quantum optical systems, the most important coarse-graining scale is the time resolution of the measuring device, which is limited by its bandwidth (e.g. Ref. \cite{da_silva_schemes_2010}, Appendix D). Although not always explicitly acknowledged, the measurement time resolution is a particularly crucial free parameter for driven nonlinear quantum systems, since the exchange of information and energy with an external drive can leave long-lasting effects on the system which propagate through different time scales and manifest in different forms due to nonlinearities. In many situations, the finite time resolution of the measurement apparatus can be captured by working with the time-coarse grained density matrix $\overline{\rho}(t)$ of the measured system, as illustrated in Fig. \ref{fig:tcg-scheme}.
\begin{equation}
\label{eq: def_of_rhobar}
\overline{\rho}(t)
=
\int_{-\infty}^\infty \tilde{f}_\tau(t - t') \rho(t') dt'
\end{equation}
where $\tilde{f}_{\tau}(t)$ is a moving-average function with width $\tau$ which we call the coarse-graining time scale.

In the time-coarse graining (TCG) framework \cite{gamel_time-averaged_2010}, the finite time resolution is actively taken into account during the formulation of effective theories. This approach has both fundamental and practical value. Fundamentally, it reflects the fact that all physical apparatus have finite response time and inevitably perform some type of time-coarse graining of information during their interactions with the measured systems. Practically, it can also offer significant simplification of the effective model by separating the experimentally relevant observables from the unresolvable ones, while at the same time keeping track of the latter's impact on former. This feature is becoming increasingly important as recent advancements in quantum devices focus on precise engineering of effective interactions via parametric driving, which often involves regimes where counter-rotating processes, typically regarded as negligible, assume significance, rendering the commonly used rotating-wave approximation (RWA) or even more sophisticated effective Hamiltonian methods unsatisfactory \cite{stefano_feynman-diagrams_2017, masuda_controls_2021, petrescu_lifetime_2020}. 
Correctly modelling the dynamics of these strongly driven systems has been a long standing problem, to which our work here offers a solution. 

In this work, we present a general solution to the time-coarse graining approach, that aims to directly capture the evolution of observables that are measurable and resolvable by a measurement device with a given temporal resolution (i.e. bandwidth). We employ this tool to analyze in detail the measurement problem in another paper \cite{wentao_transmon_tcg}. Our approach, systematic time-coarse graining method (STCG), explicitly derives a Liouvillian for $\overline{\rho}(t)$, without having to solve for the exact dynamics or assume any knowledge about the initial density matrix $\rho$ with infinite time resolution. Unlike previous attempts at deriving the TCG Liouvillian, our method is completely general, with closed-form formulas for the contributions to any order in the expansion.

STCG generalizes the effective static Hamiltonian method: we show that to any order, the long time dynamics of the system are described by an effective generalized Quantum Master Equation (QME), despite the fact that the microscopic dynamics are fully unitary. In many situations, especially when the drive amplitudes have their own time evolution, the emergent effective incoherent dynamics assume significance and have to be treated on the same footing as the coherent ones in the resulting QME. In addition to its conceptual novelty, our approach also has important numerical advantages. In fact, numerically solving the time evolution of a driven nonlinear quantum system with much higher time resolution than the measured variable dynamics can be very inefficient and often unfeasible, as their dynamics encompass vastly different time scales, making the equations highly stiff and numerically unstable; to make the situation worse, the solutions obtained in such brute-force approaches are usually difficult to interpret. Our method, {\it Systematic time-coarse graining} (STCG), gives an effective low-frequency quantum model leveraging the separation of time-scales as small parameters in a systematic expansion, giving an effective low-frequency description of the dynamics, which is long-time stable and efficient to study numerically.

The paper is organized as follows: We begin with a canonical example, the Rabi Model, illustrating the conceptual and mechanistic elements behind the TCG methodology in Section \ref{sec:rabi-model}. We then delve into the theoretical background in section \ref{sec:theory}, where we introduce the TCG method in the context of other methods, while presenting essential concepts of the theory while reserving the detailed derivation to the Appendix.  In section \ref{sec:diagrammatic expansion}, we introduce a diagrammatic approach for deriving the TCG master equation. In section \ref{sec:software} we introduce \textit{QuantumGraining.jl},  a Julia package that automates the TCG procedure, together with a short code example for the Rabi-model. Examples are available as Jupyter Notebooks on our GitHub repository. In section \ref{sec:examples}, we then showcase the method using two additional examples - \textbf{(1)} the Kerr-parametron and \textbf{(2)} the driven Duffing oscillator. In particular, we show that the TCG method predicts measurable modifications to the dynamics of these model, some of which are supported by analytical or numerical results in the literature obtained by other methods.  Finally, we conclude and discuss future prospects in section \ref{sec:conclusions}. 

\section{An illustrative example -- the time-coarse grained Rabi model}
\label{sec:rabi-model}
To demonstrate our method, we start with a simple but illustrative example -- the Rabi model. The Rabi model describes the interaction between a single linear cavity mode and an atom described by a two-level system (TLS), and can be simplified to the Jaynes-Cummings Model when the rotating-wave approximation (RWA) is justified. In the so called {\it ultrastrong coupling regime}, the Rabi model is of special interest since the RWA breaks down when the spin-cavity coupling strength becomes comparable or greater than the mode frequencies \cite{Braumüller_USC_Rabi, Casanova_DSC_JC, Ashhab_USC_Rabi}, and special approaches that go beyond the RWA are required \cite{casanova_deep_2010, wang_bridging_2015, amniat-talab_quantum_2005}.

More concretely, the model we consider here is described by the following Hamiltonian terms in the interaction picture,
\begin{equation}
\begin{split}
\label{eq:Rabi Hamiltonian}
\hat{H}
=&
\frac{g}{2} \left ( \hat{a}^\dagger \hat{\sigma}_+ e^{+i(\omega_c + \omega_a) t} + \hat{a}\hat{\sigma}_-e^{-i(\omega_c + \omega_a) t} \right )\\
&
+
\frac{g}{2} \left ( \hat{a}^\dagger \hat{\sigma_-} e^{+i(\omega_c - \omega_a) t} + \hat{a}\hat{\sigma}_+e^{-i(\omega_c - \omega_a) t} \right )
\end{split}
\end{equation}
where we $\omega_c$  $(\omega_a)$  is the cavity (atom) resonance.

\begin{figure}[!h]
    \centering
    \includegraphics[width=0.7\textwidth]{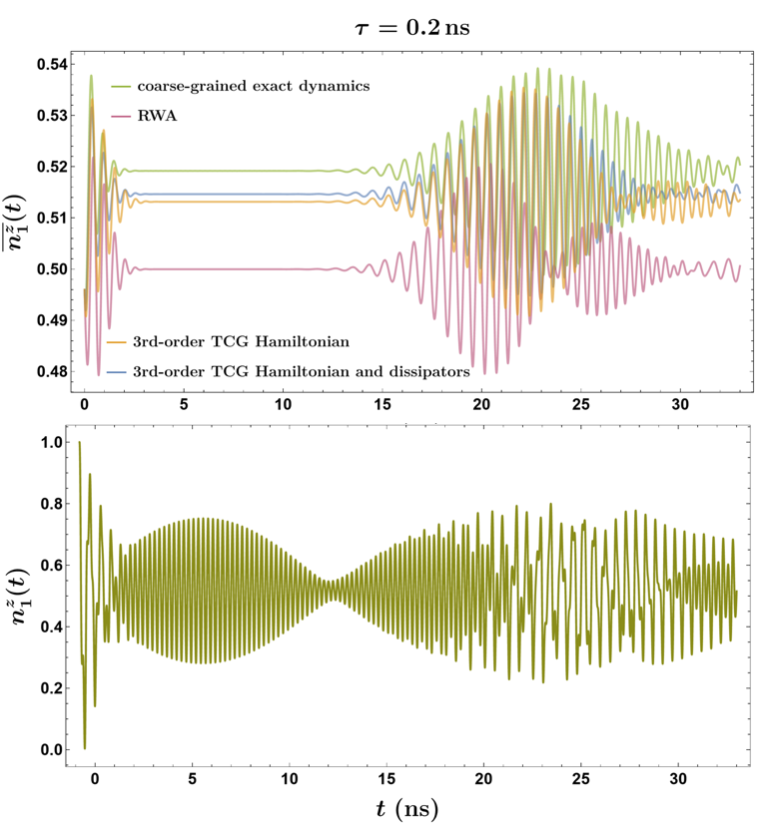}
    \caption{
    \textbf{(Bottom)} Numerical simulation of the excited TLS-state population $n^{z}_{1}(t)$, obtained by numerically solving the Schr\"odinger equations for 
    the interaction-picture Rabi Hamiltonian in Eq.(\ref{eq:Rabi Hamiltonian}). The initial state at $t_{i} = -0.8 \textrm{ns}$ is $|1\rangle_{s} |\alpha = 4.5\rangle_{c}$ with $|1\rangle$ and $|\alpha = 4.5\rangle_{c}$ representing the exited TLS state and the coherent cavity state with amplitude $\alpha = 4.5$, respectively. All TCG simulations are performed without making any further approximation to the perturbative TCG master equation.
    \textbf{(Top)} Numerical simulation of the time-coarse grained excited TLS-state population $\overline{n^{z}_{1}}(t)$ with a coarse-graining time scale of $\tau = 0.2 \textrm{ns}$. The coarse-grained exact dynamics is obtained by directly applying the Gaussian averaging function $\tilde{f}_\tau(t)$ to the exact $n^{z}_{1}(t)$ in the bottom figure, whereas the TCG results are obtained by numerically solving the corresponding TCG master equations; In all calculations the cavity Hilbert space is truncated to include the lowest $100$ levels during the simulation. The initial state of all equations is taken to be the same.}
    \label{fig:rabi-models}
\end{figure}

In order to showcase the time-coarse graining method, we focus on this ultrastrong-coupling regime where the coupling strength $g$ is comparable to the TLS and cavity frequencies $\omega_{a}$ and $\omega_{c}$. i.e. $g \approx \omega_c \ (\omega_a)$. In that regime, the counter-rotating terms assume significance, and the induced dynamics depend on the time resolution of the measurement apparatus. The numerical simulations in this subsection assume the following set of parameters:
\begin{equation}
\begin{split}
    \frac{\omega_{c}}{2\pi}
    =
    \frac{\omega_{a}}{2\pi}
    =
    2 \textrm{GHz};
    \qquad
    \frac{g}{2\pi} = 0.4 \textrm{GHz}.
\end{split}
\end{equation}
Under this set of parameters, the TLS-state population displays rapid oscillations whose envelope undergoes intricate evolution over a much longer time scale, as shown in Fig. \ref{fig:rabi-models}. When observing the dynamics of a coherent-state cavity mode interacting with a single atom, the Jaynes-Cummings model is known to show collapse-revival dynamics, due to the photon-number dispersion of a coherent state. Interestingly, we show that these collapse-revival cycles are completely absent in the full Rabi-model dynamics (if all time-scales can be resolved), and only appear under finite-time resolution. The latter is directly captured by the STCG approach, as discussed below. 

Using the STCG method, we obtain an effective description that gives us \textbf{directly} the time-averaged observables that would be obtained from a bandwidth-limited measurement apparatus, and we do not need to assume any knowledge about the density matrix $\rho(t)$ at an infinitely precise moment in time. The STCG prescription produces a set of Hamiltonian corrections, comprised of products of the original Hamiltonian operators $\hat{s}_{\vec{\mu}} = \prod_{\mu_i \in \vec\mu} \hat{h}_{\mu_i}$  (i.e. multi-body transitions or multi-mode resonances), and their corresponding coupling strengths $q_{\vec{\mu}}$. The time-coarse grained evolution is not unitary in general, and the TCG effective Hamiltonian has to be complemented with a set of pairs of operators $(\hat{L}_{\vec\mu}, \hat{J}_{\vec\nu})$ with (generally) complex coefficients $\gamma_{\vec{\mu}, \vec{\nu}}^{(k)}$ such that the time evolution of $\overline{\rho}(t)$ is given by the following TCG master equation:
\begin{equation}
\begin{split}
\partial_{t} \overline{\rho}(t)
=&
-
\big[
\hat{H}_{\textrm{TCG}}(t)
,
\overline{\rho}(t)
\big]
+
\hat{D}_{\textrm{TCG}}(t)
\overline{\rho}(t) \\
=&
-
\big[
\sum_{k=1}^{\infty}
\hat{H}_{\textrm{TCG}}^{(k)}(t)
,
\overline{\rho}(t)
\big]
+
\sum_{k=1}^{\infty}
\hat{D}_{\textrm{TCG}}^{(k)}(t)
\overline{\rho}(t)
\end{split}
\end{equation}
\begin{subequations}
    \begin{align}
    \hat{H}_{\rm{TCG}^{(k)}} &= q_{\vec{\mu}}^{(k)} e^{-i (\sum_{j} \mu_{j}) t} \prod_{\mu_{i}} h_{\mu_{i}}^{(k)} \\ 
    \hat{D}^{(k)}_{\rm{TCG}}(t) \overline{\rho} &= \gamma_{\vec{\mu}, \vec{\nu}}^{(k)}
\mathcal{D}\big[ \hat{L}_{\vec{\mu}}, \hat{J}_{\vec{\nu}} \big]
e^{-i (\sum_{j_{1}} \mu_{j_{1}} + \sum_{j_{2}} \nu_{j_{2}}) t} 
    \end{align}
\end{subequations}
In the expressions above, $k = l + r$ is the order of the perturbative expansion in the original coupling strengths, and the (pseudo-)dissipator notation $\mathcal{D}\big[\cdot, \cdot\big]$ is defined such that
\begin{equation}
\mathcal{D}\big[ \hat{L}_{\vec{\mu}}, \hat{J}_{\vec{\nu}}
\big] \overline{\rho}
\equiv
\hat{L}_{\vec{\mu}} \overline{\rho} \hat{J}_{\vec{\nu}}
-
\frac{1}{2}
\big\{
\hat{J}_{\vec{\nu}} \hat{L}_{\vec{\mu}}
,
\overline{\rho} \big\}.
\end{equation}
We explain details of the calculation in the following sections, but for now we simply assume that the TCG effective Hamiltonian and effective (pseudo-)dissipators can be perturbatively calculated as functions of the coarse-graining time scale $\tau$ and the original Hamiltonian, for example by the symbolic software package \emph{QuantumGraining.jl} we developed. The full explicit calculation up to second-order is shown in the appendix, App.\ref{app:rabi-model_contraction_coefficients}. 

In order to effectively capture the coarse-grained dynamics of the Rabi model, we apply the STCG perturbation theory up to the third order and derive the corresponding master equation. In particular, we will see that the TCG procedure reproduces the RWA Hamiltonian at the first-order (under simplifying assumptions for the filter function), and goes beyond it starting at the second-order. For example, in the weak coupling limit, realistic values of the coarse-graining time scale $\tau$ usually falls in the range $\frac{1}{\omega_{a}} \ll \tau \ll \frac{1}{\abs{\omega_{c}-\omega_{a}}}, \frac{2}{g}$, and the most significant terms in the effective Hamiltonian are found to be,
\begin{equation}
\begin{split}
\hat{H}_{\rm TCG}^{(2)}
\approx&
\hat{H}_{\rm{RWA}}
+
\frac{g^{2}}{8} \Big[
\frac{1}{2\omega_{a}}
-
\big(
\tau^{2}
+
\frac{1}{4\omega_{a}^{2}}
\big) (\omega_{a} - \omega_{c})
\Big]\\
&\qquad\qquad\cdot
(1 + 2 \hat{a}^{\dagger} \hat{a}) \cdot \hat{\sigma}_z.
\end{split}
\end{equation}
where at the first order, we get exactly the RWA Hamiltonian
\begin{equation}
\label{eq: Rabi_HRWA}
\begin{split}
\hat{H}_{\textrm{RWA}}
=&
\frac{g}{2}
e^{-\frac{(\omega_{a}-\omega_{c})^{2}\tau^{2}}{2}}
\Big(
e^{i(\omega_{a}-\omega_{c})t} \hat{a} \hat{\sigma}_{+}
+
e^{-i(\omega_{a}-\omega_{c})t} \hat{a}^{\dagger} \hat{\sigma}_{-}
\Big)\\
\approx&
\frac{g}{2}
\Big(
e^{+i(\omega_{a}-\omega_{c})t} \hat{a} \hat{\sigma}_{+}
+
e^{-i(\omega_{a}-\omega_{c})t} \hat{a}^{\dagger} \hat{\sigma}_{-}
\Big).
\end{split}
\end{equation}
Here the filter-dependent factor $e^{-\frac{(\omega_{a}-\omega_{c})^{2}\tau^{2}}{2}}$ in the first line is a major difference from the standard RWA. Unlike the RWA, we consider the coarse-graining time-scale $\tau$ as an experimentally tunable parameter, and the standard RWA is subject to significant modification once $\tau$ is large enough to be comparable to $\frac{1}{\abs{\omega_{a} - \omega_{c}}}$. In general, as the coupling strength $g$ approaches the ultrastrong-coupling regime or when the time resolution $\tau^{-1}$ becomes comparable with the detuning $\big| \omega_{a} - \omega_{c} \big|$, one needs to go beyond the simple expression in Eq.(\ref{eq: Rabi_HRWA}) and consider contributions from higher-order terms. Considering such a situation, we present the exact form of the Hamiltonian and (pseudo-) dissipator corrections calculated using the QuantumGraining.jl package in App. \ref{app:TCG-derivation}.

Apart from the Hamiltonian corrections, a unique property of the STCG method is that it also captures the effective non-unitary dynamics of the system, which would be missed by effective-static Hamiltonian methods~\cite{venkatraman_static_2022}. For example, at the second order, we get the following set of pseudo-dissipators in addition to the dispersive correction in the Hamiltonian,
\begin{subequations}
\begin{align}
\hat{L}_1 &= \hat{J}_1 = \hat{a} \hat{\sigma}_+ &
\gamma_1 &= - i \frac{g^2 \tau^2}{2} (\omega_c - \omega_a) e^{-2i(\omega_c - \omega_a) t} \\
\hat{L}_2 &= \hat{J}_2 = \hat{a}^\dagger \hat{\sigma}_- &
\gamma_2 &= i \frac{g^2 \tau^2}{2} (\omega_c - \omega_a) e^{2i(\omega_c - \omega_a) t}.
\end{align}
\end{subequations}
Notice that the coefficients $\gamma_{1}$ and $\gamma_{2}$ are purely imaginary, so the corresponding pseudo-dissipators do not break the time reversal symmetry (hence the prefix ``pseudo-''). In addition, we do not expect significant secular effects from the second-order pseudo-dissipators since their coefficients are oscillatory at frequency $\pm 2(\omega_{c} - \omega_{a})$ and vanish in the resonant limit when $\omega_{c} \rightarrow \omega_{a}$. In particular, they do not induce any secular gain or loss of the system energy, but rather only add small fluctuating corrections to the entropy and energy of the system.

However, this is not the case for higher-order corrections in general. For example, we have the following third-order contributions in the resonant limit where $\omega_{c} \rightarrow \omega_{a}$
\begin{equation}
\begin{split}
\hat{H}_{\textrm{TCG}}^{(3)}
\approx
\lim_{\omega_{c} \rightarrow \omega_{a}}
\hat{H}_{\textrm{TCG}}^{(2)}
-
\frac{g^{3}}{32 \omega_{a}^{2}} \hat{a}^{\dagger} \hat{a} \hat{a}^{\dagger} \hat{\sigma}_{-}
+
h.c.
\end{split}
\end{equation}
\begin{equation}
\begin{split}
\hat{D}_{\textrm{TCG}}^{(3)}
\approx
\frac{i g^{3}}{32 \omega_{a}^{2}}
\Big(
\mathcal{D}[\hat{a}^{2} \hat{\sigma}_{z}, \hat{a}^{\dagger} \hat{\sigma}_{+}]
-
\mathcal{D}[\hat{a}^{\dagger 2} \hat{\sigma}_{z}, \hat{a} \hat{\sigma}_{-}]
\Big)
+
h.c.
\end{split}
\end{equation}
from which we see that both the unitary and non-unitary contributions give rise to time-independent corrections to the Jaynes-Cummings model in the resonant limit.
From the numerical simulation in Fig.\ref{fig:rabi-models} (obtained with a coarse-graining time scale of $\tau = 0.2 \textrm{ns}$), we see that they both have secular effects on the collapse and revival of TLS-state population. In particular, although $\hat{D}_{\textrm{TCG}}^{(3)}$ does not cause any dissipation of energy over long periods of time due to its purely imaginary pre-factor, it does have observable secular effects on the \textbf{coherence} of the TLS, which affects the collapse-revival pattern of the TLS-state population. For example, both the RWA (first-order TCG) and the third-order TCG Hamiltonians make the false prediction of a double-revival pattern between $t=15\textrm{ns}$ and $t=35\textrm{ns}$, which is removed by including the third-order pseudo-dissipators in $\hat{D}_{\textrm{TCG}}^{(3)}$. As expected, inclusion of the pseudo-dissipators brings the TCG dynamics closer to the exact dynamics, as can be observed in Fig.\ref{fig:rabi-models}. The TCG master equations can be numerically simulated with much larger time steps without encountering any stiffness problems compared to the exact von-Neumann equation which contains fast-oscillating counter-rotating terms. Therefore, the TCG master equation not only offers analytical insights into the physics of an interacting system, but also allows much more efficient numerical simulation of the dynamics.

\section{Effective static methods and time-coarse graining }
\label{sec:theory}
In this work, we focus on time-coarse graining Hamiltonian dynamics, i.e., closed-system dynamics. In an accompanying paper \cite{wentao_transmon_tcg}, we delve into the more general problem of time-coarse graining an open quantum system, how the time-coarse graining affects the bath interaction, and the implications of this on the measurement and back-action dynamics. 
In general terms, we consider interaction-picture Hamiltonians $\hat{H}$ of the form
\begin{equation}
\label{eq: H decomp}
\hat{H}
=
\sum_{\omega \in \Omega} g_\omega \hat{h}_{\omega} e^{-i \omega t}
\end{equation}
where $\Omega$ is the set of all the frequencies appearing in the Hamiltonian. . The frequencies involved can be of vastly different scales, and we would like to obtain an effective, low-bandwidth description of the resulting dynamics. For example, the well-known rotating-wave approximation (RWA) corresponds to simply removing all Hamiltonian terms $\hat{h}_{\omega}$ with ``fast'' frequencies from the sum, and retaining those with ``slow'' frequencies.  

\begin{figure}
    \centering
    \includegraphics[width=0.9\linewidth]{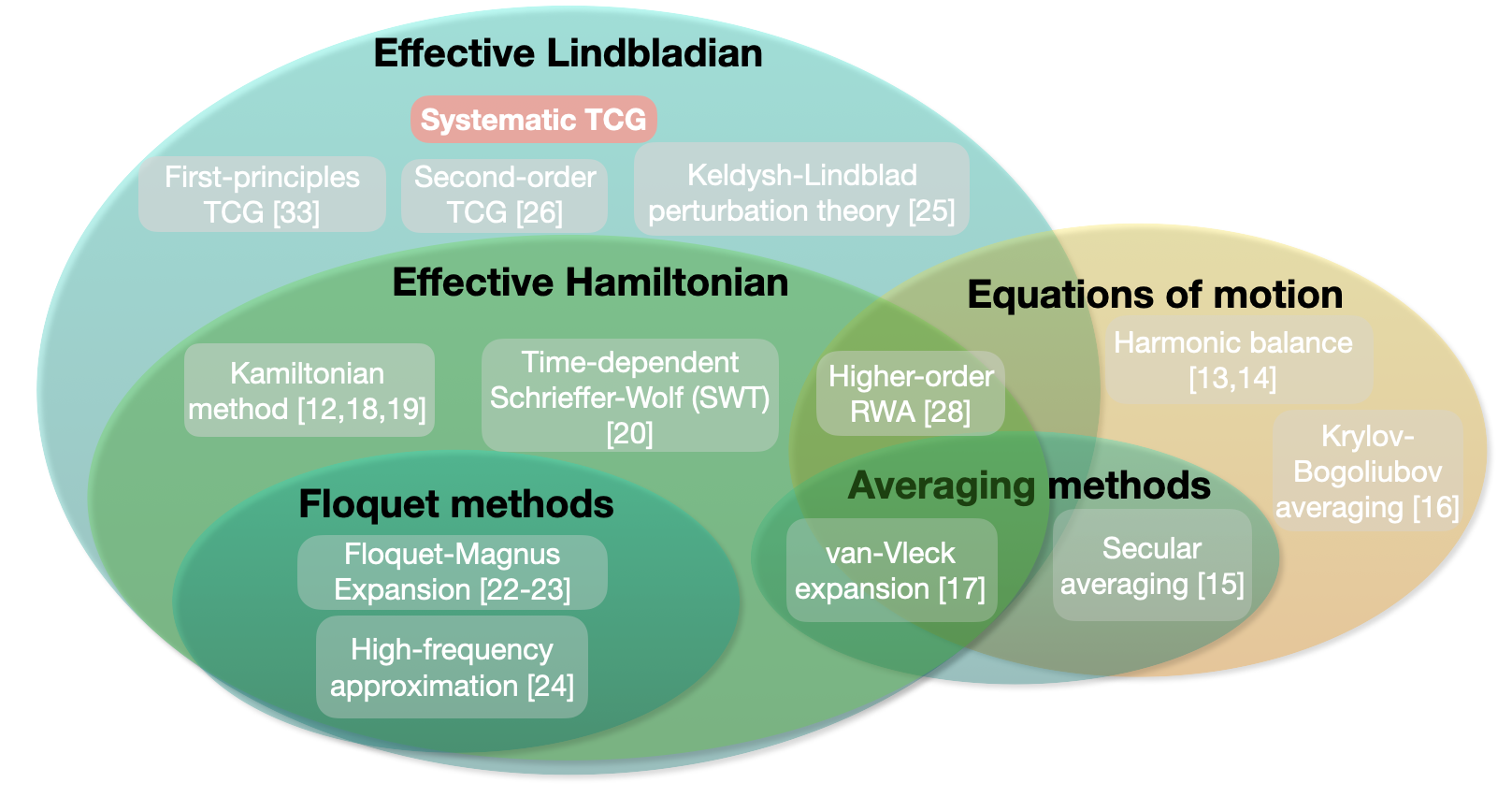}
    \caption{A Venn diagram of different methods dealing with time-dependent driven systems. The bold black titles indicate the level of operation of the method, and the grey squares indicate different methods. TCG generalizes many current methods to include non-unitary contributions. Systematic TCG (this work, in red) permits the TCG calculation to any order, and for more general regimes.}
    \label{fig:methods_comp}
\end{figure}

In the literature, the problem of fast-varying time-dependent Hamiltonians is treated by a variety of tools, as illustrated by the Venn diagram on Fig. \ref{fig:methods_comp}. Some of these methods rely directly on performing the averaging or ansatz at the equations of motion level \cite{krack_harmonic_2019,kosata_harmonicbalancejl_2022, buishvili_higher_1981, krylov_introduction_1947}, but importantly, many methods rely on producing an effective, low-frequency description of the Hamiltonian \cite{rahav_effective_2003}. The most relevant ones to this work include the ``Kamiltonian" method \cite{cary_lie_1981, grozdanov_quantum_1988, venkatraman_static_2022} inspired by similar methods in classical analytical mechanics and plasma physics, and methods based around a systematic perturbation theory leveraging the Schrieffer-Wolff transformation \cite{xiao_perturbative_2022} and the Floquet theory \cite{shirley_solution_1965, wilcox_exponential_1967, casas_floquet_2001}, as well as high-frequency expansions \cite{eckardt_high-frequency_2015}. These methods produce an effective static Hamiltonian description that encodes some of the effects of the fast dynamics in a low-frequency description. 
Other methods describe the time-coarse grained evolution in terms of a Liouville equation; Notable examples are the Keldysh-Lindblad expansion \cite{muller_deriving_2017} and the TCG method \cite{gamel_time-averaged_2010, lee_effective_2018, wentao_transmon_tcg} which can be thought of as a generalization of the effective Hamiltonian methods. The TCG method also captures non-unitary effects that accounts for the information lost during coarse-graining. This is achieved by a general approach to take a moving average over the expanded generator, rather than the Hamiltonian. Among other things, this allows us to consider off-resonant virtual transitions as well as resonant multi-body transitions, such as two-photon processes, which are omitted in the (first-order) RWA \cite{eckardt_dynamics_2015}. In addition, the resulting terms are weighted in a way that is fine-tuned to the chosen measurement resolution and the form of the filter function. STCG differs from effective Hamiltonian approaches in several ways: \textbf{(1)} Even starting with a Hamiltonian of a system with finite degrees of freedom it produces a Liouvillian that captures the non-unitary effects due to time-averaging, as illustrated in Fig.  \ref{fig:tcg-tikz}, giving a more complete description of the low-frequency dynamics. As we will show with examples, the non-unitary effects can be important in the presence of strong and/or time-varying drives, or certain initial conditions \textbf{(2)} TCG is generalizable to systems with infinitely many degrees of freedom~\cite{chruscinski_brief_2017,manzano_short_2020,cresser_coarse-graining_2017} producing an effective non-Markovian dynamics of a subsystem consistent with the coarse-graining time scale desired, \textbf{(3)} The coarse-graining time scale can be used to justify rigorously the Markov approximation when integrating out part of the system, as we discuss thoroughly in the accompanying paper \cite{wentao_transmon_tcg}, \textbf{(4)} TCG is generalizable to arbitrary filter function $f_\tau(\omega)$ applied by the measurement device. Our work, the systematic time-coarse graining (STCG), provides the first systematic framework for generalizing the RWA and other effective Hamiltonian methods into a Lindblad-like generator, and allows time-coarse graining (TCG) to be performed to arbitrary degree of accuracy when suitable conditions are satisfied. Time-coarse graining has been introduced in Refs.~\cite{james_effective_Hamiltonian, gamel_time-averaged_2010, lee_effective_2018,majenz_coarse_2013} and the time-coarse grained generator has been explicitly derived up to the second order in the coupling constants, under the assumption that the transition frequencies are either much greater or much smaller than $\tau^{-1}$. Our method, as presented in this work, is novel in the fact that it is fully systematic, supported by a diagrammatic expansion, and is capable of explicitly calculating the effective TCG Liouvillian at any order in the coupling constants, while allowing for the most general coarse-graining time scale $\tau$.

\section{Systematic time-coarse graining by diagrammatic expansion}
\label{sec:diagrammatic expansion}
Since the time-coarse graining procedure inherently erases information, the time evolution of $\overline{\rho}(t)$ cannot be generated by a Hamiltonian in general. In fact, we can only assume that the equation of motion for $\overline{\rho}(t)$ is linear and trace-preserving, since these two properties are respected by both the time-coarse graining process and the underlying von-Neumann equation. Consequently, the correct ansatz is a Liouville-like equation which we will refer to as the ``TCG master equation''. In order to obtain a perturbative formula for $\mathcal{L}$, we assume that this Liouvillian can be expanded as a Dyson series in powers of the Hamiltonian $\hat{H}$,
\begin{equation}
    i\frac{\partial}{\partial t} \overline{\rho}(t) = \mathcal{L} \overline{\rho}(t) = \sum_{k=1}^{\infty} \mathcal{L}_k(t) \overline{\rho}(t).
    \label{eq:TCG-definition}
\end{equation}
this expansion requires the assumption of small $\hat{H}$, which often implies that we need to work in a certain interaction picture, and the TCG effective master equation thus obtained depends implicitly on our choice of the interaction picture. We note that the choice of the interaction picture is partially determined by the measurement channel through which the quantum system is observed, we explain that in more detail in the appendix of the accompanying paper \cite{wentao_transmon_tcg}. The resulting dynamics of observables can be shown to be equivalent to the dynamics observed through a band-limited measurement apparatus with a filter function $\tilde{f}_\tau (t)$, with $\tau$ parameterizing the coarse-graining time-scale. The choice of the interaction picture is thus rendered consequential for the observed physics as it represents the measured dynamics by a given measurement apparatus. The way in which this is captured through STCG is through the choice determining the frequencies $\omega \in \Omega$ and the associated forms of the different operators $\hat{h}_{\omega}$ in Eq.(\ref{eq: H decomp}). These operators appear in conjugate pairs $\hat{h}_\omega = \hat{h}_{-\omega}^\dagger$ due to the hermicity of the Hamiltonian.

\begin{figure}[!h]
\label{fig:tcg-tikz}
\centering
\begin{tikzpicture}
\node[left] at (0,0) (Rho) {\larger[2]$\boldsymbol{\rho(t)}$};
\draw[very thick, ->] (Rho.east) -- ++(4.5cm,0) node[above, pos=0.5] {$-i\big[\hat{H}, \CIRCLE\big]$} node[right] (dRhodt) {\larger[2]$\boldsymbol{\dot{\rho}(t)}$};
\draw[very thick, ->] (Rho.south) -- ++(0,-3.0cm) node[left, pos=0.5] {TCG} node[below] (RhoBar) {\larger[2]$\boldsymbol{\overline{\rho}(t)}$};
\draw[very thick, ->] (dRhodt.south) -- ++(0,-3.0cm) node[right, pos=0.5] {TCG} node[below] (dRhoBardt) {\larger[2]$\boldsymbol{\dot{\overline{\rho}}(t)}$};
\draw[very thick, dashed, ->] (RhoBar.east) -- ++(4.5cm,0) node[below, pos=0.5] {$-i \mathcal{L} = -i\big[\hat{H}_{\textrm{RWA}}, \CIRCLE\big] + \cdots$};
\end{tikzpicture}
\end{figure}

Our goal here is to find a closed-form formula for the partial Liouvillian $\mathcal{L}_{k}$ at each order $k$, which gives us an effective master equation describing the TCG dynamics as long as the Hamiltonian can be assumed to be small for the part of Hilbert space we are interested in, where ``small'' means that the coupling rates $g_\omega$ are small either in comparison to all energy differences \textbf{or} in comparison to the time-coarse graining scale $\tau$.
\begin{subequations}
\begin{align}
    &g_\omega \ll \abs{\omega - \omega'} \ \ \  \forall \ \omega, \ \omega' \\ 
    &g_\omega \ll 1/\tau \ \ \  \forall \ \omega
\end{align}
\end{subequations}
This is similar to how the RWA Hamiltonian gives an approximate description of the time-coarse grained evolution. Indeed, we will later show that for an appropriate choice of the moving-average function, the von-Neumann equation with the RWA Hamiltonian is exactly the TCG master equation truncated at the first order.

Low-order formulas for $\mathcal{L}_{k}$ have been obtained by iteratively and perturbatively solving the coupled equations of the maps $\rho(0) \rightarrow \overline{\rho}(t)$ and $\overline{\rho}(t) \rightarrow \rho(0)$ \cite{gamel_time-averaged_2010, cresser_coarse-graining_2017}. The increasing complexity of the coupled equations prevents efficient calculation of $\mathcal{L}_{k}$ for $k>3$, and the resulting expressions of $\mathcal{L}_{k}$ are seemingly structure-less nested averages of the form
\begin{equation}
\label{eq: L3_nested_averages}
\begin{split}
\mathcal{L}_{3} \overline{\rho}
=
\overline{\hat{H} \hat{U}_{2}} \overline{\rho}
-
\overline{\hat{H}} \overline{\hat{U}_{2}} \overline{\rho}
-
\overline{ \hat{H} \overline{\hat{U}_{1}} \overline{\rho} \hat{U}_{1}^{\dagger} }
+
2 \overline{\hat{H}} \overline{\hat{U}_{1}} \overline{\rho} \overline{\hat{U}_{1}^{\dagger}}
+
\cdots
\end{split}
\end{equation}
where we use the overline to denote the time average of any function of $t$ defined in the same fashion as in Eq.(\ref{eq: def_of_rhobar}), and $\hat{U}_{n}(t)$ is defined as  the $n$-th order term in the Dyson series
\begin{equation}
\begin{split}
\hat{U}_{n}(t)
&=
-
i \int_0^t dt^{\prime} \hat{H}(t^{\prime})\hat{U}_{n-1}(t^{\prime}) \\
&=
\sum_{j=0}^n (-i)^j \int_0^t \cdots \int_0^{t_{j-1}} dt_1 \cdots dt_j \cdot
\hat{H}(t_1) \cdots \hat{H}(t_j).
\end{split}
\end{equation}
However, if one explicitly calculates the nested averages in Eq.(\ref{eq: L3_nested_averages}), one would immediately find massive cancellation among the terms, as shown in the appendix App. \ref{app:harmonic_dependence}. In fact, this massive cancellation becomes much more significant at higher orders, which indicates hidden structures in the nested averages that can be utilized to significantly simplify the algebra.

In this work, we develop the TCG method into a systematic perturbation theory where the structure of $\mathcal{L}_{k}$ is explored in detail, leading to an explicit closed-form formula for $\mathcal{L}_{k}$ which can be calculated by the QuantumGraining.jl package at arbitrary order.
The first crucial step towards achieving this objective is the following recursive formula for $\mathcal{L}_{k}$ the derivation of which can be found in App. \ref{app:TCG-derivation}:
\begin{equation}
\label{eq:TCG-raw}
\begin{split}
    \mathcal{L}_k(t) \overline{\rho} &=
\sum_{l=1}^{k} \left ( \overline{\hat{H} \hat{U}_{l-1} \overline{\rho} \hat{U}^\dagger_{k-l}}(t) - \overline{\hat{U}_{l-1} \overline{\rho} \hat{U}_{k-l}^\dagger \hat{H}}(t) \right ) \\
&-
\sum_{k^{\prime}=0}^{k-1} \mathcal{L}_{k^{\prime}}(t) \sum_{l=0}^{k-k^{\prime}} \overline{\hat{U}_l \overline{\rho} \hat{U}_{k-k^{\prime}-l}^\dagger}  
\end{split}
\end{equation}
where we define $\mathcal{L}_{0} \overline{\rho} \equiv 0$ for any density matrix $\overline{\rho}$. Starting from Eq. (\ref{eq:TCG-raw}), one can show (see the derivation in App. \ref{app:contraction_coefficients}) that the terms in $\mathcal{L}_{k}$ can be rearranged into ``contraction superoperators'' which allow both diagrammatic representation and closed-form formulas. The following subsection is dedicated to the discussion of these contraction superoperators.

\subsection{Diagrammatic expansion and the contraction coefficients}

In Lee et al. \cite{lee_effective_2018}, the corrections $\Scale[0.8]{\overline{\hat{H} \hat{U}_{1}}} \overline{\rho}$ and $- \Scale[0.8]{\overline{\hat{H}} \cdot \overline{\hat{U}_{1}}} \overline{\rho}$ in $\mathcal{L}_{2} \overline{\rho}$ are grouped together and denoted by the contraction $\wick{\c{\Scale[0.8]{\hat{H}}} \c{\Scale[0.8]{\hat{U}_{1}}}} \overline{\rho}$  for convenience. The same is true for two-point contractions of the type $\wick{\c{\Scale[0.8]{\hat{H}}} \overline{\rho} \c{\Scale[0.8]{\hat{U}_{1}^{\dagger}}}} \equiv \overline{\Scale[0.8]{\hat{H}} \overline{\rho} \Scale[0.8]{\hat{U}_{1}^{\dagger}}} - \overline{\Scale[0.8]{\hat{H}}} \overline{\rho} \overline{\Scale[0.8]{\hat{U}_{1}^{\dagger}}}$. Assuming a Hamiltonian of the form in Eq.(\ref{eq: H decomp}), we first observe that the cancellation pointed out in the previous section takes place within each such contraction in $\mathcal{L}_{2}$. Furthermore, each of the surviving superoperators will have a frequency equal to the sum of the frequencies of the operators involved, which is not true for the individual nested averages constituting the contraction. We show in App. \ref{app:contraction_coefficients} that contractions must be generalized to exhibit this mass-cancellation and homogeneous time-dependence at all orders. An unexpected byproduct of this finding is that a closed-form expression can be found that is amenable to symbolic computation in a generic manner. A detailed derivation is presented in Appendices App. \ref{app:contraction_coefficients} and App. \ref{app:harmonic_dependence}. Here we summarize the key steps in the derivation. 

According to Eq. (\ref{eq:TCG-raw}), the nested averages in $\mathcal{L}_{k} \overline{\rho}$ always have $\hat{H}(t)$ on either the left or the right end of the product. Therefore, we make the ansatz that the partial Liouvillian $\mathcal{L}_{k}$ at order $k \ge 1$ can be expressed as a sum over all such generalized contractions with weights $(l,r)$ such that $l+r=k$:
\begin{subequations}
\label{eq:TCG-Lindblad}
\begin{align}
\mathcal{L}_{k} \overline{\rho}(t)
&
=
\sum_{r = 0}^{k} \mathcal{W}_{k-r,r}(t)[\overline{\rho}]
-
h.c.
\end{align}
\end{subequations}
where each contraction $\mathcal{W}_{l,r}(t)[\overline{\rho}]$, whose form is to be found, is the sum of all nested averages that have $l$ ($r$) operators to the left (right) of $\overline{\rho}$ with the left-most operator being $\hat{H}$. As shown in App. \ref{app:contraction_coefficients}, these contractions can always be written as a sum of operator products with a simple homogeneous harmonic time-dependence.

\begin{figure}[!h]
    \centering
    \includegraphics[width=0.6\textwidth]{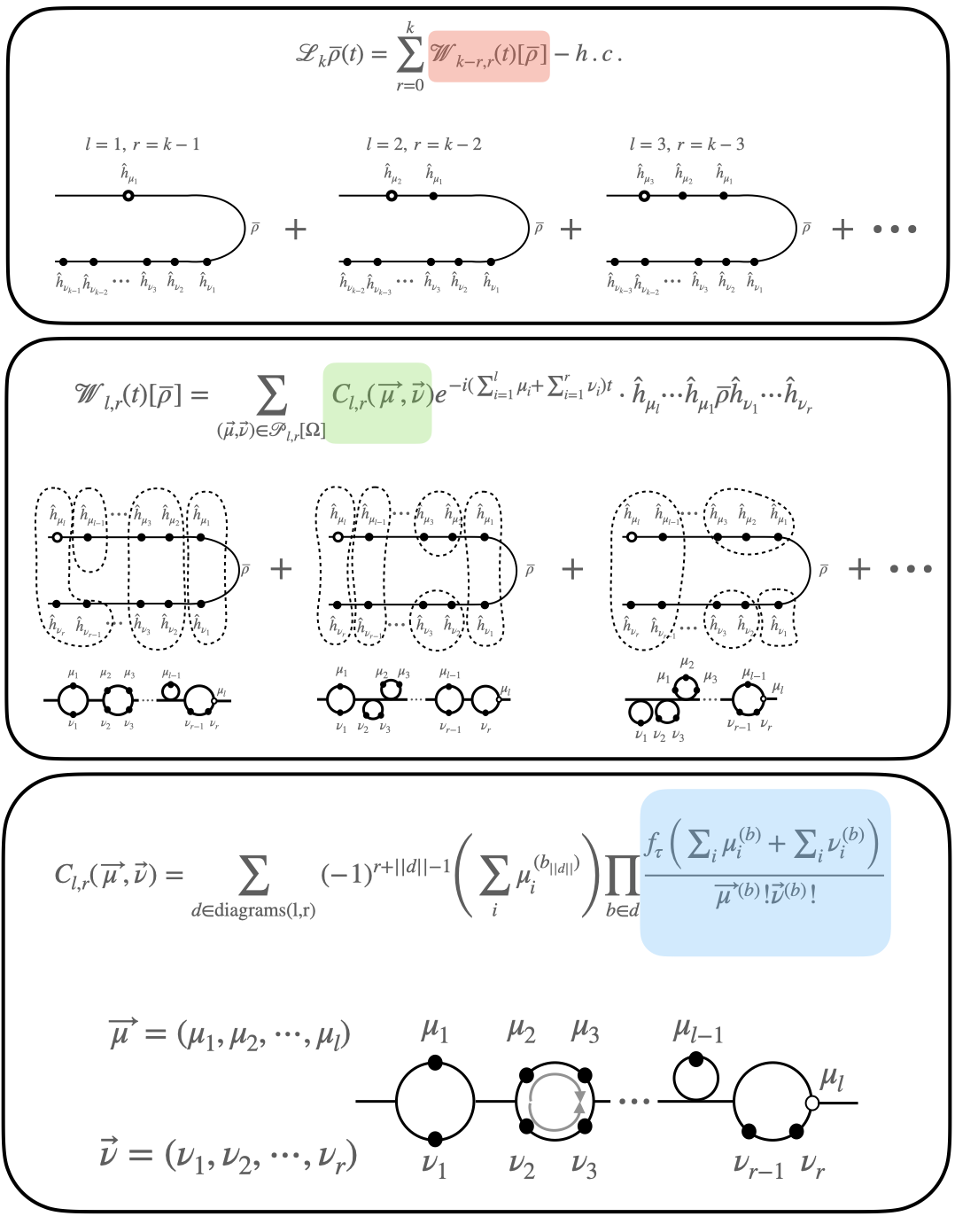}
    \caption{\textbf{(top)} The TCG Liouvillian can be written as a sum of contraction superoperators $\mathcal{W}_{k-r,r}$, each representing a superoperator with $l$ operators to the left and $r$ operators to the right of $\overline{\rho}$, such that $l+r=k$. \textbf{(middle)} The contribution of each superoperator term is weighted by the contraction coefficients $C_{l,r}$, which encode the contributions from all the different ways one can partition the operators into groups. We represent these partitions with "bubble" diagrams (note that the order is inverted). \textbf{(bottom)} The contribution of each diagram is computed in STCG through a closed-form formula, where each bubble contributes a factor to a product.}
    \label{fig:tcg-anatomy}
\end{figure}

We next write each contraction in the following form:
\begin{equation}
\begin{split}
    \mathcal{W}_{l,r}(t)[\overline{\rho}]
&=
\sum_{(\vec{\mu}, \vec{\nu}) \in \mathcal{P}_{l,r} [\Omega]} C_{l,r}(\vec{\mu}, \vec{\nu}) \cdot \hat{h}_{\mu_{l}} \cdots \hat{h}_{\mu_{1}} \overline{\rho} \hat{h}_{\nu_1} \cdots \hat{h}_{\nu_r} \\ 
&\qquad\cdot
e^{-i(\sum_{i=1}^l \mu_i + \sum_{i=1}^r \nu_i) t } .
\end{split}
\label{eq:contraction_expansion}
\end{equation}
where we split the frequencies into two vectors $\vec \mu$ ($\vec \nu$) for the frequencies of the modes to the left (right) of the density matrix, while the notation $\mathcal{P}_{l,r}[\Omega]$ denotes the set of all possible pairs of vectors $\vec{\mu}$ and $\vec{\nu}$ whose elements are chosen from the Hamiltonian frequencies. Apart from the combinatorics involved in the summation, calculation of the \textbf{contraction coefficients} $C_{l,r}(\vec{\mu}, \vec{\nu})$ is the central element of STCG as each of the resulting coefficients encodes contributions from the many different nested time-averages in $\mathcal{L}_{k}$ to a particular superoperator in the sum, as detailed in the appendix.

The harmonic expansion in Eq.(\ref{eq:contraction_expansion}) allows us to derive the main result of the STCG method. In fact, using the harmonic ansatz, we can interpret every diagram as contributing a simple term to the contraction coefficients for each combination of mode frequencies involved. And we are able to obtain the following closed-form expression for the contraction coefficient $C_{l,r}(\vec\mu, \vec\nu)$:
\begin{equation}
\label{eq:contraction-coeff}
\begin{split}
    C_{l,r}(\vec\mu, \vec\nu)
&=
\sum_{d \in \rm{diagrams}(l, r)} (-1)^{r + \norm{d} - 1} \left ( \sum_i \mu_i^{(b_{\norm{d}})} \right) \\ &\times \prod_{b \in d} 
\frac{f_\tau\left(\sum_i \mu_i^{(b)} + \sum_i \nu_i^{(b)} \right ) }{\vec{\mu}^{(b)}! \vec{\nu}^{(b)}!}.  
\end{split}
\end{equation}
where $d = \big( b_{1}, b_{2}, \cdots, b_{\norm{d}} \big)$ represents a particular diagram $d$ by the ordered set of bubbles in $d$ with $\norm{d}$ being the total bubble number, and the vector factorials $\vec{\mu}^{(b)}!$ and $\vec{\nu}^{(b)}!$ are defined in Eq.(\ref{eq:vector-factorial}). In other words, we enumerate over diagrams which partition the frequencies into bubbles $\{b\}$ with frequencies $\vec\mu^{(b)}$ and $\vec\nu^{(b)}$; for each diagram, we then calculate the product of all the bubble factors, and multiply the result by an extra factor of $\sum_i \mu_i^{(b_{\norm{d}})}$ and an overall sign. For a detailed proof of Eq.(\ref{eq:contraction-coeff}), see App. \ref{app:harmonic_dependence}. Note that the contraction coefficients are always real if the frequencies of all the original Hamiltonian terms are real. This property ensures that the corresponding TCG effective Hamiltonian and effective (pseudo-)dissipator in subsection \ref{sec:Ham and diss decomp} are Hermitian at any order $k$.

Plugging the expansion of the contraction superoperators (Eq.(\ref{eq:contraction_expansion}) into the TCG master equation \ref{eq:TCG-Lindblad}, we can write the Liouvillian in terms of the contraction coefficients:
\begin{equation}
\begin{split}
    \mathcal{L}_k(t) \overline{\rho}
    &=
    \sum_{l=1}^{k} \sum_{\vec{\mu}, \vec{\nu} \in \mathcal{P}_{l, k-l}[\Omega]}
    \Big[
    C_{l, k-l}(\vec{\mu}, \vec{\nu}) e^{-i(\sum_i \mu_i + \sum_i \nu_i)t} \\ 
    &\times \hat{h}_{\mu_{l}} \cdots \hat{h}_{\mu_{1}} \rho \hat{h}_{\nu_1} \cdots \hat{h}_{\nu_r}
    -
    h.c.
    \Big].
\end{split}
\end{equation}
Note that in general, if we have a number of $\abs{\Omega}$ unique frequency-operator pairs in the original Hamiltonian, then for each $l \in \{1,2,\cdots,k\}$, the contraction would be a sum of $|\mathcal{P}_{l,r}[\Omega]| = |\Omega|^{l+r} = |\Omega|^{k}$ unique terms. To a large extent, this is where the complexity of STCG stems from, and why we complement the analytical formula with a symbolic computational tool.

\subsection{The effective TCG Hamiltonian and pseudo-dissipators}
\label{sec:Ham and diss decomp}

The structure of the contraction superoperators allows us to separate the different terms in the Liouvillian into Hamiltonian terms and pseudo-dissipator terms, akin to a Lindblad-type equation. In fact, with the help of the symmetry relation in Eq.(\ref{eq: C shifting}), we can write the TCG partial Liouvillian at order $k$ as $\mathcal{L}_{k} = \mathcal{L}_k^{H} + \mathcal{L}_k^{\gamma}$ with
\begin{subequations}
\label{eq: Lk_H}
\begin{align}
\mathcal{L}_k^{H} \overline{\rho}
&
=
[\hat{\mathcal{H}}_{k}, \overline{\rho}] \\
\begin{split}
    \mathcal{L}_k^{\gamma} \overline{\rho}
&
\equiv
i \hat{D}_{TCG}^{(k)} \overline{\rho}\\ 
&= \sum_{l=1}^{k-1} \sum_{\vec{\mu}, \vec{\nu} \in \mathcal{P}_{l,k-l}[\Omega]}
i \gamma_{\vec\mu, \vec\nu}^{(k)}
e^{-i\sum_j (\mu_j + \nu_j)t} \mathcal{D}[\hat{L}_{\vec{\mu}}, \hat{J}_{\vec{\nu}}] \overline{\rho}
\end{split}
\end{align}
\end{subequations}
where $\hat{H}_{k}$ and $\hat{D}_{TCG}^{(k)}$ can both be explicitly written in terms of the contraction coefficients $C_{l,r}(\vec\mu, \vec\nu)$ in Eq.(\ref{eq:contraction-coeff}).
As usual, this decomposition of $\mathcal{L}_{k}$ into Hermitian Hamiltonian terms and trace-preserving (pseudo-)dissipators is not the only option, and alternative representations also exist. One option is to absorb the anti-commutator terms into the Hamiltonian, portraying the dynamics as a (generally non-Hermitian) Hamiltonian with virtual jumps. While this perspective may prove convenient for specific problems, it doesn't alter the underlying physics. 

The Hamiltonian terms in Eq.(\ref{eq: Lk_H}) are one-sided, having only operators multiplied to the left or right of the density matrix. In the rest of this work, we represent $\hat{\mathcal{H}}_{k}$ as three vectors of equal length, where the $n$-th element of the vectors represent the coefficient, the operator, and the frequency of the $n$-th contraction term in $\hat{\mathcal{H}}_{k}$ respectively:
\begin{subequations}
\label{eq:TCG-hamiltonian}
\begin{align}
q^{(k)}_{\vec\mu}
&
= 
\Big\{ \frac{1}{2} \big( C_{k,0}(\vec{\mu}) + C_{k,0}(-\vec{\mu}^{\,\textrm{rev}}) \big) \prod_{\omega \in \vec{\mu}} g_\omega  \quad \vert \quad \vec{\mu} \in \mathcal{P}_{k,0}[\Omega] \Big\} \\
\hat{s}^{(k)}_{\vec{\mu}}
&
=
\Big\{ \hat{h}_{\mu_{k}} \hat{h}_{\mu_{k-1}} \cdots \hat{h}_{\mu_{1}} \quad \vert \quad \vec{\mu} \in \mathcal{P}_{k,0}[\Omega] \Big\}
\\
\Omega^{(k)}_{\vec{\mu}}
&
= \Big\{\sum_{\mu_i \in \vec{\mu}} \mu_i \quad \vert \quad \vec{\mu} \in \mathcal{P}_{k,0}[\Omega] \Big\}.
\end{align}
\end{subequations}
Here $\vec{\mu}^{\,\textrm{rev}} := (\mu_{k}, \mu_{k-1}, \cdots, \mu_{1})$ denotes the reverse of $\vec{\mu} = (\mu_{1}, \mu_{2}, \cdots, \mu_{k})$, and $\mathcal{P}_{l,r}[\Omega]$ is defined as the set of two-vector pairs of lengths $l$ and $r$ where each component is drawn from $\Omega$ with repetition. The total effective Hamiltonian to order $k$ can then be written as
\begin{equation}
\begin{split}
\hat{H}_{\rm{TCG}}^{(k)}(t)
&=
\sum_{k^{\prime}=1}^{k} \hat{\mathcal{H}}_{k^{\prime}}(t)
=
\sum_{k^{\prime}=1}^{k} \sum_{\vec{\mu} \in \mathcal{P}_{k^{\prime},0}[\Omega]} g^{(k^{\prime})}_{\vec{\mu}} e^{-i \sum_{j=1}^{k^{\prime}} \mu_{j} t}\\
&\qquad\qquad\qquad\quad\cdot
\hat{h}_{\mu_{k^{\prime}}} \hat{h}_{\mu_{k^{\prime}}-1} \cdots \hat{h}_{\mu_{1}}.
\end{split}
\end{equation}
The (pseudo-)dissipator terms, on the other hand, are generalizations of Lindblad dissipators which assume a more general form and can be written as $\mathcal{D}[\hat{L}, \hat{J}]$ for some operators $\hat{L}$ and $\hat{J}$ so that it acts on any density matrix $\rho$ as 
\begin{equation}
\label{eq:dissipator def}
    \mathcal{D}[\hat{L}, \hat{J}] \rho = \hat{L}\rho\hat{J} - \frac{1}{2} \{ \hat{J}\hat{L}, \rho \}.
\end{equation}
Notice that these (pseudo-)dissipators become the usual Lindblad dissipators if $\hat{J} = \hat{L}^{\dagger}$.  While the Hamiltonian terms only appear on one side of the density matrix, the generalized Lindblad dissipators $\mathcal{D}\big[\hat{L}_{\vec{\mu}}, \hat{J}_{\vec{\nu}}\big]$ act on both sides of $\overline{\rho}(t)$. Therefore, for each dissipator term, we need to specify a pair of operators $\big(\hat{L}_{\vec{\mu}}, \hat{J}_{\vec{\nu}}\big)$. Using similar notations as before, we can represent the $k$-th order effective dissipator terms by three vectors of the coefficients, the operator pairs $\big\{ \big(\hat{L}_{\vec{\mu}}, \hat{J}_{\vec{\nu}}\big) \big\}$, and the corresponding frequencies respectively:
\begin{subequations}
\label{eq:TCG-dissipators}
\begin{align}
\begin{split}
   i \gamma^{(k)}_{\vec{\mu}, \vec{\nu}}
&=
\Big\{ \left ( C_{l,r}(\vec{\mu}, \vec{\nu}) - C_{r,l}(-\vec{\nu}^{\,\textrm{rev}}, -\vec{\mu}^{\,\textrm{rev}}) \right ) \\ 
&\times \prod_{\mu_i \in \vec{\mu}} g_{\mu_i} \cdot \prod_{\nu_i \in \vec{\nu}} g_{\nu_i} \quad \vert \quad (\vec{\mu}, \vec{\nu}) \in \mathcal{P}_{l,r}[\Omega]  \Big\}  
\end{split} \\
\begin{split}
    \big( \hat{L}_{\vec{\mu}}, \hat{J}_{\vec{\nu}} \big)
    &
    =
    \Big\{ \big( \hat{h}_{\mu_{l}} \hat{h}_{\mu_{l-1}} \cdots \hat{h}_{\mu_{1}}, \hat{h}_{\nu_{1}} \hat{h}_{\nu_{2}} \cdots \hat{h}_{\nu_{r}} \big) \\ 
    &\quad \vert \quad (\vec{\mu}, \vec{\nu}) \in \mathcal{P}_{l,r}[\Omega]  \Big\}  
\end{split}
\\
\begin{split}
    \Omega_{\vec\mu, \vec\nu}
    &
    =
    \Big\{\sum_{\mu_i \in \vec\mu} \mu_i + \sum_{\nu_i \in \vec\nu} \nu_i \quad \vert \quad (\vec{\mu}, \vec{\nu}) \in \mathcal{P}_{l,r}[\Omega] \Big\}    
\end{split}
\end{align}
\end{subequations}
where we range over all positive integers $(l,r)$ such that $l+r=k$. 

According to the GKLS theorem \cite{lindblad76, gorini_completely_1976}, a necessary condition for the evolution of  $\overline{\rho}(t)$ to be a completely positive map, is that the matrix $\big[ \gamma_{ij} \big]$ is positive-definite. It is important to note that this condition is not satisfied by the TCG effective dissipators in general. 
This is not a problem as not all Hermitian positive semi-definite density matrices are possible in the coarse-grained world. For example, when the interaction-picture Hamiltonian $\hat{H}$ is time-dependent, there is no time-independent pure state $\rho$ that commutes with $\hat{H}$, and consequently the resulting coarse-grained $\overline{\rho}(t)$ cannot be a pure state in general. Indeed, one must keep in mind that $\overline{\rho}(t)$ does not represent the quantum state of the system at any particular moment in time, rather, it is a phenomenological object that allows one to calculate the time-coarse grained observables. Therefore, even when $\big[ \gamma_{ij} \big]$ is not positive-definite, the corresponding TCG master equation can still describe well-behaved time evolution without generating negative probabilities if the initial state is allowed in the time-coarse grained picture \cite{pechukas_reduced_1994, alicki_comment_1995, pechukas_pechukas_1995}. 

Summing over all these terms up to order $k$, we obtain the following total effective (pseudo-)dissipator:
\begin{equation}
\begin{split}
    \hat{D}_{\rm{TCG}}^{(k)}[\overline{\rho}](t) 
    &\equiv
    \sum_{l=1}^{k-1} \sum_{(\vec{\mu}, \vec{\nu}) \in \mathcal{P}_{l,r}[\Omega]}
    \gamma^{(k)}_{\vec{\mu}, \vec{\nu}} \cdot \mathcal{D}[\hat{L}_{\vec\mu}, \hat{J}_{\vec\nu}]\overline{\rho} \\
    &\qquad\cdot e^{-i (\sum_{j=1}^{l} \mu_{j} + \sum_{j=1}^{r} \nu_{j}) t}. 
\end{split}
\end{equation}
Note that for $k=1$, we get no effective dissipators since there are no double-sided terms, and the corresponding effective Hamiltonian is simply the time-coarse grained original Hamiltonian:
\begin{equation}
\begin{split}
\hat{H}^{(1)}_{\textrm{TCG}}(t)
=
\sum_{\omega} f_\tau(\omega) e^{-i\omega t} \hat{h}_{\omega}
=
\int_{-\infty}^{\infty} dt^{\prime} \tilde{f}_\tau(t-t^{\prime}) \hat{H}(t^{\prime}).
\end{split}
\end{equation}
For a physical coarse-graining window function $\tilde{f}_\tau (t)$, the Fourier transform $f_\tau(\omega) \equiv \int_{-\infty}^{\infty} dt \cdot \tilde{f}_{\tau}(t) e^{i\omega t}$ can be considered as a low-pass filter in the interaction picture, which manifestly suppresses the coefficients of the high-frequency terms in $\hat{H}(t)$. In the lab frame, however, the function $f_{\tau}(\omega)$ effectively applies a band-pass filter to all the terms in the Hamiltonian, as stated earlier in the introduction.
Therefore, ignoring the small terms in $\hat{H}^{(1)}_{\textrm{TCG}}(t)$ (rather than the ``high-frequency'' terms by some ambiguous standard) would give us the RWA Hamiltonian. In this sense, the RWA can be considered as the lowest-order TCG with suitable choices of the coarse-graining time scale to suppress the high-frequency terms.

At higher-orders, one typically begins to see the appearance of effective (pseudo-)dissipators. Depending on the quantum system, these (pseudo-)dissipators may either account for micro fluctuations in the entropy and energy due to time-coarse-graining, or represent secular energetic loss or gain through high-frequency channels or non-adiabatic effects, depending also on whether the rates are real or imaginary. According to our experience, the latter situation usually occurs in systems with resonant drive, parametric time-dependence, or external heat reservoirs. In fact, these (pseudo-)dissipators are an important novelty of the STCG method, since they cannot be properly accounted for by any effective Hamiltonian approaches commonly employed in such situations, such a Floquet or Kamiltonian methods. Rather, they represent corrections of a different type, which are usually ignored in effective theories but can prove to be important for modeling the long-time dynamics of certain quantum systems.

\section{Software framework}
\label{sec:software}
STCG is general and analytic, but it requires extensive symbolic manipulation since the number of terms grows rapidly with the order of approximation, making manual calculation of the TCG Lindladian inconvenient at best and infeasible at worst. To this end, we developed \textbf{`QuantumGraining.jl'}, a Julia package that automates the TCG process efficiently for any order, calculating a symbolic expression for the effective Liouvillian in the form of a Hamiltonian and generalized Lindblad operators. We opt for \textbf{Julia} \cite{bezanson_julia_2012} due to its high performance, robust symbolic packages, and thriving scientific community.  Our package can be used directly to aid analytic methods and pen-and-paper calculations, and can also be seamlessly integrated with other software packages such as \textit{QuantumCumulants.jl} \cite{plankensteiner_quantumcumulantsjl_2022} or \textit{QuantumOptics.jl} \cite{kramer_quantumopticsjl_2018} for efficient numerical computation.

At the core of the STCG method is the calculation of the contraction coefficients $C_{l,r}$ defined in the previous section. Remarkably, this can be done independently from the operator algebra of the system, and is therefore fully encoded in a separate TCG process. The operator algebra is used for expressing the original Hamiltonian in a simple way, and more importantly, for writing the final Hamiltonian and pseudo-dissipators. In the current version of the code, the operator algebra is handled using the \textit{QuantumCumulants.jl} package, while other symbolic functions are handled using the \textit{Symbolics.jl} package. The details of calculating the STCG contributions, such as calculating singular contributions and enumerating the contributing diagrams, are detailed in App. \ref{app:enumerating} and \ref{app:singular_corrections}.

Taking the Rabi model as an example, we first define the Hilbert space and the original model Hamiltonian as,
\begin{lstlisting}[language=Julia, mathescape=true]
    @variables g $\omega$c $\omega$a
    $\Omega$ = [-$\omega$c - $\omega$a, $\omega$c + $\omega$a, -$\omega$c + $\omega$a, $\omega$c - $\omega$a]
    gvec = (g/2).*[1, 1, 1, 1]
    
    # Hilbert space definitions (QuantumCumulants.jl)
    h_cav = FockSpace(:cavity)
    h_atom = NLevelSpace(:atom, (:g,:e))
    h = tensor(h_cav, h_atom)
    
    # Operator definitions
    @qnumbers a::Destroy(h) $\sigma$::Transition(h)
    $\sigma$m = $\sigma$(:e, :g); $\sigma$p = $\sigma$(:g, :e)
    hvec = [a*$\sigma$m, a'*$\sigma$p, a*$\sigma$p, a'*$\sigma$m]
\end{lstlisting}

With the Hamiltonian defined, we proceed to calculate the TCG Hamiltonian and dissipators,
\begin{lstlisting}[language=Julia, mathescape=true]
    g_eff, $\Omega$_eff = effective_hamiltonian(hvec, gvec, $\Omega$, order=2; as_dict=true)
    $\gamma$_eff, $\Omega$_eff = effective_dissipator(hvec, gvec, $\Omega$, order=2)
\end{lstlisting}
Behind the scenes, these two lines of code calculate all relevant contractions and automatically sum up their contributions. The objects returned are dictionaries, with the new operators as keys, and coupling strengths and frequencies as values. The results can either be printed out, for aiding analytical calculations, or converted to other packages for direct numerical calculation.

For example, for integrating with \textit{QuantumOptics.jl}, first define a new set of compatible operators,
\begin{lstlisting}[language=Julia, mathescape=true]
    # Units
    $\mu$s = 1; MHz = 1/$\mu$s
    
    ### QuantumOptics.jl definitions
    ha_qo = SpinBasis(1//2); hc_qo = FockBasis(100); 
    h_qo = hc_qo $\otimes$ ha_qo
    
    # Operator definitions
    $\sigma$p_qo = sigmap(ha_qo); $\sigma$m_qo = sigmam(ha_qo)
    a_qo = destroy(hc_qo)
    I_a = identityoperator(ha_qo); I_c = identityoperator(hc_qo)
    
    p_sym = [g, $\omega$c, $\omega$a]
\end{lstlisting}

With the operators defined, we can generate a `QuantumOptics.jl' object Hamiltonian and solve directly.
\begin{lstlisting}[language=Julia, mathescape=true]
    tspan = [0:0.01:120$\mu$s;]
    $\psi$0 = coherentstate(hc_qo, 4.5) $\otimes$ spinup(ha_qo)
    
    base_qc = [a, a', $\sigma$m, $\sigma$p, $\sigma$(:e, :e)]
    Id = [I_c, I_a]
    base_qo = [a_qo, a_qo', $\sigma$m_qo, $\sigma$p_qo, $\sigma$p_qo*$\sigma$m_qo, Id...]
    
    H = hamiltonian_function(g_eff, $\Omega$_eff, base_qc, base_qo, p_sym)
    
    args = [2$\pi$*0.2MHz, 2$\pi$*2MHz, 2$\pi$*2.1MHz, 0.2$\mu$s]   
    tout, $\psi$t = timeevolution.schroedinger_dynamic(tspan, $\psi$0, (t, $\psi$) -> H(t, $\psi$; args=args));
    
    plot(tout, real(expect(1, a_qo'*a_qo, $\psi$t)), lw=2.5, label="a'*a - 1");
    
    xlabel!("Time [$\mu$s]")
    ylabel!(L"$\langle n \rangle$")
\end{lstlisting}

A full example notebook, including integration with `QuantumCumulants.jl' and `QuantumOptics.jl' is available on the Github repository \cite{quantumgraining-github}.

\section{Examples}
\label{sec:examples}
To showcase the capabilities and use cases of STCG, we demonstrate it on two select examples. The driven Kerr-parametron, where we show how it can be used to analyze adiabatic and non-adiabatic effects, and the driven Duffing oscillator, where we show STCG produces all the terms produced by the Kamiltonian methods, while complementing them with additional non-unitary corrections.

\subsection{Driven Kerr-parametron}
\label{subsec: parametron}
The Kerr parametron is a phase-locked parametric oscillator, which bifurcates between two possible opposite phases when driven by an oscillating pump field at approximately twice their natural frequency. In the quantum regime, the parametron can exist in a superposition of these two phase-states, known as a cat-state, acting as an effective biased-error qubit \cite{grimm_stabilization_2020}. The cat-state generated is fragile, and the parametron decay causes it to quickly decohere, necessitating rapid controls. These rapid controls must adhere to two conflicting requirements. On the one hand, they require a large pump field to prevent unwanted non-adiabatic transitions. On the other hand, large pump fields degrade qubit coherence by introducing unwanted non-resonant rapidly oscillating terms (NROTs). This is a prime use-case for the time-coarse graining method, which allows us to generally consider the contribution of these NROTs. Previously, this trade-off and optimization has been studied in Masuda et al. \cite{masuda_controls_2021}, and we show here that our method reproduces their numerical results and generalizes their effective model to capture unitary and non-unitary corrections.

The parametron is composed of a SQUID-array resonator with $N$ SQUIDs, which can be represented in the interaction picture in the rotating-frame at $\omega_p/2$ by an effective Hamiltonian,
\begin{equation}
\begin{split}
        \hat{H} &= \big( \Delta + \chi \big) \hat{a}^\dagger \hat{a} - \frac{\chi}{12} \left ( \hat{a} e^{-i\frac{\omega_p}{2}t} + \hat{a}^\dagger e^{+i\frac{\omega_p}{2} t} \right)^4 \\
    &+ 2\beta (\hat{a}e^{-i\frac{\omega_p}{2}t} + \hat{a}^\dagger e^{i\frac{\omega_p}{2}t})^2 \cos(\omega_p t) 
\end{split}
    \label{eq:parametron_hamiltonian}
\end{equation}
where $\Delta \equiv \omega_c^{(0)} - \chi - \omega_p/2$ is the detuning from the``bare'' resonance frequency $\omega_c^{(0)} = \sqrt{8E_C E_J/N}$. In addition, $\chi = E_C/N^2$ is the nonlinear coefficient strength and $\beta = \omega_c^{(0)} \frac{\delta E_J}{8E_J}$ is the pump strength. This form of the Hamiltonian is correct in the regime $\chi \beta \ll \big( 
\omega_c^{(0)} \big)^{2}$.  As explained in \cite{masuda_controls_2021}, both the pump strength $\beta$ and the detuning $\Delta$ can be made time-dependent by simultaneously tuning $\omega_{c}^{0}$ and $\delta E_{J}$. In particular, Masuda et al. considers a particular driving schedule for $\beta$ and $\Delta$ for $0 \le t \le T$:
\begin{equation}
\begin{split}
\beta(t)
=
\beta_{0} \frac{t}{T}
;\qquad
\Delta(t)
=
\Delta_{0} \Big(
1 - \frac{t}{T}
\Big).
\end{split}
\end{equation}
 Although it is shown that the fidelity of the process depends significantly on the non-RWA terms in the Hamiltonian (also referred to as the ``NROTs''), there is no analytical formula which quantifies their long-term effects over a time scale much greater than $\omega_{p}^{-1}$. Moreover, in addition to the non-RWA terms, non-adiabatic effects also play a significant role in the performance of the device, and are considered in the literature to be completely independent from the non-RWA effects. Here we perform the analysis using the STCG, and show how both effects be captured by STCG, and that the two are not entirely separate phenomena. Following  Masuda et al. \cite{masuda_controls_2021},  we assume the following numerical values of the parameters, 
\begin{equation}
\label{eq: parametron parameter values}
\begin{split}
&\omega_{p} / 2\pi = 16 \textrm{GHz}
;\qquad
\Delta_{0} / 2\pi =-67 \textrm{MHz}
;\\
&\beta_{0} / 2\pi
=
200 \textrm{MHz}
;\qquad
\chi / 2\pi = 68 \textrm{MHz}
\end{split}
\end{equation}
with $T= 50 \textrm{ns}$. Although it may seem that the linear time dependence in the coefficients $\Delta$ and $\beta$ is not immediately compatible with STCG, which assumes that the Hamiltonian can be written in the form of Eq.(\ref{eq: H decomp}). However,  time-dependent coefficients of the form $g t e^{i\omega t}$ can be obtained from the limit,
\begin{equation}
\begin{split}
g t e^{i\omega t}
=
\lim_{\delta \rightarrow 0}
\frac{
g
e^{i( \omega + \delta )t}
-
g
e^{i \omega t}
}{i \delta}.
\end{split}
\end{equation}
Using that expansion, $\hat{H}$ can be represented by the following three lists of coefficients, rotating-frame frequencies, and Hamiltonian operators:
\begin{subequations}
\begin{align}
\begin{split}
    &g_\Omega = \big\{
    \frac{\beta_{0}}{i \delta T},
    -\frac{\beta_{0}}{i \delta T}, 
    \frac{\beta_{0}}{i \delta T},
    -\frac{\beta_{0}}{i \delta T},
    -\frac{\beta_{0}}{i \delta T},
    \frac{\beta_{0}}{i \delta T},  
    -\frac{\beta_{0}}{i \delta T},
    \\&\qquad\quad
    \frac{\beta_{0}}{i \delta T},
    -\frac{2\beta_{0}}{i \delta T},
    \frac{2\beta_{0}}{i \delta T}, 
    \frac{2\beta_{0}}{i \delta T},
    -\frac{2\beta_{0}}{i \delta T}, 
    \Delta_{0},
    -\frac{\Delta_{0}}{2i \delta T},
    \frac{\Delta_{0}}{2i \delta T},
    \\&\qquad\quad
    -\frac{\chi}{2},
    -\frac{\chi}{2},
    -\frac{\chi}{3},
    -\frac{\chi}{2},
    -\frac{\chi}{3},
    -\frac{\chi}{12},
    -\frac{\chi}{12}\big\}
\end{split} \\
\begin{split}
    &\Omega =\big\{\delta,
    0,
    2\omega_{p}+\delta,
    2\omega_{p},
    -\delta,
    0,
    -2\omega_{p}-\delta, 
    \\&\qquad\quad
    -2\omega_{p},
    -\omega_{p}-\delta,
    -\omega_{p}, 
    \omega_{p}+\delta,
    \omega_{p},
    0, 
    \delta,
    \\ &\qquad\quad 
-\delta,
0,
-\omega_{p},
-\omega_{p},
\omega_{p},
\omega_{p},
-2\omega_{p},
2\omega_{p}\big\}    
\end{split} \\
\begin{split}
    &h_\Omega
    =\big\{\hat{a}^{\dagger 2},
    \hat{a}^{\dagger 2},
    \hat{a}^{\dagger 2},
    \hat{a}^{\dagger 2},
    \hat{a}^{2},
    \hat{a}^{2},
    \hat{a}^{2},
    \hat{a}^{2},
    \hat{a}^{\dagger} \hat{a},
    \hat{a}^{\dagger} \hat{a},
    \hat{a}^{\dagger} \hat{a},
    \hat{a}^{\dagger} \hat{a}, \\
    &\qquad\quad 
    \hat{a}^{\dagger} \hat{a},
    \hat{a}^{\dagger} \hat{a},
    \hat{a}^{\dagger} \hat{a},
    \hat{a}^{\dagger 2} \hat{a}^{2},
    \hat{a}^{2},
    \hat{a}^{\dagger} \hat{a}^{3},
    \hat{a}^{\dagger 2},
    \hat{a}^{\dagger 3} \hat{a},
    \hat{a}^{4},
    \hat{a}^{\dagger 4}
    \big\}.    
\end{split}
\end{align}  
\end{subequations}
The number of TCG contributions is huge, however, the highly oscillatory terms at frequencies of order $\omega_{p}$ are exponentially suppressed, and can therefore be ignored in the TCG master equation. The long-term effects of the high-frequency transitions are captured by slow-varying corrections to the RWA Hamiltonian as well as emergent high-order pseudo-dissipators. Therefore, we set a threshold and keep only the TCG contribution with coefficients greater than $0.08 \textrm{MHz}$ for a coarse-graining time scale of $\tau = 0.125 \textrm{ns}$. The TCG effective Hamiltonian and dissipators can be written as,
\begin{equation}
\label{eq: HTCG parametron}
\begin{split}
    \hat{H}_{\textrm{TCG}}(t)
    =&
    g_{11} \hat{a}^{\dagger} \hat{a}
    +
    g_{22} \hat{a}^{\dagger 2} \hat{a}^{2}
    +
    g_{33} \hat{a}^{\dagger 3} \hat{a}^{3}
    \\ &+ \big(
    g_{20} \hat{a}^{\dagger 2}
    +
    g_{31} \hat{a}^{\dagger 3} \hat{a}
    +
    h.c.
    \big)
\end{split}
\end{equation}
\begin{equation}
\label{eq: DTCG parametron}
\begin{split}
\hat{D}_{\textrm{TCG}}(t)
=&
\Gamma_{2,0;0,2} \mathcal{D}[\hat{a}^{\dagger 2}, \hat{a}^{2}]
+
\Gamma_{0,2;2,0} \mathcal{D}[\hat{a}^{2}, \hat{a}^{\dagger 2}]
\\ 
+& \Big(
\Gamma_{2,0;2,0} \mathcal{D}[\hat{a}^{\dagger 2}, \hat{a}^{\dagger 2}]
+
\Gamma_{2,0;2,2} \mathcal{D}[\hat{a}^{\dagger 2}, \hat{a}^{\dagger 2} \hat{a}^{2}]\\
&+
\Gamma_{0,2;2,2} \mathcal{D}[\hat{a}^{2}, \hat{a}^{\dagger 2} \hat{a}^{2}]
+
\Gamma_{1,1;2,0} \mathcal{D}[\hat{a}^{\dagger} \hat{a}, \hat{a}^{\dagger 2}] \\
&+ \Gamma_{1,1;0,2} \mathcal{D}[\hat{a}^{\dagger} \hat{a}, \hat{a}^{2}]
 + h.c. \Big)
\end{split}
\end{equation}
respectively, with the notation $\mathcal{D}[\cdot, \cdot]$ used as defined in Eq.(\ref{eq:dissipator def}). The analytical expressions for the (super)operator coefficients are given in appendix \ref{app: parametron}. As a baseline, we note that after truncating terms exponentially suppressed by factors of $e^{-\frac{\omega_{p}^{2}\tau^{2}}{2}}$, the first-order TCG master equation is identical to the von-Neumann equation under RWA. The corresponding RWA Hamiltonian can then be written as,
\begin{equation}
\begin{split}
\hat{H}_{\textrm{RWA}}(t)
=&
g^{\textrm{RWA}}_{11} \hat{a}^{\dagger} \hat{a}
+
g^{\textrm{RWA}}_{22} \hat{a}^{\dagger 2} \hat{a}^{2}
+
\big(
g^{\textrm{RWA}}_{20} \hat{a}^{\dagger 2}
+
h.c.
\big)
\end{split}
\end{equation}
with
\begin{equation}
\begin{split}
g^{\textrm{RWA}}_{11}
=
\Delta(t)
;\qquad
g^{\textrm{RWA}}_{22}
=
-
\frac{\chi}{2}
;\qquad
g^{\textrm{RWA}}_{20}
=
\beta(t).
\end{split}
\end{equation}
The additional corrections in Eq.(\ref{eq: HTCG parametron}) and Eq.(\ref{eq: DTCG parametron}) come from the higher-order TCG perturbative expansion. Let us focus our discussion on three representative Hamiltonian coefficients for the re-normalized detuning. We focus on these three because they represent three important cases; $g_{11}$ - a Hamiltonian term with non-adiabatic and non-RWA contributions; $\Gamma_{2,0;0,2}$ - a fully non-adiabatic dissipator; and $\Gamma_{1,1;2,0}$ - a pseudo-dissipator with both non-RWA and non-adiabatic contributions. Looking at the analytical expressions for the corrections elucidates the fact that the non-RWA effects and the non-adiabatic effects are not completely separable. By considering the dependencies on $T$ and $\omega_p$, we can identify which contributions come from non-adiabatic effects and which come from non-RWA effect, respectively. 

\begin{equation}
\begin{split}
g_{11}
=&
\Delta(t)
-
\frac{2 \big( \beta(t)^{2} + \beta(\tau)^{2} \big)}{\omega_{p}}
-
\frac{4\chi^{2}}{\omega_{p}} \\
&+ 
\Delta_{0} \beta(\tau)^{2} \Big( 20 \tau^{2} + \frac{11}{2 \omega_{p}^{2}} \Big) 
-
\Delta(t) \beta(\tau)^{2} \Big( 26 \tau^{2} + \frac{13}{\omega_{p}^{2}} \Big) \\
&-
\Big( \frac{\tau^{2} \Delta_{0}}{T} \Big)^{2} \Delta(t)
+
\frac{3}{2} \frac{\beta_{0} \Delta_{0}}{T^{2}\omega_{p}^{4}} \beta(t)\\
&
+
2 \Big( \frac{\beta(t)}{\omega_{p}} \Big)^{2} \Delta(t)
-
\frac{51}{2} \Big(\frac{\beta_{0}}{T\omega_{p}^{2}}\Big)^{2} \chi \\
-& \Big(
6 \big( \frac{\beta(t)}{\omega_{p}} \big)^{2} 
- 4 \big( \frac{\beta(\tau)}{\omega_{p}} \big)^{2}
\Big) \chi
+
8 \Big( \frac{\chi}{\omega_{p}} \Big)^{2} \Delta(t)
-
\frac{151 \chi^{3}}{6 \omega_{p}^{2}}
\end{split}
\end{equation}

Let us consider $g_{11}$ as an example: the correction $-\frac{2\beta(t)^{2}}{\omega_{p}}$ in $g_{11}$ is purely a non-RWA effect as it comes from the high-frequency Hamiltonian terms $\beta(t) \hat{a}^{\dagger 2} e^{2i \omega_{p} t}$ and $\beta(t) \hat{a}^{2} e^{-2i \omega_{p} t}$, and does not vanish in the adiabatic limit $T \rightarrow \infty$; However, the correction $20 \Delta_{0} \beta(\tau)^{2} \tau^{2}$ is purely a non-adiabatic effect as it involves no high-frequency Hamiltonian terms and vanishes as $\beta^2(\tau) \propto T^{-2}$ in the adiabatic limit; the correction $-\frac{2\beta(\tau)^{2}}{\omega_{p}}$, however, is a joint effect of the non-RWA and non-adiabatic terms in the Hamiltonian since it is produced by the contraction of the high-frequency terms $\beta(t) \hat{a}^{\dagger 2} e^{2i \omega_{p} t}$ and $\beta(t) \hat{a}^{2} e^{-2i \omega_{p} t}$ but vanishes as $T^{-2}$ in the adiabatic limit.
The dissipator terms are particularly interesting, since they are unique to the TCG approach. 
\begin{equation}
\begin{split}
\Gamma_{2,0;0,2}
=
\frac{\beta_{0}}{T} \Big( 2\tau^{2} + \frac{1}{2\omega_{p}^{2}} \Big) \beta(t)
\end{split}
\end{equation}
Interestingly, $\Gamma_{2,0;0,2} \mathcal{D}[\hat{a}^{\dagger 2}, \hat{a}^{2}]$ is a fully non-adiabatic effect which receives corrections from the counter-rotating terms in the Hamiltonian. Importantly, it has the standard Lindblad form together with a positive transition rate. Therefore, it represents effective dissipative phenomena in the form of two-photon absorption which cannot be analyzed efficiently using only effective Hamiltonian approaches.
\begin{equation}
\begin{split}
\Gamma_{1,1;2,0}
=&
-
\frac{45 \beta_{0}}{2T\omega_{p}^{3}} \Big( \beta(t)^{2} + \beta(\tau)^{2} \Big) \\
&-
\frac{2\tau^{2}\beta_{0}}{T \omega_{p}} \Big( 3\beta(t)^{2} + 2\beta(\tau)^{2} \Big) \\
&-
\frac{5}{2} \Big( 2\beta(t) - \beta_{0} \Big) \frac{\tau^{2}\Delta_{0}\chi}{T \omega_{p}}\\
&
-
\frac{2\Delta_{0}\chi}{T \omega_{p}^{3}} \beta(t)
-
\frac{19\beta_{0}\chi^{2}}{3T\omega_{p}^{3}}
-
\frac{4\tau\chi^{2}}{\omega_{p}} \beta(\tau) \\
&+
i \Big[
3 \tau^{2} \Big( \Big( \frac{\tau\Delta_{0}}{T} \Big)^{2} - 4 \beta(\tau)^{2} \Big) \beta(t)
-
4 \Big( \frac{\tau^{2}\Delta_{0}}{T} \Big)^{2} \beta_{0}\\
&
+
2 \tau \beta(\tau) \Big( \frac{\tau^{2}\Delta_{0}}{T} \Big) \chi
+
\frac{183}{4} \Big(\frac{\beta_{0}}{T \omega_{p}^{2}}\Big)^{2} \beta(t) \\
&-
17 \frac{\beta(t)^{3}}{\omega_{p}^{2}}
-
39 \Big(\frac{\beta(\tau)}{\omega_{p}}\Big)^{2} \beta(t)
-
\frac{2\beta(t) \Delta(t)}{\omega_{p}^{2}} \chi\\
&
+
\frac{13\tau^{2}\beta_{0}\Delta_{0}\chi}{4T^{2}\omega_{p}^{2}}
-
\Big(\frac{\chi}{\omega_{p}}\Big)^{2} \beta(t)
\Big].
\end{split}
\end{equation}
 The effective dissipator $\Gamma_{1,1;2,0} \mathcal{D}[\hat{a}^{\dagger} \hat{a}, \hat{a}^{\dagger 2}]$, on the other hand, does not assume the standard Lindblad form, and has both non-vanishing real and imaginary parts in general. In fact, the real part of $\Gamma_{1,1;2,0}$ originates from non-adiabatic effects and vanishes in the $T\rightarrow \infty$ limit, whereas the imaginary part of $\Gamma_{1,1;2,0}$ contains purely non-RWA contributions as well -- the third-order terms $-17i \frac{\beta(t)^{3}}{\omega_{p}^{2}}$, $- i\big(\frac{\chi}{\omega_{p}}\big)^{2} \beta(t)$, and $- i\frac{2\beta(t) \Delta(t)}{\omega_{p}^{2}} \chi$ survive the $T\rightarrow \infty$ limit. These generalized (pseudo-)dissipators are important in the strongly driven regime, but most of them vanish when $\beta_{0} \rightarrow 0$, leaving only some remnants from the time-dependence of $\Delta(t)$.

We emphasize that the symbolic calculation of these coefficients is highly non-trivial, even with the closed-form formula in Eq.(\ref{eq:contraction-coeff}).  It is therefore important that the computation is done fully automatically by `QuantumGraining.jl', in a process that is completely generic and applicable to a very wide range of problems. 

\begin{figure}[!h]
    \centering
    \includegraphics[width=0.8\textwidth]{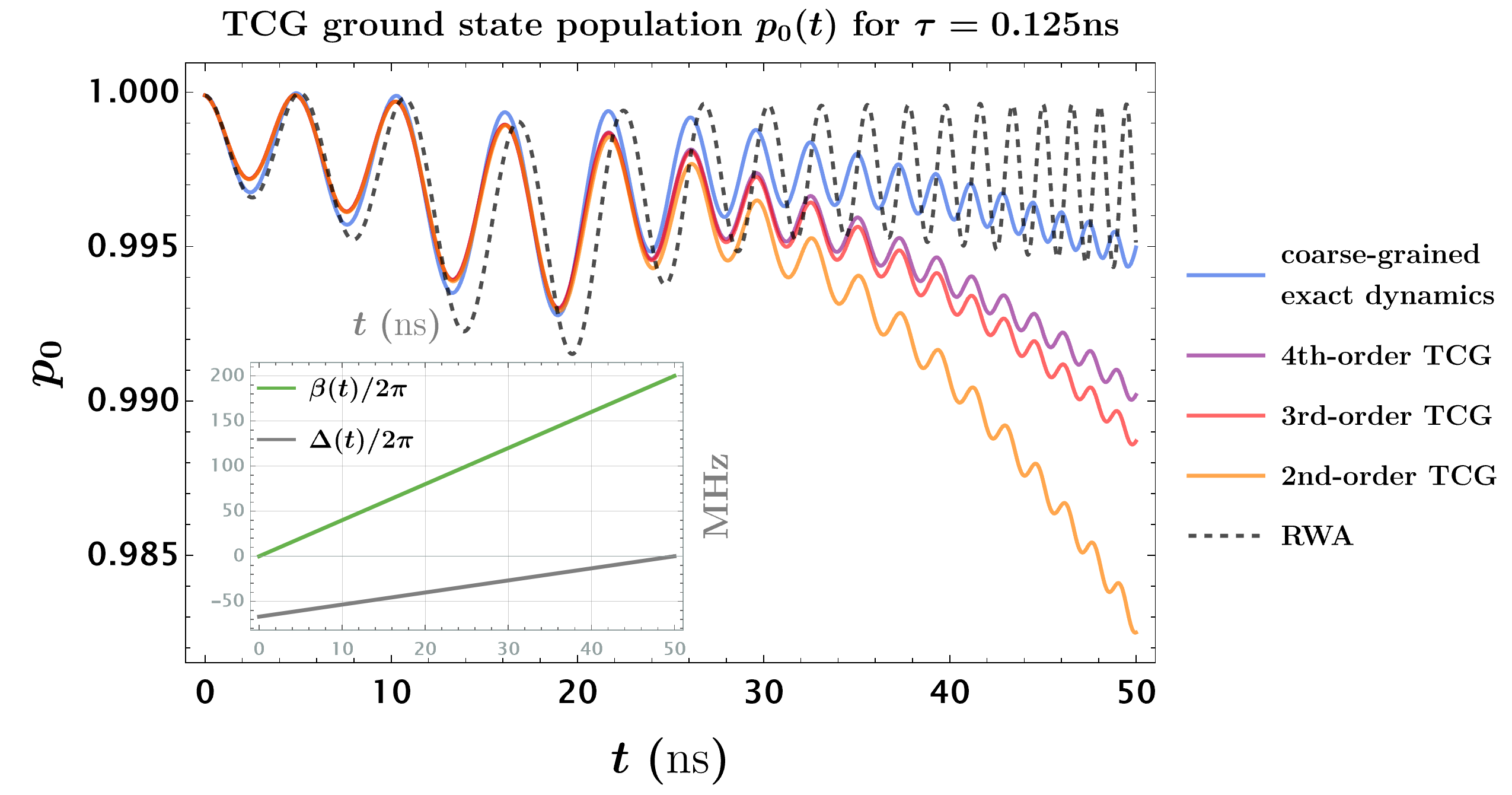}
    \caption[Caption for LOC]{The instantaneous ground state population $p_{0}$ as a function of time, using  $\overline{\rho}$ at different orders in TCG. $p_{0}(50\textrm{ns})$ is defined as the fidelity of the cat-state generation process. The exact $\rho$ is assumed to be in the pure ground state at time $t=-1\textrm{ns}$, and the initial condition for $\overline{\rho}(t)$ at $t=0$ is obtained by time-averaging the numerical solution to the non-TCG von-Neumann equation from $t=-1\textrm{ns}$ to $t=1\textrm{ns}$.}
    \label{fig:parametron p0}
\end{figure}

An important benefit of the TCG effective master equation is that it allows efficient numerical simulation of the system dynamics, since the effects of virtual transitions to and from the high-energy states are captured by the effective Hamiltonian corrections and emergent dissipators, without requiring detailed knowledge about the dynamics of the those high-energy states and their coherences. As a result, accurate numerical simulation can be done at much lower time resolutions (on the order of $\tau$ rather than $\omega_{p}^{-1}$), and truncation of the Hilbert space in the TCG framework is in general more forgiving.

Assuming the parameter values in Eq.(\ref{eq: parametron parameter values}) and taking the coarse-graining time scale to be $\tau=0.125\textrm{ns}$, we can truncate the the Hilbert space at the $20$-th level and numerically solve the TCG master equation at different orders. Note that truncating at much higher levels will not increase the accuracy since the quartic expansion of the cosine potential starts to break down above the $20$-th level for the given parameters. The coarse-grained exact dynamics, on the other hand, is obtained by applying the Gaussian window function on the numerical solution to the von-Neumann equation with the original Hamiltonian $\hat{H}(t)$. We emphasize that, compared with the TCG master equations, much smaller time steps have to be taken in order to overcome the stiffness of the exact equation of motion due to the fast oscillating nonlinear terms in $\hat{H}(t)$. In addition, solutions to the TCG master equations are more accurate than the RWA solution when benchmarked against the coarse-grained exact dynamics, with the accuracy increasing with the order of the TCG perturbation theory.

We plot the coarse-grained instantaneous ground state population $p_{0}$ as a function of time in Fig.  \ref{fig:parametron p0}. In particular, we notice that contrary to the RWA result, higher-order TCG predicts approximately linear decay in $p_{0}(t)$ during the later half of the time evolution. In fact, for the relatively short ramping time $T$ and large initial detuning $\abs{\Delta_{0}}$, this linear decay is the dominant source of the loss of cat-state fidelity. This has been observed in existing high time-resolution simulations of the cat-state fidelity (e.g. in Fig. 2 of \cite{masuda_controls_2021}), although to our knowledge, no explanation for the different behaviors of $p_{0}$ during the earlier half and the later half of the time evolution has been provided in the literature.

\begin{figure}[!h]
    \centering
    \includegraphics[width=0.85\textwidth]{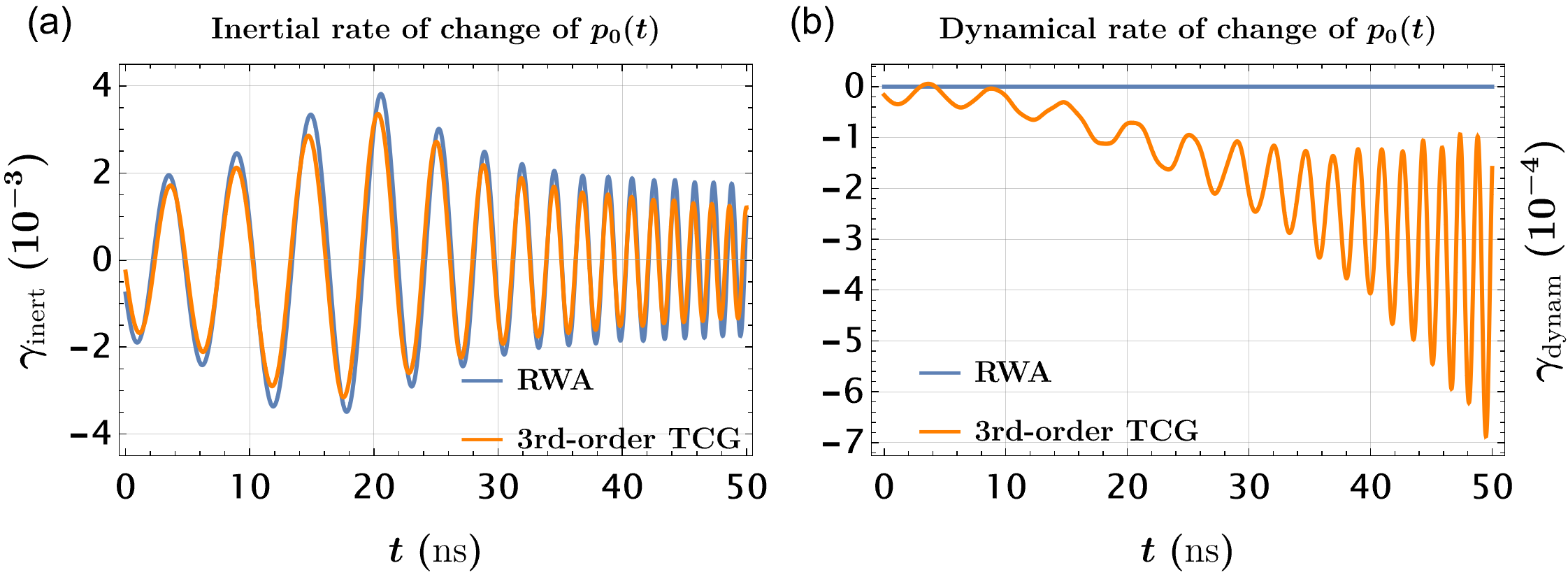}
    \caption{Time dependence of the inertial and dynamical rates of change for the instantaneous ground state population $p_{0}(t)$.}
    \label{fig:parametron gamma}
\end{figure}

Using our analytical TCG results, using leading-order perturbation theory for the estimation of $\partial_{t} |0(t)\rangle$, we obtain the following approximate expression for $\dot{p}_{0}(t)$:
\begin{equation}
\begin{split}
\dot{p}_{0}(t)
\equiv&
\gamma_{\textrm{inert}}(t)
+
\gamma_{\textrm{dynam}}(t) \\
&=
\textrm{Tr}\Big[
\partial_{t} \big(|0(t)\rangle\langle 0(t)|\big) \overline{\rho}(t)
\Big]
+
\textrm{Tr}\Big[
|0(t)\rangle\langle 0(t) | \partial_{t} \overline{\rho}(t)
\Big]\\
\approx&
\sum_{n\ge 1} \Big(
\frac{\langle n(t)|\dot{H}_{\textrm{TCG}}(t)|0(t)\rangle}{E_{0}(t) - E_{n}(t)} \langle 0(t)| \overline{\rho}(t) | n(t)\rangle
+
c.c.
\Big) \\
&+ 
\langle 0(t)| \hat{D}_{\textrm{TCG}}(t) \overline{\rho}(t) | 0(t)\rangle
\end{split}
\end{equation}
where $E_{n}(t)$ is the energy of the $n$-th instantaneous Hamiltonian eigenstate $| n(t)\rangle$ for $\hat{H}_{\textrm{TCG}}(t)$, and we have used the immediate corollary that $\langle 0(t)| \big[ - i \hat{H}_{\textrm{TCG}}(t), \overline{\rho}(t) \big] | 0(t)\rangle = 0$. We observe that the change in $p_{0}$ during the earlier half of the time evolution is dominated by the time evolution of the instantaneous ground state $|0(t)\rangle$ (with its contribution to $\dot{p}_{0}(t)$ defined as the inertial rate of change $\gamma_{\textrm{inert}}(t)$). During the later half of the time evolution, change in $p_{0}$ is dominated by the emergent TCG dissipators (with the corresponding contribution to $\dot{p}_{0}(t)$ defined as the dynamical rate of change $\gamma_{\textrm{dynam}}(t)$), as shown in Fig.  \ref{fig:parametron gamma}.
\begin{equation}
\begin{split}
&
\gamma_{\textrm{inert}}(t)
\approx
\sum_{n\ge 1} \Big(
\frac{\langle n(t)|\dot{H}_{\textrm{TCG}}(t)|0(t)\rangle}{E_{0}(t) - E_{n}(t)} \langle 0(t)| \overline{\rho}(t) | n(t)\rangle
+
c.c.
\Big)\\
&
\gamma_{\textrm{dynam}}(t)
=
\langle 0(t)| \hat{D}_{\textrm{TCG}}(t) \overline{\rho}(t) | 0(t)\rangle,
\end{split}
\end{equation}

In particular, we see in Fig.  \ref{fig:parametron gamma} that the inertial rate $\gamma_{\textrm{inert}}$ is approximately symmetrical around the zero axis, and is responsible for the initial slow oscillations in $p_{0}(t)$. It dominates over $\gamma_{\textrm{dynam}}$ during the early time, but decays towards zero as the pumping power increases into the strong-drive regime. The dynamical rate $\gamma_{\textrm{dyanm}}$ on the other hand, is induced by the emergent TCG dissipator. We notice that $\gamma_{\textrm{dyanm}}$ is almost always negative with its magnitude increasing during the early time and oscillating with increasing frequency around a certain negative mean value during the late time. Since the mean value of $\gamma_{\textrm{dyanm}}$ is negative and comparable with the magnitude of the oscillating $\gamma_{\textrm{inert}}$, its secular effects dominate over that of $\gamma_{\textrm{inert}}$ in the late time. In fact, from either the analytical formula for $\hat{D}_{\textrm{TCG}}(t)$ or numerical simulations, one can predict that increasing the non-linearity $\chi$ will significantly enhance the secular effects of $\gamma_{\textrm{dyanm}}$ while suppressing the transient oscillations due to $\gamma_{\textrm{inert}}$; having more rapid state preparation by decreasing $T$, on the other hand, will enhance both $\gamma_{\textrm{dyanm}}$ and $\gamma_{\textrm{inert}}$.

\subsection{The driven Duffing oscillator}
\label{subsec: duffing}
The Duffing oscillator is a damped oscillator driven by a non-linear force, cubic in the displacement of the particle. Despite its simple description, it shows rich phenomena, such as bi-stability, chaos, and hysteresis. The quantum Duffing oscillator, similarly can show a variety of rich quantum phenomena, such as tunneling, squeezing, and entanglement. It has been studied extensively in recent years as a model for a variety of physical systems, including superconducting circuits, trapped ions, and optical cavities, and it is also a popular model system for studying the effects of quantum mechanics on nonlinear systems. Importantly, it has been studied in the supplementary of a paper by Venkatraman et al \cite{venkatraman_static_2022}, showcasing the capabilities of an effective static Hamiltonian method, based on Lie-algebraic approach known as the ``Kamiltonian'' approach. We use that same example to demonstrate here that the STCG algorithm is not only able to reproduce the effective static Hamiltonian in Venkatraman et al. as the effective TCG Hamiltonian, but also capable of calculating the non-Hamiltonian corrections resulting from time-coarse graining. 

The quantum Duffing oscillator is described by the following Hamiltonian,
\begin{equation}
\begin{split}
\hat{H}
=
\delta \hat{a}^{\dagger} \hat{a}
+
g_{4} \big( e^{-i \frac{5}{6} \omega_{d} t} \hat{a} + e^{i \frac{5}{6} \omega_{d} t} \hat{a}^{\dagger} + e^{-i \omega_{d} t} \Pi + e^{i \omega_{d} t} \Pi^{\ast} \big)^{4}.
\end{split}
\end{equation}
In order to compare with the results in \cite{venkatraman_static_2022}, we define $\omega_{d} \equiv 6\omega$ and round numerical coefficients to the nearest integers. Assuming that $\frac{1}{g_{4}}, \frac{1}{\delta} \gg \tau \gg \frac{1}{\omega}$ and ignoring terms exponentially suppressed by factors of $e^{-\frac{\omega^{2}\tau^{2}}{2}}$, we can write the fourth-order effective TCG Hamiltonian as,
\begin{equation}
\begin{split}
\hat{H}_{\textrm{TCG}}^{(4)}(t)
&\equiv
K^{(4)}_{1} \hat{a}^{\dagger} \hat{a}
+
K^{(4)}_{2} \hat{a}^{\dagger 2} \hat{a}^{2}
+
K^{(4)}_{3} \hat{a}^{\dagger 3} \hat{a}^{3} \\
&+
K^{(4)}_{4} \hat{a}^{\dagger 4} \hat{a}^{4}
+
K^{(4)}_{5} \hat{a}^{\dagger 5} \hat{a}^{5}
\end{split}
\end{equation}
with,
\begin{fleqn}
\begin{equation}
\begin{split}
K^{(4)}_{1}
\approx&
\frac{g_{4}^{2}}{\omega}
\big(
-
58
+
625 \abs{\Pi}^{2}
+
531 \abs{\Pi}^{4}
\big) \\
&+
\frac{g_{4}^{2}}{\omega^{2}} \delta
\big(
12
+
907 \abs{\Pi}^{2}
+
665 \abs{\Pi}^{4}
\big)\\
&
+
\frac{g_{4}^{3}}{\omega^{2}}
\big(
573
+
13815 \abs{\Pi}^{2}
+
43258 \abs{\Pi}^{4}
+
21832 \abs{\Pi}^{6}
\big)\\
&
+
\frac{g_{4}^{4}}{\omega^{3}}
\big(
-
7834
+
246852 \abs{\Pi}^{2}
+
2101485 \abs{\Pi}^{4} \\
&+
3034553 \abs{\Pi}^{6}
+
1079394 \abs{\Pi}^{8}
\big)\\
&
+
\frac{g_{4}^{3} \delta}{\omega^{3}}
\big(
-
229
+
35144 \abs{\Pi}^{2}
+
106765 \abs{\Pi}^{4}
+
51312 \abs{\Pi}^{6}
\big)\\
&
+
\frac{g_{4}^{2} \delta^{2}}{\omega^{3}}
\big(
-
2
+
855 \abs{\Pi}^{2}
+
645 \abs{\Pi}^{4}
\big)
\end{split}
\end{equation}
\end{fleqn}
\begin{fleqn}

\begin{equation}
\begin{split}
K^{(4)}_{2}
\approx&
\frac{g_{4}^{2}}{\omega}
\big(
-
61
+
312 \abs{\Pi}^{2}
\big)
+
\frac{g_{4}^{2}}{\omega^{2}} \delta
\big(
12
+
453 \abs{\Pi}^{2}
\big) \\
&+
\frac{g_{4}^{3}}{\omega^{2}}
\big(
1007
+
17919 \abs{\Pi}^{2}
+
21629 \abs{\Pi}^{4}
\big)\\
&
+
\frac{g_{4}^{4}}{\omega^{3}}
\big(
-
20529
+
655974 \abs{\Pi}^{2} \\
&+
2723568 \abs{\Pi}^{4} 
+
1517277 \abs{\Pi}^{6}
\big)\\
&
+
\frac{g_{4}^{3}\delta}{\omega^{3}}
\big(
-
403
+
46418 \abs{\Pi}^{2}
+
53383 \abs{\Pi}^{4}
\big) \\
&+
\frac{g_{4}^{2}\delta^{2}}{\omega^{3}}
\big(
-
2
+
427 \abs{\Pi}^{2}
\big)
\end{split}
\end{equation}

\end{fleqn}
\begin{fleqn}

\begin{equation}
\begin{split}
K^{(4)}_{3}
\approx&
-
\frac{g_{4}^{2}}{\omega} \cdot 14
+
\frac{g_{4}^{2}}{\omega^{2}} \delta \cdot 3
+
\frac{g_{4}^{3}}{\omega^{2}}
\big(
480
+
3982 \abs{\Pi}^{2}
\big) \\
&+
\frac{g_{4}^{4}}{\omega^{3}}
\big(
-
15964
+
355032 \abs{\Pi}^{2}
+
605237 \abs{\Pi}^{4}
\big)\\
&
+
\frac{g_{4}^{3}\delta}{\omega^{3}}
\big(
-
192
+
10315 \abs{\Pi}^{2}
\big)
-
\frac{g_{4}^{2}\delta^{2}}{\omega^{3}}
\end{split}
\end{equation}

\end{fleqn}
\begin{fleqn}

\begin{equation}
\begin{split}
K^{(4)}_{4}
\approx&
\frac{g_{4}^{3}}{\omega^{2}} \cdot 60
+
\frac{g_{4}^{4}}{\omega^{3}} \big(
-
4276
+
44379 \abs{\Pi}^{2}
\big)
-
\frac{g_{4}^{3}\delta}{\omega^{3}} \cdot 24
\end{split}
\end{equation}

\end{fleqn}
\begin{fleqn}

\begin{equation}
\begin{split}
K^{(4)}_{5}
\approx&
-
\frac{g_{4}^{4}}{\omega^{3}} \cdot 342
\end{split}
\end{equation}
\end{fleqn}
if we round the numerical coefficients to the nearest integers. Similar to the models discussed previously, the TCG effective Hamiltonian (as well as the pseudo-dissipators) is dependent on the coarse-graining time scale $\tau$ in general, and what we present here should be understood as the ``IR''-limit results where $\tau$ is much greater than $\frac{1}{\omega}$. The exact analytical formulas can be found in appendix \ref{subsec: The driven Duffing}. In particular, we are able to reproduce all the classical and quantum corrections in the effective Hamiltonian derived in the supplementary material of Venkatraman et al. \cite{venkatraman_static_2022}, demonstrating that the STCG method generalizes the Kamiltonian method, yielding the same time-averaged Hamiltonian for a simple filter function with sufficiently large $\tau$. In fact, the agreement between the high-order Hamiltonian corrections obtained by the two distinct methods lead us to make the conjecture that the effective Kamiltonian can \textit{in general} be obtained by taking the IR limit of the TCG effective Hamiltonian.

\begin{figure}[!h]
    \centering
    \includegraphics[width=0.7\textwidth]{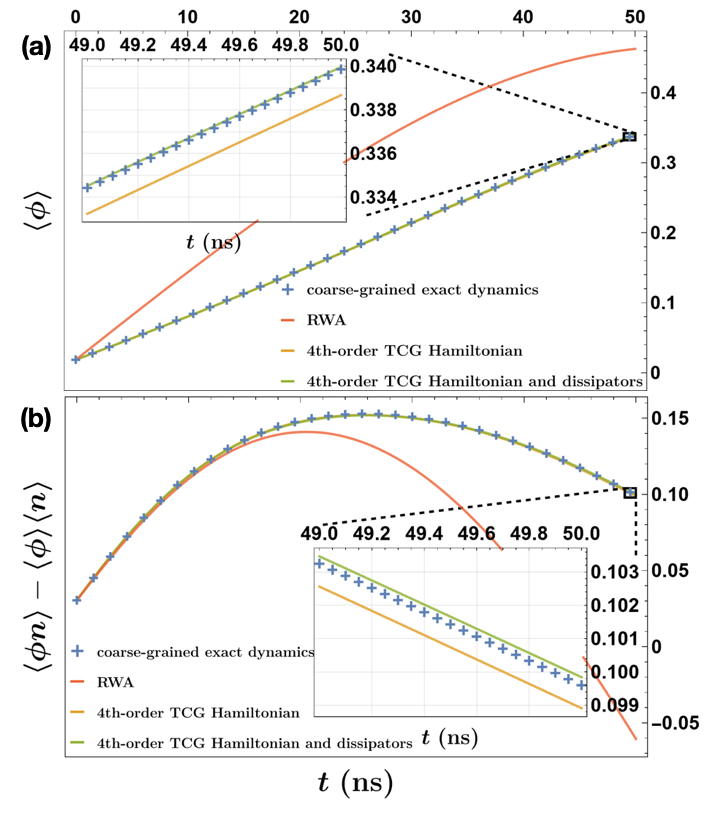}
    \caption{Time evolution of the quadrature variable $\langle \phi \rangle$ and the correlation $\langle\phi n\rangle - \langle\phi\rangle \langle n\rangle$ simulated with the lowest 25 levels and an initial microscopic (as opposed to coarse-grained) pure coherent state with displacement $\alpha = - 0.48 i$ at time $t=-3\textrm{ns}$. The inset in each panel shows the late-time dynamics in order to show the subtle differences between the different approximations.}
    \label{fig:Duffing dynamics}
\end{figure}

Unlike the parametron control model discussed in the previous subsection, here the coefficients are purely imaginary and in general do not have significant impact on the system dynamics since the drive on the Duffing oscillator is not resonant with any internal transitions, and there is no parametric time-dependence of any sort. In this case, the TCG pseudo-dissipators only provide quantitative corrections in their range of validity (i.e. when the condition $\frac{1}{g_{4}}, \frac{1}{\delta} \gg \tau \gg \frac{1}{\omega}$ is satisfied). For example, if we assume the following values of the parameters
\begin{equation*}
\label{eq: Duffing parameter values}
\begin{split}
&g_{4} / 2\pi = 0.5 \textrm{MHz} 
;\quad
\delta / 2\pi = -58.4 \textrm{MHz}
;\quad 
\omega_{d} / 2\pi = 12 \textrm{GHz}
\end{split}
\end{equation*}
with driving amplitude $\Pi = 2i$ and time-coarse graining scale $\tau = 0.5 \textrm{ns}$.
then numerical simulation of the microscopic von-Neumann equation and the TCG master equation gives us the following dynamics of the quadrature variable $\langle \phi \rangle$ and the second-order correlation $\langle\phi n\rangle - \langle\phi\rangle \langle n\rangle$ where $\phi \equiv \frac{a + a^{\dagger}}{\sqrt{2}}$ and $n \equiv \frac{a - a^{\dagger}}{\sqrt{2} i}$.
When compared directly to the coarse-grained dynamics with infinite time resolution, the fourth-order TCG QME gives much better predictions in comparison with the RWA method. However, differently from the parametron example where the TCG effective dissipators significantly modify the late-time fidelity, here the dynamics is mostly generated by the effective Hamiltonian, with the (pseudo-)dissipators only giving minor improvements during later periods of the time evolution. This kind of situation where the effective dynamics is dominated by its coherent part generated by some effective Hamiltonian usually occurs when the external drives have amplitudes that are time-independent and frequencies which do not resonate with any internal (virtual) transition processes until very high order in the nonlinearities.

\section{Conclusions and outlook}
\label{sec:conclusions}
In the lab, observable dynamics are always time-coarse grained, due to the finite time resolution of the measurement apparatus. In order to effectively model these observable dynamics, we introduced the systematic time-coarse graining (STCG) framework, a perturbative series expansion that directly describes the time-coarse grained observable dynamics of a quantum system. STCG generalizes existing TCG methods to any order, and extends beyond the description given by static effective Hamiltonian approaches ~\cite{Bukov_2016, petrescu_lifetime_2020, venkatraman_static_2022} by including non-unitary contributions to the dynamics, which we have shown to be crucial for a correct effective model of the system. 

STCG allows for more accurate description of quantum systems in non-RWA regimes and when effective Hamiltonian methods become inadequate, e.g., under strong and fast parametric drives. It gives a description of the time coarse-grained dynamics in the form of multi-body Hamiltonian corrections and (pseudo-)dissipators, providing us with an analytical method to identify and study the important emergent processes in regimes where the counter-rotating and parametric time-dependent terms become significant. In addition, the diagrammatic representation developed in this work can also help with the design and engineering of the nonlinearities and parametric drives in order to achieve certain effective long-time dynamics. Importantly, the time-coarse grained description is not only important for analytical studies, but also crucial for numerical analyses, since deep non-RWA regimes are not numerically stable. If the counter-rotating terms are included, the equations of motion can require very high-time resolutions, with strict constraints on the truncation of the Hilbert space, that can make solving them computationally prohibitive. In addition, these equations can become highly-stiff and unstable due to the involvement of vastly different time-scales and numerous modes. The STCG description, on the other hand, effectively filters the dynamics around a certain frequency band, and accounts for the effects of high-frequency (virtual) transitions in the Hilbert space without needing to keep track of their intricate time evolution. This allows for very efficient numerical study of driven nonlinear quantum systems deep into the non-RWA regime or even when more sophisticated effective Hamiltonian methods fail.

In order to showcase the STCG method, we used it to study a couple of problems - the Rabi model, the quantum parametron and the quantum Duffing oscillator. We have shown that STCG generalizes the existing treatments of these problems, reproducing well-known corrections while revealing new, non-unitary contributions that are not captured by other methods. Moreover, we have shown that our results match the exact time-coarse graining of the non-RWA solutions, with the agreement improving for higher orders of approximation. Specifically, we have used the examples to showcase important properties of the STCG method. Using the quantum parametron example, we demonstrated the importance of the non-unitary contributions captured by STCG, which prove to be crucial for correct modelling of the system fidelity and for capturing non-adiabatic effects; using the quantum Duffing oscillator, we showed that the effective Hamiltonian produced by STCG completely reproduces the effective static Hamiltonian generated by perturbative Lie series approaches, which suggests that effective Hamiltonian approaches may be considered as certain limits of the STCG approach.

The STCG calculation is complex and involved, since it requires the calculation and bookkeeping of many different superoperators which can easily be in the hundreds for a third-order expansion. To aid this situation, we envision the STCG framework as part of a computational framework, where the repetitive, intricate procedure is done automatically by a symbolic computation package, aiding the numeric and analytical study of the system. We accompany STCG with the Julia software package `QuantumGraining.jl' \cite{quantumgraining-github}, which automatically calculates the emergent terms, their coupling strengths or pseudo-dissipation rates to help automate analytical and numerical calculations. We hope that the utility and generality of STCG, in addition to the low bar-of-entry by use of `QuantumGraining.jl', can make it a valuable tool for theorists and experimentalists alike.

\section{Acknowledgements}
We are grateful to Emanuele Dalla Torre for his comments on the manuscript. This work was supported by the US Department of Energy, Office of Basic Energy Sciences, Division of Materials Sciences and Engineering, under Award No. DESC0016011. Support was also provided by the Army Research Office under contract W911NF-23-1-0252.

\bibliography{references,main}

\clearpage

\onecolumngrid
\appendix

\title{Supplemental Material for "Systematic time-coarse graining for driven quantum systems"}
\maketitle

\section{TCG master equation derivation}
\label{app:TCG-derivation}
As we discuss in the main text, we want to derive a closed formula for the TCG master equation. We start by expanding  the unitary evolution in the following Dyson series:
\begin{equation}
\begin{split}
&\hat{U}(t)
\equiv
\sum_{k=0}^{\infty} \hat{U}_k(t)
=
\hat{I} + \sum_{k=1}^{\infty} (-i)^{k} \int_{0}^t \cdots \int_{0}^{t_{k-1}} dt_1 \cdots dt_k \hat{H}(t_1) \cdots \hat{H}(t_k)  
\end{split}
\end{equation}
The TCG procedure represents a low-pass filter of the system dynamics in the interaction picture can then be implemented by working exclusively with the time-averaged density matrix, 
\begin{equation}
\label{eq:rho_Dyson}
\overline{\rho}(t)
=
\overline{U(t) \rho_0 U^\dagger(t)}
=
\sum_{k=0}^{\infty} \sum_{l=0}^{k} \overline{\hat{U}_l \rho_0 \hat{U}^\dagger_{k-l}} (t)
\end{equation}
where the time average of a time-dependent operator $\hat{O}(t)$ is defined as
\begin{equation}
    \overline{\hat{O}}(t) := \int_{-\infty}^{\infty} dt^{\prime} \tilde{f}_\tau(t-t^{\prime})\hat{O}(t^{\prime})
    \label{eq:time-average}
\end{equation}
for some window function $\tilde{f}_\tau(t)$ normalized so that $\int_{-\infty}^{\infty} \tilde{f}_\tau(t) dt = 1$; the coarse-graining time scale $\tau$, on the other hand, defines the shape of the filter. In what follows, we will denote its frequency-domain representation by $f_\tau(\omega)$.

Starting from Eq.(\ref{eq:rho_Dyson}), we would like to obtain the generator of the time evolution of $\overline{\rho}(t)$. This requires finding at least an approximate inverse of the TCG operation, and we show in the following paragraphs that it can be done by perturbatively reversing Eq.(\ref{eq:rho_Dyson}). Along this line of thought, the first thing to notice from the form of Eq.(\ref{eq:rho_Dyson}) is that the time evolution of $\overline{\rho}(t)$ cannot be generated by a Hamiltonian in general, as discussed in the main text. Consequently, the correct ansatz is a general master equation generated by a quantum Liouvillian \cite{lindblad_generators_1976, chruscinski_brief_2017, gorini_completely_1976, lidar_lecture_2019},
\begin{equation}
\label{eq:Lindbladian def}
    i\dot{\overline{\rho}} = \mathcal{L} \overline{\rho} 
\end{equation}
Let us define the time-averaged state,
\begin{equation}
    \overline{\rho}(t) = \sum_{k=0}^{\infty} \sum_{l=0}^{k} \overline{\hat{U}_l \rho \hat{U}^\dagger_{k-l}}(t) \equiv \sum_k \mathcal{E}_k(t) \rho
\end{equation}
where we denote the time-averaging operation  $\mathcal{E}(t) = \sum_k \mathcal{E}_k(t)$. By collecting powers of $k$ we find,
\begin{equation}
\mathcal{E}_k(t) \rho
=
\sum_{l=0}^{k} \overline{\hat{U}_l \rho U^\dagger_{k-l}} (t)
\end{equation}
Using the definition of the time-average in Eq.(\ref{eq:time-average}) and taking the derivative of $\overline{\rho}(t)$, the LHS of Eq.(\ref{eq:TCG-definition}) becomes
\begin{equation}
\begin{split}
i\partial_{t} \overline{\rho}(t)
&=
\sum_{k = 1}^{\infty} \sum_{l=1}^{k} \big( \overline{\hat{H} \hat{U}_{l-1} \rho U^\dagger_{k-l}}(t) -  \overline{\hat{U}_{l-1} \rho U^\dagger_{k-l} \hat{H}}(t)\big)
\\
&
= \sum_{j=1}^{\infty} \mathcal{L}_j(t) \overline{\rho}(t)\\
&
= \sum_{j=1}^{\infty} \mathcal{L}_j(t) \sum_{k=0}^{\infty} \mathcal{E}_k(t) \rho = \sum_{j=1}^{\infty} \mathcal{L}_j(t) \sum_{k-j=0}^{\infty} \mathcal{E}_{k-j}(t)\rho\\
&
= \sum_{k=1}^{\infty} \sum_{j=1}^{k} \mathcal{L}_j(t) \mathcal{E}_{k-j}(t)\rho
\end{split}
\end{equation}
Matching terms at order $k$ and expanding $\mathcal{E}_{k-j}(t)$, we find the following recurrence relation for the Liouvillian expansion
\begin{equation}
\begin{split}
            &\mathcal{L}_k(t) \rho = \sum_{l=1}^{k} \left ( \overline{\hat{H}\hat{U}_{l-1} \rho \hat{U}_{k-l}}(t) - \overline{\hat{U}_{l-1} \rho \hat{U}_{k-l} \hat{H}}(t) \right ) - \sum_{k=1}^{\infty} \sum_{j=1}^{k} \mathcal{L}_j(t) \mathcal{E}_{k-j}(t) \rho.
\end{split}
\end{equation}
where we define $\mathcal{L}_{0} \overline{\rho} \equiv 0$ for any density matrix $\overline{\rho}$. This formula relates $\mathcal{L}_k$ to all the partial Liouvillians of order $r<k$.

\section{Derivation of the contraction coefficients}
\label{app:contraction_coefficients}
\subsection{The contraction superoperators}
Due to their complicated time dependence, directly calculating the averages in Eq.(\ref{eq:TCG-raw}) is difficult. For example, even with only two operators we would have,
\begin{equation*}
\begin{split}
    \overline{\hat{H} \hat{U}_{1}}(t)
&=
\sum_{l_{1}, l_{2}}
\frac{f_\tau(\omega_{l_1} + \omega_{l_2}) e^{-i(\omega_{l_1} + \omega_{l_2})t} - f_\tau(\omega_{l_{2}}) e^{-i\omega_{l_{2}}t} }{\omega_{l_{1}} } \hat{h}_{l_{2}} \hat{h}_{l_{1}}  
\end{split}
\end{equation*}
which contains two terms at different frequencies. However, Eq.(\ref{eq:TCG-raw}) can be used to show that such a time average is always accompanied by the following product of time averages,
\begin{equation*}
\begin{split}
    \overline{\hat{H}}(t) \overline{\hat{U}_{1}}(t) &=
\sum_{l_{1},l_{2}}
\frac{
f_\tau(\omega_{l_{1}}) f_\tau(\omega_{l_{2}}) e^{-i(\omega_{l_{1}}+\omega_{l_{2}})t} -
f_\tau(\omega_{l_{2}}) e^{-i \omega_{l_{2}}t}
}{\omega_{l_{1}}}\hat{h}_{l_{2}} \hat{h}_{l_{1}}.
\end{split}
\end{equation*}
In previous work by Lee et al.\cite{lee_effective_2018}, it has been shown that the time-coarse grained dynamics to the second-order are generated by two-point ``contractions'' which happen to be equivalent to covariances
\begin{subequations}
    \begin{align}
        \wick{\c{P}\c{Q}} &= \overline{PQ} - \overline{P}\cdot\overline{Q} \\
        \wick{\c{P}\overline{\rho}\c{Q}} &= \overline{P \overline{\rho} Q} - \overline{P}\cdot\overline{\rho}\cdot\overline{Q}.
    \end{align}
\end{subequations}
Note that by plugging in the time-averaged terms into the contraction, the terms at frequency $\omega_{l_{2}}$ always cancel each other out, and we have the following expression where the time dependence is homogeneous for each combination of $\hat{h}_{l_{1}}$ and $\hat{h}_{l_{2}}$:
\begin{equation}
\label{eq:2nd order contraction}
\begin{split}
\wick{\c{H} \c{U}_1}
&=
\sum_{l_1, l_2}
\frac{f_\tau(\omega_{l_1} + \omega_{l_2}) - f_\tau(\omega_{l_1}) f_\tau(\omega_{l_2}) }{\omega_{l_{1}}} e^{-i(\omega_{l_1} + \omega_{l_2})t} 
\hat{h}_{l_{2}} \hat{h}_{l_1}.
\end{split}
\end{equation}
In fact, at high orders in the perturbative expansion, cancellations of this kind take place on a very large scale (see App. \ref{app:harmonic_dependence}), which makes Eq.(\ref{eq:TCG-raw}) inefficient for computation. In fact, we show in App. \ref{app:harmonic_dependence} that after the mass cancellation takes place, all operator products of the form $\hat{h}_{\mu_{l}} \cdots \hat{h}_{\mu_{1}} \rho \hat{h}_{\nu_1} \cdots \hat{h}_{\nu_r}$ have the same harmonic time-dependence $\omega = \sum_i \mu_i + \sum_j \nu_j$. 

In what follows, we generalize the two-point contractions defined in Lee et al. \cite{lee_effective_2018} to multi-point contractions. According to Eq.(\ref{eq:TCG-raw}), $\mathcal{L}_{k} \overline{\rho}$ can always be written as the sum of a collection of nested averages of $\hat{H}(t)$, $\hat{U}_{i}(t)$, and $\hat{U}_{j}^{\dagger}(t)$ multiplied to either sides of $\overline{\rho}$, where $\hat{U}_{i}(t)$ and $\hat{U}_{j}^{\dagger}(t)$ can only appear on the left and right side of $\overline{\rho}$ respectively, while $\hat{H}$ is always present and can appear on either end of the product. Therefore, as discussed in the main text, we can group terms in $\mathcal{L}_{k}$ by the shape of the nested averages so that
\begin{subequations}
\begin{align}
\mathcal{L}_{k} \overline{\rho}(t)
&
=
\sum_{r = 0}^{k-1} \mathcal{W}_{k-r,r}(t)[\overline{\rho}]
-
h.c.
\end{align}
\end{subequations}
for any $k \ge 1$, where the contraction $\mathcal{W}_{l,r}(t)[\overline{\rho}]$ is the sum of all nested averages that have $l$($r$) operators to the left(right) of $\overline{\rho}$ with the first operator being $\hat{H}$.

Plugging Eq.(\ref{eq:TCG-Lindblad}) into the recurrence relation \ref{eq:TCG-raw} and grouping together all terms that have the weight $(l,r)$ with $l+r=k$, we find the following recurrence relation for these contraction superoperators
\begin{subequations}
\label{eq:contraction_recursion}
\begin{align}
&
\mathcal{W}_{0,0}
\equiv
0\\
&
\mathcal{W}_{k,0}[\overline{\rho}]
=
\overline{H U_{k-1}}\overline{\rho}
-
\sum_{k^{\prime}=1}^{k-1} \mathcal{W}_{k^{\prime}, 0}
\Big[
\overline{U_{k - k^{\prime}}}\overline{\rho}
\Big]\\
&\mathcal{W}_{l,k-l}(t)[\overline{\rho}]
= 
\overline{H U_{l-1} \overline{\rho} U_{k-l}^\dagger}
-
\sum_{k^{\prime}=1}^{k - 1}
\sum_{l^{\prime}=\max(0,l-k^{\prime})}^{\min(l-1, k-k^{\prime})}
\mathcal{W}_{l-l^{\prime},k^{\prime}-(l-l^{\prime})} \Big[\;
\overline{U_{l^{\prime}} \overline{\rho} U^\dagger_{k-k^{\prime}-l^{\prime}}}\;
\Big]
\end{align}
\end{subequations}
where the second equation is just a special case of the third equation, given here for clarity only. The contraction in the second sum of the last two equations are to be understood as operating on the inner time-averaged product, multiplying operators to the left and/or to the right of the time-average. The recurrence relation in Eq.(\ref{eq:contraction_recursion}) can be confirmed by plugging it back into Eq.(\ref{eq:TCG-raw}) together with the decomposition in Eq.(\ref{eq:TCG-Lindblad}).

\begin{figure}[!h]
\centering
\includegraphics[width=0.8\textwidth]{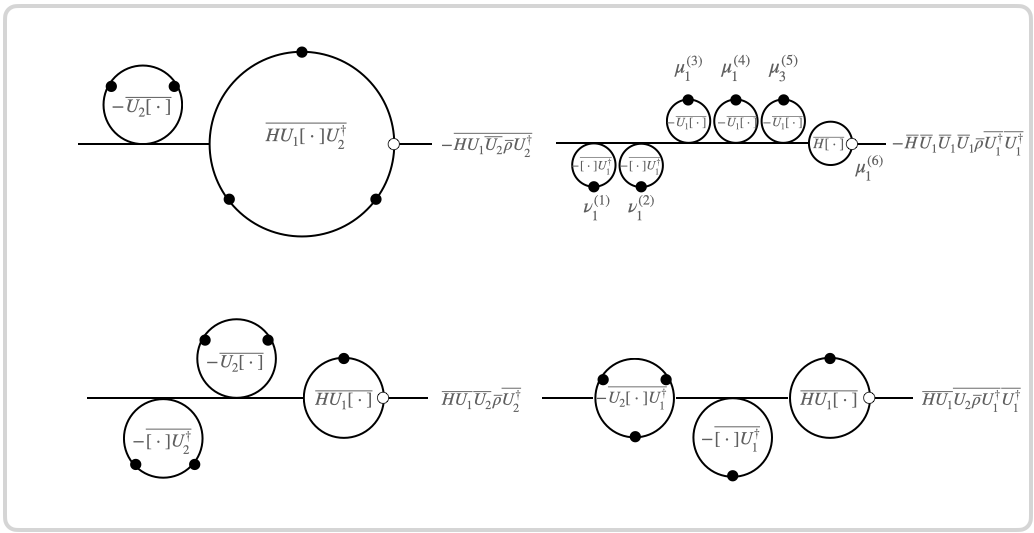}
\caption{A couple of example bubble diagrams contributing to the contraction superoperator $\mathcal{W}_{4,2}[\overline{\rho}]$. The time-averaged (super)operator inside the bubble represents the factor the bubble gives rise to, and the nested average to the right of the bubble shows the total contribution from that diagram. The diagram is read left to right, with the leftmost bubbles operating closest to $\overline{\rho}$ and the subsequent bubbles operating on the result of previous ones, appearing further away from $\overline{\rho}$.}
\label{fig:tcg-bubbles}
\end{figure}

Looking at the contractions as defined through the recurrence relation in Eq.(\ref{eq:contraction_recursion}), we see that they are always of the following form,
\begin{equation}
\label{eq:contractions_in_averages}
\begin{split}
&
\mathcal{W}_{l,r}[\overline{\rho}]
=
\sum_{d \in \text{diagrams}(l,r)} \overline{\hat{H} \hat{U}_{l_{\norm{d}}-1} [\cdot] \hat{U}_{r_{\norm{d}}}^\dagger}
\prod_{n=1}^{\norm{d}-1}
\Big(
-
\overline{\hat{U}_{l_{n}} [\cdot] \hat{U}_{r_{n}}^\dagger}
\,
\Big) \overline{\rho}
=
\overline{\hat{H} \hat{U}_{l-1} \overline{\rho} \hat{U}^\dagger_{r}}
-
\overline{\hat{H} \hat{U}_{l-2}\overline{\hat{U}_{1}} \overline{\rho} \hat{U}_{r}^\dagger} + \cdots  
\end{split}
\end{equation}
where ``$\textrm{diagrams}(l,r)$'' is a partition-valued function which represents the different ways in which the averages may be arranged by the different natural-number partitions of the $2$-tuple $(l,r)$. These different partitions can be represented visually by ``bubble diagram'', where each bubble corresponds to the $n$-th part of a particular partition $d$. Namely, for the $n$-th bubble we would have a $2$-tuple $(l_{n}, r_{n})$ such that $\sum_{n=1}^{\norm{d}} l_{n} = l$ and $\sum_{n=1}^{\norm{d}} r_{n} = r$.

Fig. \ref{fig:tcg-bubbles}, shows a couple of examples of the bubble-diagram visualization, where the dots in a bubble represent the Hamiltonian operators appearing in the corresponding time average. In particular, dots in the upper (lower) half of the diagram represent operators to the left (right) of the density matrix. A contraction with weight $(l,r)$ can then be represented by the set of all bubble diagrams with $l$ upper operators and $r$ lower operators inserted into the bubbles. Finally, the hollow dot represents the leftmost $\hat{H}$ term in the product.

Note that the contractions are not symmetric with respect to the weights, since there is always an $\hat{H}$ term on the left. For example, a $\mathcal{W}_{3,1}[\overline{\rho}]$ contraction superoperator would be a sum over the following time-averaged terms
\begin{align*}
    &\mathcal{W}_{3,1}[\overline{\rho}] \equiv \wick{\c{H} \hat{U}_2 \overline{\rho} \c{U}_1^\dagger} = \\ &\overline{HU_2\overline{\rho} U_1^\dagger} 
    - \overline{H U_2} \overline{\rho} \overline{U_1^\dagger} 
    -\overline{H \overline{U}_2 \overline{\rho} \overline{U}_1^\dagger} 
    - \overline{\overline{H} U_2 \overline{\rho} U_1^\dagger} 
    + 2 \overline{H} \overline{U}_2 \overline{\rho} \overline{U_1^\dagger} \\
    &- \overline{H U_1 \overline{U}_1 \overline{\rho} U_1^\dagger}
    - \overline{H U_1} \overline{U_1 \overline{\rho} U_1^\dagger}
    + 2 \overline{HU_1} \cdot \overline{U_1} \overline{\rho} \overline{U_1^\dagger} 
    + \overline{H \overline{U}_1 \overline{U}_1 \overline{\rho} U_1^\dagger} \\
    &+ \overline{H}\overline{U}_1 \overline{U_1 \overline{\rho} U_1^\dagger} 
    + \overline{H}{\overline{U_1 \overline{U_1} \overline{\rho} U_1^\dagger}}
    - 3 \overline{H}\overline{U}_1 \overline{U}_1 \overline{\rho} \overline{U_1^\dagger}
\end{align*}
where we have generalized the contraction notation in Lee et al. \cite{lee_effective_2018} as another representation of $\mathcal{W}_{3,1}[\overline{\rho}]$ in the first equality.
These are all the contributions appearing in the TCG master equation with weight $(3,1)$, where the weight implies the number of Hamiltonian operators in $\hat{H}$ appearing on either side of the density matrix. 

\subsection{Proving the harmonic time-dependence of the contractions}
\label{app:harmonic_dependence}
In Eq.(\ref{eq:contraction_expansion}), we made the ansatz that the contraction superoperators can be written as a sum of operator product with only a harmonic time-dependence,
\begin{equation*}
    \mathcal{W}_{l,r}(t)[\overline{\rho}] = \sum_{(\vec{\mu}, \vec{\nu}) \in \mathcal{P}_{l,r} [\Omega]} C_{l,r}(\vec{\mu}, \vec{\nu}) e^{-i(\sum_i^l \mu_i + \sum_i^r \nu_i) t } \cdot \hat{h}_{\mu_1} \cdots \hat{h}_{\mu_l} \hat{\rho} \hat{h}_{\nu_1} \cdots \hat{h}_{\nu_r}
\end{equation*}
where the the contraction coefficients $C_{l,r}(\vec{\mu}, \vec{\nu})$ are time-independent functions of the inserted frequencies, and they effectively encode all the diagram contributions.

Additionally, in Equation \ref{eq:contractions_in_averages} we have shown that the contraction superoperators can be written as a sum of products of averages:
\begin{equation*}
\mathcal{W}_{l,r}[\overline{\rho}]
=
\sum_{d \in \rm{diagrams}(l,r)} \alpha_d \cdot \overline{H U_{l_{\norm{d]}}-1} [\cdot] U_{r_{\norm{d}}}^\dagger}
\prod_{n=1}^{\norm{d}-1}
\overline{U_{l_{n}} [\cdot] U_{r_{n}}^\dagger} \overline{\rho}
=
\overline{H U_{l-1} \overline{\rho} U^\dagger_{r}} + \overline{H U_{l-2}\overline{U_{1}} \overline{\rho} U_{r}^\dagger} + \cdots
\end{equation*}
More specifically, each nested average in the sum can be obtained by repeatedly acting superoperators of the form $\overline{U_l [\cdot] U_r^\dagger}$ on $\overline{\rho}$ before finally acting the last superoperator $\overline{H U_{l-1} [\cdot] U_r^\dagger}$ onto the preceding result. Using these two assumptions, and by calculating these averages, we can derive an expression for the contraction coefficients.

We start by assuming that the Hamiltonian can be written as a discrete sum of Harmonic terms $\hat{H} = \sum_j \hat{h}_j e^{-i\omega_j t}$, where the coupling strengths have been absorbed into the Hamiltonian operators $\hat{h}_{j}$. We can derive an analytical expression for $\hat{U}_k$ in the expression of the unitary evolution operator:
\begin{equation}
\begin{split}
\hat{U}_n(t)
&=
-
i \int_0^t dt^{\prime} \hat{H}(t^{\prime})\hat{U}_{n-1}(t^{\prime})
=
(-i)^{n} \int_0^t \int_0^{t_{1}} \cdots \int_0^{t_{n-1}} dt_{1} \cdots dt_{n} \cdot
\hat{H}(t_{1}) \cdots \hat{H}(t_{n}).
\end{split}
\end{equation}
Plugging in the expansion of the Hamiltonian gives a simpler integral and allows us to decouple the limits of integration
\begin{equation}
\label{Eq: U_n integral}
\hat{U}_{n}(t)
=
\sum_{\omega_1, \cdots, \omega_n \in \Omega} \hat{h}_{\omega_1} \cdots \hat{h}_{\omega_n} \cdot
(-i)^{n}  \int_{0}^{t} dt_1 e^{-i\omega_1 t_1} \int_{0}^{t_{1}} \cdots \int_0^{t_{n-1}} dt_{n} \cdot
e^{-i \omega_{n} t_{n}}.
\end{equation}
The last integral over $t_{n}$  is fully decoupled, allowing us to calculate it explicitly:
\begin{equation}
\int_0^{t_{n-1}} dt_{n} \cdot
e^{-i\omega_{n} t_{n}}
=
-
\frac{1}{i\omega_{n}} \left [ e^{-i\omega_{n} t_{n-1}} -1 \right ]
\end{equation}
Plugging it back into Eq.(\ref{Eq: U_n integral}), we see that there is now a sum of two terms that have the same structure of integrals but with different frequencies. If we keep doing that $n$ times, we end up with
\begin{equation}
\begin{split}
U_{l}
=
\sum_{\vec{\mu}} \sum_{j_{l}=0}^{l}
\frac{
(-1)^{l+j_{l}} e^{ -i \sum_{k=0}^{j_{l}-1} \mu_{l-k} t }
}{
\vec{\mu}[l-j_{l}+1:l] ! \vec{\mu}[l-j_{l}:1] !
}
h_{\mu_{l}} \cdots h_{\mu_{1}}
\end{split}
\end{equation}

\begin{equation}
\begin{split}
U_{r}^{\dagger}
=&
\sum_{\vec{\nu}} \sum_{j_{r}=0}^{j_{r}}
\frac{
(-1)^{r+j_{r}} e^{ -i \sum_{k=0}^{j_{r}-1} \nu_{r-k} t }
}{
( - \vec{\nu}[r-j_{r}+1:r]) ! ( - \vec{\nu}[r-j_{r}:1] ) !
}
h_{\nu_{1}} \cdots h_{\nu_{r}}\\
=&
\sum_{\vec{\nu}} \sum_{j_{r}=0}^{r}
\frac{
(-1)^{j_{r}} e^{ -i \sum_{k=0}^{j_{r}-1} \nu_{r-k} t }
}{
\vec{\nu}[r-j_{r}+1:r] ! \vec{\nu}[r-j_{r}:1] !
}
h_{\nu_{1}} \cdots h_{\nu_{r}}
\end{split}
\end{equation}
and
\begin{equation}
\begin{split}
H U_{l}
=
\sum_{\vec{\mu}} \sum_{j_{l}=0}^{l}
\frac{
	(-1)^{l+j_{l}} e^{ -i \sum_{k=0}^{j_{l}} \mu_{l+1-k} t }
}{
	\vec{\mu}[l-j_{l}+1:l] ! \vec{\mu}[l-j_{l}:1] !
}
h_{\mu_{l+1}} h_{\mu_{l}} \cdots h_{\mu_{1}}
\end{split}
\end{equation}
where $\vec{\mu}[i, j] \equiv (\mu_{i}, \mu_{i+1}, \cdots, \mu_{j})$ and we define $\vec{\mu}!$ to be the ``vector-factorial'' so that
\begin{equation}
\label{eq:vector-factorial}
\vec{\mu}!
\equiv
\big( \mu_{1}, \mu_{2}, \cdots, \mu_{\norm{v}} \big) !
=
(\mu_{1} + \mu_{2} + \cdots + \mu_{\norm{v}}) \cdots (\mu_{1} + \mu_{2}) \cdot \mu_{1},
\end{equation}
with $\norm{\mu}$ indicating the number of elements in $\vec{\mu}$. So for example, $(\mu_{1}, \mu_{2}, \mu_{3})! = (\mu_{1} + \mu_{2} + \mu_{3}) \cdot (\mu_{1} + \mu_{2}) \cdot \mu_{1}$.
Therefore, we have the following expression for a typical bubble factor (here ``typical'' means that it is not the special bubble at the right end of the diagram):
\begin{equation}
\label{eq: large typical bubble}
\begin{split}
-
\overline{ U_{l} \big[\cdot\big] U_{r}^{\dagger} }
=&
-
\sum_{\vec{\mu},\vec{\nu}} \sum_{j_{l}=0}^{l} \sum_{j_{r}=0}^{r}
\frac{
(-1)^{l+j_{l}+j_{r}} f_\tau( \sum_{k=0}^{j_{l}-1} \mu_{l-k} + \sum_{k=0}^{j_{r}-1} \nu_{r-k} )
e^{-i (\sum_{k=0}^{j_{l}-1} \mu_{l-k} + \sum_{k=0}^{j_{r}-1} \nu_{r-k}) t}
}{
\vec{\mu}[l-j_{l}+1:l] ! \vec{\mu}[l-j_{l}:1] ! \vec{\nu}[r-j_{r}+1:r] ! \vec{\nu}[r-j_{r}:1] !
}\\
&\qquad\cdot
h_{\mu_{l}} \cdots h_{\mu_{1}}
\big[\cdot\big]
h_{\nu_{1}} \cdots h_{\nu_{r}}\\
=&
\sum_{\vec{\mu},\vec{\nu}}
\sum_{j_{l}=0}^{l} \sum_{j_{r}=0}^{r}
\Big(
-
B^{j_{l},j_{r}}( \vec{\mu}, \vec{\nu} )
\Big)
h_{\mu_{l}} \cdots h_{\mu_{1}}
\big[\cdot\big]
h_{\nu_{1}} \cdots h_{\nu_{r}}
\end{split}
\end{equation}
where,
\begin{equation}
    B^{j_{l},j_{r}}( \vec{\mu}, \vec{\nu} ) \equiv \frac{
    	(-1)^{l+j_{l}+j_{r}} f_\tau( \sum_{k=0}^{j_{l}-1} \mu_{l-k} + \sum_{k=0}^{j_{r}-1} \nu_{r-k} )
    	e^{-i (\sum_{k=0}^{j_{l}-1} \mu_{l-k} + \sum_{k=0}^{j_{r}-1} \nu_{r-k}) t}
    }{
    	\vec{\mu}[l-j_{l}+1:l] ! \vec{\mu}[l-j_{l}:1] ! \vec{\nu}[r-j_{r}+1:r] ! \vec{\nu}[r-j_{r}:1] !
    }    
\end{equation}
with $l$ and $r$ being the lengths of $\vec{\mu}$ and $\vec{\nu}$ respectively; similarly, the special bubble at the right end gives rise to the factor,
\begin{equation}
\begin{split}
\overline{ H U_{l} \big[\cdot\big] U_{r}^{\dagger} }
=&
\sum_{\vec{\mu},\vec{\nu}} \sum_{j_{l}=0}^{l} \sum_{j_{r}=0}^{r}
\frac{
	(-1)^{l+j_{l}+j_{r}} f_\tau( \sum_{k=0}^{j_{l}} \mu_{l+1-k} + \sum_{k=0}^{j_{r}-1} \nu_{r-k} )
	e^{-i (\sum_{k=0}^{j_{l}} \mu_{l+1-k} + \sum_{k=0}^{j_{r}-1} \nu_{r-k}) t}
}{
	\vec{\mu}[l-j_{l}+1:l] ! \vec{\mu}[l-j_{l}:1] ! \vec{\nu}[r-j_{r}+1:r] ! \vec{\nu}[r-j_{r}:1] !
}\\
&\quad\cdot
h_{\mu_{l+1}} h_{\mu_{l}} \cdots h_{\mu_{1}}
\big[\cdot\big]
h_{\nu_{1}} \cdots h_{\nu_{r}}\\
\equiv&
\sum_{\vec{\mu},\vec{\nu}}
\sum_{j_{l}=0}^{l} \sum_{j_{r}=0}^{r}
\tilde{B}^{j_{l},j_{r}}( \vec{\mu}, \vec{\nu} )
\cdot
h_{\mu_{l}} \cdots h_{\mu_{1}}
\big[\cdot\big]
h_{\nu_{1}} \cdots h_{\nu_{r}}.
\end{split}
\end{equation}
where we define,
\begin{equation}
 \tilde{B}^{j_{l},j_{r}}( \vec{\mu}, \vec{\nu} ) \equiv \frac{
	(-1)^{l+j_{l}+j_{r}} f_\tau( \sum_{k=0}^{j_{l}} \mu_{l+1-k} + \sum_{k=0}^{j_{r}-1} \nu_{r-k} )
	e^{-i (\sum_{k=0}^{j_{l}} \mu_{l+1-k} + \sum_{k=0}^{j_{r}-1} \nu_{r-k}) t}
}{
	\vec{\mu}[l-j_{l}+1:l] ! \vec{\mu}[l-j_{l}:1] ! \vec{\nu}[r-j_{r}+1:r] ! \vec{\nu}[r-j_{r}:1] !
}.   
\end{equation}

\begin{figure}[!h]
\centering
\includegraphics[width=0.9\textwidth]{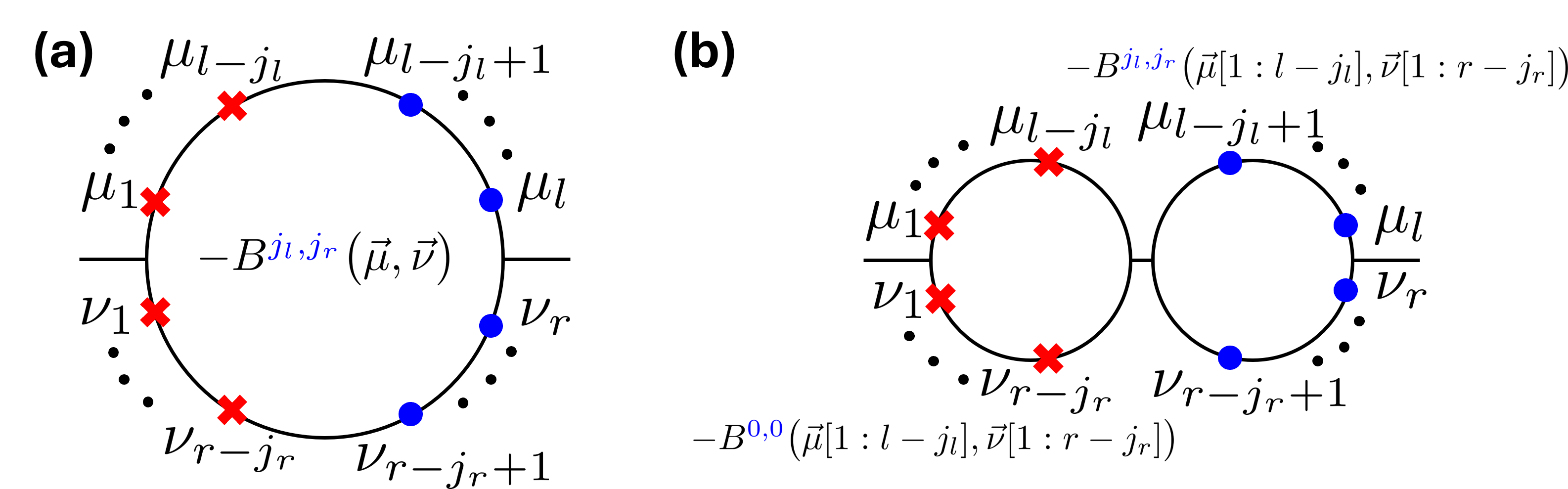}
	\caption{These color-coded bubbles correspond to individual contributing terms oscillating at the total frequency of the blue dots on the right side of the bubbles. As shown in Eq.(\ref{eq: large typical bubble}) and Eq.(\ref{eq: small typical bubble}), one has the identity $B^{j_{l},j_{r}}\big(\vec{\mu}, \vec{\nu}\big) = B^{0,0}\big( \vec{\mu}[1:l-j_{l}], \vec{\nu}[1:r-j_{r}] \big) B^{j_{l},j_{r}}\big( \vec{\mu}[1:l-j_{l}], \vec{\nu}[1:r-j_{r}] \big)$, which implies that the bubble contributions in panel (a) and panel (b) cancel each other out exactly.}
	\label{fig:bubble splitting}
\end{figure}

Therefore, we see that for each combination of inserted operators, the contributing factor of a particular bubble can be written as a sum of terms each of which oscillates at the total frequency of a subset of neighboring frequencies inserted on the right side of the bubble, as indicated by the blue dots in Fig.  \ref{fig:bubble splitting}.

On the other hand, we also have the following formula for the contributions of two adjacent typical bubbles:
\begin{equation}
\label{eq: small typical bubble}
\begin{split}
&
\Big(
-
\overline{U_{j_{l}} \big[\cdot\big] U_{j_{r}}^{\dagger}}
\,
\Big)
\Big(
-
\overline{U_{l-j_{l}} \big[\cdot\big] U_{r-j_{r}}^{\dagger}}
\,
\Big)\\
=&
\sum_{\vec{\mu},\vec{\nu}}
\frac{
	(-1)^{l-j_{r}}
}{
	\vec{\mu}[l-j_{l}:1] ! \vec{\nu}[r-j_{r}:1] !
}
\cdot
\frac{
	(-1)^{j_{l}} f_\tau( \sum_{k=0}^{j_{l}-1} \mu_{l-k} + \sum_{k=0}^{j_{r}-1} \nu_{r-k} )
	e^{-i( \sum_{k=0}^{j_{l}-1} \mu_{l-k} + \sum_{k=0}^{j_{r}-1} \nu_{r-k} )t}
}{
	\vec{\mu}[l-j_{l}+1:l] ! \vec{\nu}[r-j_{r}+1:r] !
}\\
&
+
\sum_{\vec{\mu},\vec{\nu}} \sum_{(n_{l},n_{r},m_{l},m_{r})\neq(0,0,j_{l},j_{r})}^{(l-j_{l},r-j_{r},j_{l},j_{r})}
\frac{
(-1)^{l-j_{r}+n_{l}+n_{r}} f_\tau( \sum_{k=0}^{n_{l}-1} \mu_{l-j_{l}-k} + \sum_{k=0}^{n_{r}-1} \nu_{r-j_{r}-k} )
}{
\vec{\mu}[l-j_{l}-n_{l}+1:l-j_{l}] ! \vec{\mu}[l-j_{l}-n_{l}:1] !
}\\
&\quad\cdot
\frac{
e^{-i( \sum_{k=0}^{n_{l}-1} \mu_{l-j_{l}-k} + \sum_{k=0}^{n_{r}-1} \nu_{r-j_{r}-k} )t}
}{
\vec{\nu}[r-j_{r}-n_{r}+1:r-j_{r}] ! \vec{\nu}[r-j_{r}-n_{r}:1] !
}\\
&\quad\cdot
\frac{
(-1)^{j_{r}+m_{l}+m_{r}} f_\tau( \sum_{k=0}^{m_{l}-1} \mu_{l-k} + \sum_{k=0}^{m_{r}-1} \nu_{r-k} )
e^{-i( \sum_{k=0}^{m_{l}-1} \mu_{l-k} + \sum_{k=0}^{m_{r}-1} \nu_{r-k} )t}
}{
\vec{\mu}[l-m_{l}+1:l] ! \vec{\mu}[l-m_{l}:l-j_{l}+1] ! \vec{\nu}[r-m_{r}+1:r] ! \vec{\nu}[r-m_{r}:r-j_{r}+1] !
}\\
&\quad\cdot
h_{\mu_{l}} \cdots h_{\mu_{1}}
\big[\cdot\big]
h_{\nu_{1}} \cdots h_{\nu_{r}}\\
=&
\sum_{\vec{\mu},\vec{\nu}}
B^{j_{l},j_{r}}( \vec{\mu}, \vec{\nu} )
\cdot
h_{\mu_{l}} \cdots h_{\mu_{1}}
\big[\cdot\big]
h_{\nu_{1}} \cdots h_{\nu_{r}}\\
&
+
\sum_{\vec{\mu},\vec{\nu}} \sum_{(n_{l},n_{r},m_{l},m_{r})\neq(0,0,j_{l},j_{r})}
B^{n_{l},n_{r}}\big( \vec{\mu}[1:l-j_{l}], \vec{\nu}[1:r-j_{r}] \big)
\cdot
B^{m_{l},m_{r}}\big( \vec{\mu}[l-j_{l}+1 : l], \vec{\nu}[r-j_{r}+1 : r] \big)\\
&\qquad\qquad\qquad\qquad\qquad\qquad
\cdot
h_{\mu_{l}} \cdots h_{\mu_{1}}
\big[\cdot\big]
h_{\nu_{1}} \cdots h_{\nu_{r}}
\end{split}
\end{equation}
and similarly, for a typical bubble adjacent to the special bubble at the right end we have
\begin{equation}
\begin{split}
&
\Big(
-
\overline{H U_{j_{l}} \big[\cdot\big] U_{j_{r}}^{\dagger}}
\,
\Big)
\Big(
-
\overline{U_{l-j_{l}} \big[\cdot\big] U_{r-j_{r}}^{\dagger}}
\,
\Big)\\
=&
\sum_{\vec{\mu},\vec{\nu}}
\frac{
	(-1)^{l-j_{r}}
}{
	\vec{\mu}[l-j_{l}:1] ! \vec{\nu}[r-j_{r}:1] !
}
\cdot
\frac{
	(-1)^{j_{l}} f_\tau( \sum_{k=0}^{j_{l}-1} \mu_{l-k} + \sum_{k=0}^{j_{r}-1} \nu_{r-k} )
	e^{-i( \sum_{k=0}^{j_{l}-1} \mu_{l-k} + \sum_{k=0}^{j_{r}-1} \nu_{r-k} )t}
}{
	\vec{\mu}[l-j_{l}+1:l] ! \vec{\nu}[r-j_{r}+1:r] !
}\\
&
+
\sum_{\vec{\mu},\vec{\nu}} \sum_{(n_{l},n_{r},m_{l},m_{r})\neq(0,0,j_{l},j_{r})}^{(l-j_{l},r-j_{r},j_{l},j_{r})}
\frac{
(-1)^{l-j_{r}+n_{l}+n_{r}} f_\tau( \sum_{k=0}^{n_{l}-1} \mu_{l-j_{l}-k} + \sum_{k=0}^{n_{r}-1} \nu_{r-j_{r}-k} )
}{
\vec{\mu}[l-j_{l}-n_{l}+1:l-j_{l}] ! \vec{\mu}[l-j_{l}-n_{l}:1] !
}\\
&\quad\cdot
\frac{
e^{-i( \sum_{k=0}^{n_{l}-1} \mu_{l-j_{l}-k} + \sum_{k=0}^{n_{r}-1} \nu_{r-j_{r}-k} )t}
}{
\vec{\nu}[r-j_{r}-n_{r}+1:r-j_{r}] ! \vec{\nu}[r-j_{r}-n_{r}:1] !
}\\
&\quad\cdot
\frac{
	(-1)^{j_{r}+m_{l}+m_{r}} f_\tau( \sum_{k=0}^{m_{l}-1} \mu_{l-k} + \sum_{k=0}^{m_{r}-1} \nu_{r-k} )
	e^{-i( \sum_{k=0}^{m_{l}-1} \mu_{l-k} + \sum_{k=0}^{m_{r}-1} \nu_{r-k} )t}
}{
	\vec{\mu}[l-m_{l}+1:l] ! \vec{\mu}[l-m_{l}:l-j_{l}+1] ! \vec{\nu}[r-m_{r}+1:r] ! \vec{\nu}[r-m_{r}:r-j_{r}+1] !
}\\
&\quad
\cdot
h_{\mu_{l}} \cdots h_{\mu_{1}}
\big[\cdot\big]
h_{\nu_{1}} \cdots h_{\nu_{r}}\\
=&
\sum_{\vec{\mu},\vec{\nu}}
\tilde{B}^{j_{l},j_{r}}( \vec{\mu}, \vec{\nu} )
\cdot
h_{\mu_{l}} \cdots h_{\mu_{1}}
\big[\cdot\big]
h_{\nu_{1}} \cdots h_{\nu_{r}}\\
&
+
\sum_{\vec{\mu},\vec{\nu}} \sum_{(n_{l},n_{r},m_{l},m_{r})\neq(0,0,j_{l},j_{r})}
B^{n_{l},n_{r}}\big( \vec{\mu}[1:l-j_{l}], \vec{\nu}[1:r-j_{r}] \big)
\\
&\qquad
\cdot
\tilde{B}^{m_{l},m_{r}}\big( \vec{\mu}[l-j_{l}+1 : l+1], \vec{\nu}[r-j_{r}+1 : r] \big)\cdot
h_{\mu_{l}} \cdots h_{\mu_{1}}
\big[\cdot\big]
h_{\nu_{1}} \cdots h_{\nu_{r}}
\end{split}
\end{equation}

In fact, according to Fig.  \ref{fig:bubble splitting}, only two kinds of typical bubble contributions cannot be canceled in this fashion: the contributing term is either a constant in time (only red crosses are inserted) or oscillates at the total bubble frequency (only blue dots are inserted), which we refer to as ``empty'' or ``full'' bubbles respectively. Following the same procedure, one can also show that the contributions from the special bubble at the right end can also be canceled by bubble-splitting unless it is either full or empty (here the contribution corresponding to the ``empty'' special bubble oscillates at the frequency $\mu_{l}$ of the special operator at the right end, instead of being a constant in time).

\begin{figure}[!h]
\centering
\includegraphics[width=0.5\textwidth]{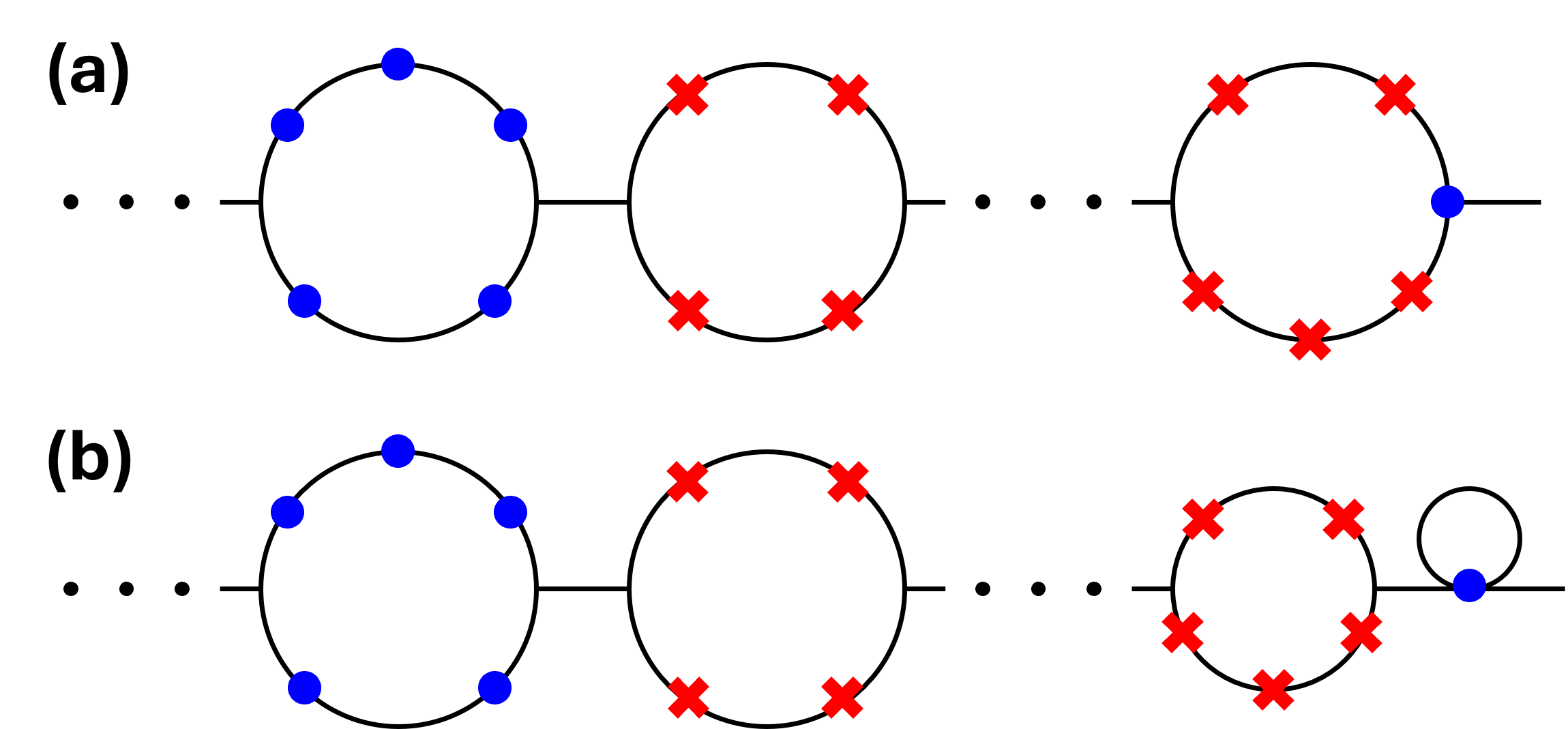}
\caption{For a diagram not to be canceled by splitting partially filled bubbles or by merging adjacent empty and full bubbles, it has to contain no partially filled bubbles, and all its empty bubbles have to be on the right side of the diagram. However, by our construction, the special operator on the right end of a diagram comes from the un-integrated Hamiltonian $H$ and is therefore always filled (i.e., it always contributes the time-dependent factor $e^{-i \mu_{l} t}$ corresponding to its frequency $\mu_{l}$). Therefore, the diagrams shown in panel (a) and panel (b) cancel each other out.}
\label{fig:special bubble splitting}
\end{figure}

On the flip side of cancellation by bubble splitting, we see that an empty bubble right to the left of a full bubble can be canceled by merging the two bubbles together. Therefore, the only bubble diagrams surviving this cancellation process are those with full bubbles on the left and empty bubble on the right. However, this implies that there are no empty bubbles, since the special operator inserted on the right end of a diagram is always filled (represented by a blue dot) by construction, as shown in Fig.  \ref{fig:special bubble splitting}.

Consequently, according to Eq.(\ref{eq:contractions_in_averages}), we have
\begin{equation}
\begin{split}
\mathcal{W}_{l,r}[\overline{\rho}]
=&
\sum_{d \in \rm{diagrams}(l,r)}
\Big(
\tilde{B}^{ l_{\norm{d}}, r_{\norm{d}} }( \vec{\mu}^{(b_{\norm{d}})}, \vec{\nu}^{(b_{\norm{d}})} )
\Big)
\prod_{n=1}^{\norm{d}-1}
\Big(
-
B^{ l_{n}, r_{n} }( \vec{\mu}^{(b_{n})}, \vec{\nu}^{(b_{n})} )
\Big)
\cdot \hat{h}_{\mu_{l}} \cdots \hat{h}_{\mu_{1}} \overline{\rho} \hat{h}_{\nu_1} \cdots \hat{h}_{\nu_r}\\
=&
\sum_{d \in \rm{diagrams}(l,r)}
\Big( \sum_{i} \mu_{i}^{(b_{\norm{d}})} \Big)
\frac{
	(-1)^{r_{\norm{d}}} f_\tau( \sum_{i} \mu_{i}^{(b_{\norm{d}})} + \sum_{i} \nu_{i}^{(b_{\norm{d}})} )
	e^{-i ( \sum_{i} \mu_{i}^{(b_{\norm{d}})} + \sum_{i} \nu_{i}^{(b_{\norm{d}})} ) t}
}{
	\vec{\mu}^{(b_{\norm{d}})} ! \vec{\nu}^{(b_{\norm{d}})} !
}\\
&\qquad\qquad\qquad\cdot
(-1)^{\norm{d}-1} \prod_{n=1}^{\norm{d}-1}
\frac{
	(-1)^{r_{n}} f_\tau( \sum_{i} \mu_{i}^{(b_{n})} + \sum_{i} \nu_{i}^{(b_{n})} )
	e^{-i (\sum_{i} \mu_{i}^{(b_{n})} + \sum_{i} \nu_{i}^{(b_{n})}) t}
}{
	\vec{\mu}^{(b_{n})} ! \vec{\nu}^{(b_{n})} !
}\\
&\qquad\qquad\qquad\cdot
\hat{h}_{\mu_{l}} \cdots \hat{h}_{\mu_{1}} \overline{\rho} \hat{h}_{\nu_1} \cdots \hat{h}_{\nu_r}\\
=&
\Big[
\sum_{d \in \rm{diagrams}(l,r)}
(-1)^{r + \norm{d} - 1}
\Big( \sum_{i} \mu_{i}^{(b_{\norm{d}})} \Big)
\prod_{b\in d} \frac{ f_\tau( \sum_{i} \mu_{i}^{(b)} + \sum_{i} \nu_{i}^{(b)} ) }{ \vec{\mu}^{(b)} ! \vec{\nu}^{(b)} ! }
\Big]
e^{-i ( \sum_{i=1}^{l} \mu_{i} + \sum_{i=1}^{r} \nu_{i} ) t}\\
&\quad\cdot
\hat{h}_{\mu_{l}} \cdots \hat{h}_{\mu_{1}} \overline{\rho} \hat{h}_{\nu_1} \cdots \hat{h}_{\nu_r}
\end{split}
\end{equation}
Or equivalently,
\begin{equation*}
\mathcal{W}_{l,r}(t)[\overline{\rho}]
=
\sum_{(\vec{\mu}, \vec{\nu}) \in \mathcal{P}_{l,r} [\Omega]} C_{l,r}(\vec{\mu}, \vec{\nu}) e^{-i(\sum_{i=1}^l \mu_i + \sum_{i=1}^r \nu_i) t } \cdot \hat{h}_{\mu_{l}} \cdots \hat{h}_{\mu_{1}} \overline{\rho} \hat{h}_{\nu_1} \cdots \hat{h}_{\nu_r}.
\end{equation*}
with
\begin{equation*}
C_{l,r}(\vec\mu, \vec\nu)
=
\sum_{d \in \rm{diagrams}(l, r)} (-1)^{r + \norm{d} - 1} \left ( \sum_i \mu_i^{(b_{\norm{d}})} \right) \prod_{b \in d} 
\frac{f_\tau\left(\sum_i \mu_i^{(b)} + \sum_i \nu_i^{(b)} \right ) }{\vec{\mu}^{(b)}! \vec{\nu}^{(b)}!},
\end{equation*}
which proves Eq.(\ref{eq:contraction_expansion}) and Eq.(\ref{eq:contraction-coeff}) in the main text. Noticeably, the mass cancellation of the different frequency terms is due to the summation of all diagram contributions, leaving only a single harmonic contribution for each diagram; in particular, the special operator due to $H$ inserted on the right end of each diagram is crucial for this mass cancellation to take place.

\subsection{Symmetries and separation into Hamiltonian and pseudo-dissipators}
We notice that Eq.(\ref{eq:contraction-coeff}) reveals some important symmetries of the contraction coefficients. These symmetries keep the last element of $\vec{\mu}$ constant as we permute the vectors,
\begin{equation}
    C_{l, r} = C_{l,r} \left( (\mu_1, \cdots, \mu_{l-1}, \mu_l), (\nu_1, \cdots, \nu_r) \right )
\end{equation}
To respect the symmetries of $C_{l,r}$, permutations of the frequencies must keep $\omega_1$  as the last element of $\vec{\mu}$. For example, with $\vec{\mu} = (\mu_{1}, \mu_{2}, \cdots, \mu_{l})$ and $\vec{\nu} = (\nu_{1}, \nu_{2}, \cdots, \nu_{r})$, one has
\begin{subequations}
\begin{align}
    &C_{l,r}(-\vec{\mu}, -\vec{\nu}) = (-1)^{l+r-1} C_{l,r}(\vec{\mu}, \vec{\nu}) \label{eq: C parity}\\
    &C_{l,r}(\vec{\mu}, \vec{\nu}) = C_{r+1,l-1}\big( -(\nu_{1}, \nu_{2}, \cdots, \nu_{r}, \mu_l), -(\mu_{1}, \mu_{2}, \cdots, \mu_{l-1}) \big) \label{eq: C shifting}.
\end{align}
\end{subequations}
where the symmetry in the second identity reflects swapping the upper modes and lower modes, while keeping the special Hamiltonian mode contribution $\mu_l$ unaffected. 
\begin{equation}
\label{Eq: mirror_symmetry_condensed}
C_{l,r}(\vec{\mu}, \vec{\nu})
=
C_{r+1,l-1} \big( -(\vec{\nu} \oplus_{r+1} \mu_{l}), -(\vec{\mu} \ominus_{l} \mu_{l}) \big).
\end{equation}
where $\oplus_{r+1}$ denotes concatenation at index $r+1$ and $\ominus_l$ denotes removing from position $l$, producing vectors with length $r+1$ and $l-1$ respectively.

\begin{figure}[!h]
    \centering
    \includegraphics[width=0.8\textwidth]{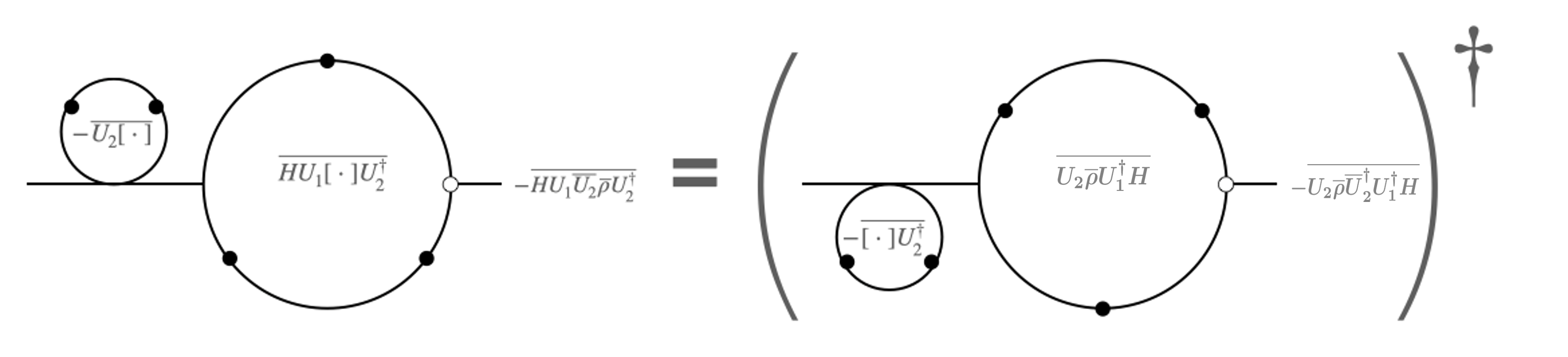}
    \caption{The symmetries of the contraction coefficients reflect the underlying time reversal and Hermitian symmetries of the dynamics in the absence of spontaneous symmetry breaking. In terms of the diagrams, they manifest as a symmetry for reflection around the main axis of the diagram, and the fact that the special frequency $\mu_{l}$ from the $H$ operator does not participate in the vector factorials in Eq.(\ref{eq:contraction-coeff}) and may therefore be moved from the right of $\overline{\rho}$ to the left of $\overline{\rho}$ together with an extra minus sign. However, the time reversal symmetry can be spontaneously broken if some of the inserted operators assume nonzero expectation values or if the regularization of apparently singular bubbles produces imaginary prefactors.}
    \label{fig:mirror_symmetry}
\end{figure}

This symmetry reflects the mirror symmetry of the diagrams around the main axis, as shown in Fig.  \ref{fig:mirror_symmetry}, which in turn is a manifestation of the Hermicity of the terms.
In particular, the second identity is not only useful for deriving the analytical expression of $C_{l,r}$ from that of $C_{r+1,l-1}$, but also necessary for the partial Liouvillian $\mathcal{L}_{k}$ to be trace-preserving at all orders in the perturbation theory, by virtue of the Lindblad-like form of $\mathcal{L}_{k}$  in subsection \ref{sec:Ham and diss decomp}.  Let us write the TCG Liouvillian again,
\begin{align}
\mathcal{L}_k(t)\overline{\rho}
&=
\sum_{l=1}^k \mathcal{W}_{l,k-l}[\overline{\rho}] - h.c.\\
&=
\sum_{l=1}^{k} \sum_{\vec{\mu}, \vec{\nu} \in \mathcal{P}_{l,k-l}[\Omega]} \left ( C_{l,r}(\vec{\mu}, \vec{\nu}) s_{\vec\mu^{\,\textrm{rev}}} \overline{\rho} s_{\vec\nu} - C_{l,r}(\vec\mu, \vec\nu) s^\dagger_{\vec\nu} \overline{\rho} s_{\vec\mu^{\,\textrm{rev}}}^\dagger \right )
\label{Eq: L_k with C coefficients}
\end{align}
where we have absorbed the harmonic time-dependent factors as well as the coupling strengths into the operator $s_{\vec{\nu}}$ for brevity:
\begin{align}
s_{\vec\nu}
=
\Big(
\prod_{\omega \in \vec{\nu}} e^{-i\omega t}
\Big)
h_{\nu_{1}}
h_{\nu_{2}}
\cdots 
h_{\nu_{r}}
.
\end{align}
Since we sum over all frequencies in the set $\Omega$ and all vector lengths
\[
(l,r)
\in
\big\{
(1, k-1), (2,k-2), \cdots, (k, 0)
\big\}
,
\]
we can make the following dummy variable substitution in the second term of the right-hand side of Eq.(\ref{Eq: L_k with C coefficients}) for all $1\le l \le k-1$:
\begin{align}
\vec\mu \rightarrow -\vec\nu 
\\
\vec\nu \rightarrow -\vec\mu
\end{align}
whereas for $l=k$ we make the substitution $\vec\mu \rightarrow -\vec\mu^{\,\textrm{rev}}$.
This allows us to write the sum in terms of a single operator,
\begin{equation}
\begin{split}
\label{eq: HD_seperation_step1}
\mathcal{L}_k(t)\overline{\rho}
=&
\sum_{\vec{\mu}, \vec{\nu} \in \mathcal{P}_{k,0}[\Omega]}
C_{k,0}(\vec{\mu}, \emptyset) s_{\vec\mu^{\,\textrm{rev}}} \overline{\rho}
-
\sum_{\vec{\mu}, \vec{\nu} \in \mathcal{P}_{k,0}[\Omega]}
C_{k,0}(-\vec{\mu}^{\,\textrm{rev}}, \emptyset) \overline{\rho} s_{\vec\mu^{\,\textrm{rev}}}\\
&
+
\sum_{l=1}^{k-1} \sum_{\vec{\mu}, \vec{\nu} \in \mathcal{P}_{l,k-l}[\Omega]}
\Big(
C_{l,r}(\vec{\mu}, \vec{\nu})
-
C_{r, l}(-\vec{\nu} , -\vec{\mu} )
\Big)
s_{\vec{\mu}^{\,\textrm{rev}}} \overline{\rho} s_{\vec{\nu}}
\end{split}
\end{equation}
where we have used the identity $s_{-\vec{\mu}}^\dagger = s_{\vec{\mu}}$, and then divided the sum into $l=k$ and $l\neq k$ terms. Now we note that we can rearrange the terms into dissipators and commutators, with the help of the following identities:
\begin{align}
&
s_{\vec{\mu}^{\,\textrm{rev}}} \overline{\rho} s_{\vec{\nu}}
=
\mathcal{D}[s_{\vec{\mu}^{\,\textrm{rev}}}, s_{\vec{\nu}}] \overline{\rho}
+
\frac{1}{2} \left \{ s_{\vec{\nu}} s_{\vec{\mu}^{\,\textrm{rev}}} , \overline{\rho} \right \}\\
&
s_{\vec{\mu}^{\,\textrm{rev}}} \overline{\rho}
=
\frac{1}{2} \left ( [s_{\vec{\mu}^{\,\textrm{rev}}}, \overline{\rho}] + \{s_{\vec{\mu}^{\,\textrm{rev}}}, \overline{\rho} \} \right )\\
&
\overline{\rho} s_{\vec{\mu}^{\,\textrm{rev}}}
=
\frac{1}{2} \left (
-
[s_{\vec{\mu}^{\,\textrm{rev}}}, \overline{\rho}]
+
\{s_{\vec{\mu}^{\,\textrm{rev}}}, \overline{\rho} \} \right )
.
\end{align}
Plugging the identities above into Eq.(\ref{eq: HD_seperation_step1}), we have
\begin{equation}
\label{eq: HD_seperation_step2}
\begin{split}
\mathcal{L}_k(t)\overline{\rho}
=&
\sum_{\vec{\mu}, \vec{\nu} \in \mathcal{P}_{k,0}[\Omega]}
\frac{
C_{k,0}(\vec{\mu}, \emptyset)
+
C_{k,0}(-\vec{\mu}^{\,\textrm{rev}}, \emptyset)
}{2}
[s_{\vec{\mu}^{\,\textrm{rev}}}, \overline{\rho}]\\
&
+
\sum_{l=1}^{k-1} \sum_{\vec{\mu}, \vec{\nu} \in \mathcal{P}_{l,k-l}[\Omega]}
\Big(
C_{l,r}(\vec{\mu}, \vec{\nu})
-
C_{r, l}(-\vec{\nu}, -\vec{\mu} )
\Big)
\mathcal{D}[s_{\vec{\mu}^{\,\textrm{rev}}}, s_{\vec{\nu}}] \overline{\rho}\\
&
+
\sum_{\vec{\mu}, \vec{\nu} \in \mathcal{P}_{k,0}[\Omega]}
\frac{
C_{k,0}(\vec{\mu}, \emptyset)
-
C_{k,0}(-\vec{\mu}^{\,\textrm{rev}}, \emptyset)
}{2}
\{s_{\vec{\mu}^{\,\textrm{rev}}}, \overline{\rho} \}\\
&
+
\sum_{l=1}^{k-1} \sum_{\vec{\mu}, \vec{\nu} \in \mathcal{P}_{l,k-l}[\Omega]}
\frac{
C_{l,r}(\vec{\mu}, \vec{\nu})
-
C_{r, l}(-\vec{\nu}, -\vec{\mu} )
}{2}
\left \{ s_{\vec{\nu}} s_{\vec{\mu}^{\,\textrm{rev}}} , \overline{\rho} \right \}
\end{split}
\end{equation}
where by the symmetry relation in Eq.(\ref{Eq: mirror_symmetry_condensed}), we can rewrite the anti-commutator terms in the last two lines of Eq.(\ref{eq: HD_seperation_step2}) as
\begin{equation}
\begin{split}
&
\sum_{\vec{\mu}, \vec{\nu} \in \mathcal{P}_{k,0}[\Omega]}
\frac{
C_{k,0}(\vec{\mu}, \emptyset)
-
C_{k,0}(-\vec{\mu}^{\,\textrm{rev}}, \emptyset)
}{2}
\{ s_{\vec{\mu}^{\,\textrm{rev}}}, \overline{\rho} \}\\
&
+
\sum_{l=1}^{k-1} \sum_{\vec{\mu}, \vec{\nu} \in \mathcal{P}_{l,k-l}[\Omega]}
\frac{
C_{l,r}(\vec{\mu}, \vec{\nu})
-
C_{l+1, r-1}( \vec\mu \oplus_{l} \nu_{r}, \vec\nu \ominus_{r} \nu_{r} )
}{2}
\left \{ s_{\vec\nu} s_{\vec\mu^{\,\textrm{rev}}} , \overline{\rho} \right \}\\
=&
-
\sum_{\vec{\mu}, \vec{\nu} \in \mathcal{P}_{k,0}[\Omega]}
\frac{
	C_{k,0}(-\vec{\mu}^{\,\textrm{rev}}, \emptyset)
}{2}
\{ s_{\vec{\mu}^{\,\textrm{rev}}}, \overline{\rho} \}
+
\sum_{l=1}^{k} \sum_{\vec{\mu}, \vec{\nu} \in \mathcal{P}_{l,k-l}[\Omega]}
\frac{C_{l,r}(\vec{\mu}, \vec{\nu})}{2}
\left \{ s_{\vec\nu} s_{\vec\mu^{\,\textrm{rev}}} , \overline{\rho} \right \}\\
&
-
\sum_{l = 2}^{k} \sum_{\vec{\mu}, \vec{\nu} \in \mathcal{P}_{l,k-l}[\Omega]}
\frac{C_{l,r}(\vec{\mu}, \vec{\nu})}{2}
\left \{ s_{\vec\nu} s_{\vec\mu^{\,\textrm{rev}}} , \overline{\rho} \right \}\\
=&
-
\sum_{\vec{\mu}, \vec{\nu} \in \mathcal{P}_{k,0}[\Omega]}
\frac{
C_{k,0}(-\vec{\mu}^{\,\textrm{rev}}, \emptyset)
}{2}
\{ s_{\vec{\mu}^{\,\textrm{rev}}}, \overline{\rho} \}
+
\sum_{\vec{\mu}, \vec{\nu} \in \mathcal{P}_{1,k-1}[\Omega]}
\frac{C_{1,k-1}(\vec{\mu}, \vec{\nu})}{2}
\left \{ s_{\vec\nu} s_{\vec\mu^{\,\textrm{rev}}} , \overline{\rho} \right \}\\
=&
-
\sum_{\vec{\mu}, \vec{\nu} \in \mathcal{P}_{k,0}[\Omega]}
\frac{
C_{k,0}(-\vec{\mu}^{\,\textrm{rev}}, \emptyset)
}{2}
\{ s_{\vec{\mu}^{\,\textrm{rev}}}, \overline{\rho} \}
+
\sum_{\vec{\mu}, \vec{\nu} \in \mathcal{P}_{1,k-1}[\Omega]}
\frac{C_{k,0}(-\vec{\nu} \oplus_{k-1} \mu_{1}, \emptyset)}{2}
\left \{ s_{\vec\nu \oplus_{k-1} \mu_{1}} , \overline{\rho} \right \}
=
0
\end{split}
\end{equation}
where one can make the dummy variable substitution $\vec{\nu} \oplus_{k-1} \mu_{1} \rightarrow \vec{\mu}$ in the second term of the last time in order to make the cancellation manifest.
Now that the anti-commutator terms vanish, we are left with
\begin{align}
\mathcal{L}_k(t)\overline{\rho} =&
\sum_{\vec{\mu}, \vec{\nu} \in \mathcal{P}_{k,0}[\Omega]}
\frac{
C_{k,0}(\vec{\mu}, \emptyset)
+
C_{k,0}(-\vec{\mu}^{\,\textrm{rev}}, \emptyset)
}{2}
[\hat{s}_{\vec\mu^{\,\textrm{rev}}}, \overline{\rho}]\\
&
+
\sum_{l=1}^{k-1} \sum_{\vec{\mu}, \vec{\nu} \in \mathcal{P}_{l,k-l}[\Omega]} \left ( C_{l,r}(\vec{\mu}, \vec{\nu}) - C_{r, l}(-\vec\nu, -\vec\mu ) \right )
\mathcal{D}[\hat{s}_{\vec{\mu}^{\,\textrm{rev}}}, \hat{s}_{\vec{\nu}}] \overline{\rho}
\end{align}
which is separated into dissipator and Hamiltonian terms, proving Eq.(\ref{eq: Lk_H}, \ref{eq:TCG-hamiltonian}, \ref{eq:TCG-dissipators}).

\section{Algorithms and implementation}
\subsection{Enumeration of the diagrams}
\label{app:enumerating}
To calculate a contraction coefficient $C_{l,r}$, we first need to identify all the diagrams associated with such a contraction, and then calculate the TCG corrections associated with each diagram. For example, the contraction coefficient $C_{l,r}$ receives contributions from all diagrams that have $l$ left modes and $r$ right modes. 

\begin{figure}[h!]
    \centering
    \includegraphics[width=0.5\textwidth]{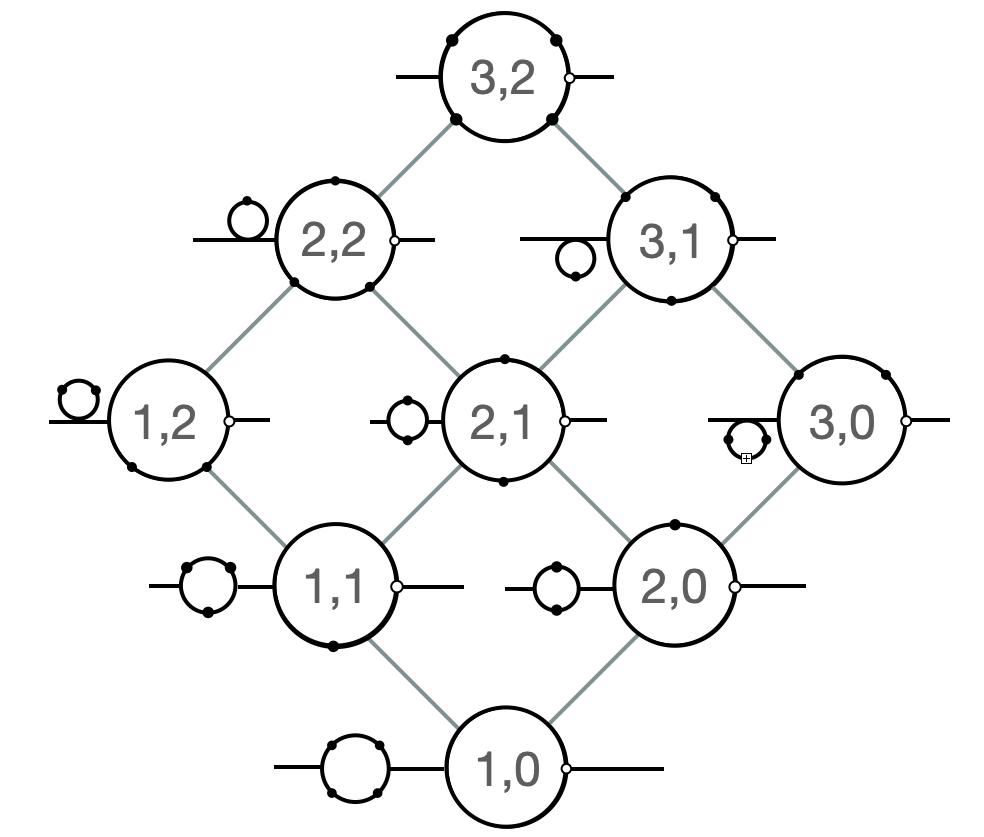}
    \caption{A single level (the first one in this case) in a decomposition tree. It covers all the different ways in which a single-bubble diagram of the form $(3,2)$ can be decomposed into two-bubble diagrams. The process is repeated on each non-trivial bubble until none is left.}
    \label{fig:tcg-tree}
\end{figure}

\begin{algorithmic}[1]
\label{alg:tree}
\Procedure{diagram\_node}{root\_bubble, new\_bubble}
    \If{$\rm{root\_bubble}.left \ge 1, root\_bubble.right \ge 0$}
        \State $\rm{left\_node} \gets  \rm{diagram\_node}(root\_bubble, (new\_bubble.left + 1, new\_bubble.right)) $
        \State $\rm{right\_node} \gets  \rm{diagram\_node}(root\_bubble, (new\_bubble.left, new\_bubble.right + 1))$ \\
        \ \ \ \ \ \ \ \ \ \Return diagram\_node(root\_bubble, new\_bubble, left\_node, right\_node)
    \EndIf
\EndProcedure

\Procedure{decompose\_diagram}{node, $\rm{decomp\_list}$}
    \If{node can be broken \textbf{and} is not in decomp\_list}
        \State{decomp\_list $\gets$ node.left, node.right}
        \State{decompose\_diagram(node.left, decomp\_list)}
        \State{decompose\_diagram(node.right, decomp\_list)}
    \EndIf \\
    \ \ \ \ \,\Return decomp\_list
\EndProcedure
\end{algorithmic}

Eq.(\ref{eq:contraction_recursion}) establishes a recurrence relation between a contraction coefficient of weight $(l,r)$ and those of lower weights, as elaborated upon in appendix \ref{app:contraction_coefficients}. This recursive structure bears resemblance to a binary tree, which we refer to as the decomposition tree of the coefficient, as depicted in Fig.  \ref{fig:tcg-tree}. For the purpose of systematically examining all possible diagrams, we arrange them in a tree-like structure, with a single-bubble diagram at the pinnacle. A rightward (leftward) step down the tree is equivalent to rupturing the rightmost bubble in the diagram, removing a right-mode (left-mode) from it and incorporating it into a bubble to its left, as illustrated in Fig.  \ref{fig:tcg-tree}. This yields all the distinct ways in which we can break a bubble into two bubbles while preserving the same number of left-modes and right-modes. By iterating this process over the newly added bubble, generating another tree, we obtain all the bubble diagrams. The algorithm for this process is outlined in Algorithm \ref{alg:tree} and illustrated in Fig.  \ref{fig:tcg-tree}. 

One caveat is worth noting regarding the handling of the $\hat{H}$ term in the contraction defined by Eq.(\ref{eq:contraction_recursion}), which we refer to as the ``special mode'' and represent by a hollow dot in the diagrams. This term has to be treated in a different manner since the time averages in Eq.(\ref{eq:TCG-raw}) are asymmetrical around the density matrix -- the only $\hat{H}$ term always appears as the left-most operator in the contraction. Therefore the mode associated with the $\hat{H}$ term must remain within the right-most bubble in order to preserve this structure, and is treated differently from the other modes in the calculation of their contributions to the diagram.

\subsection{Calculating singular diagram corrections}
\label{app:singular_corrections}
Once we have all the different diagrams, calculating the contraction coefficient is a simple matter of summing up the corrections due to each diagram, according to the formula in Eq.( \ref{eq:contraction-coeff}). Although the formula is simple enough, it proves problematic to directly implement numerically and symbolically, since the different bubble corrections can diverge when any factor in the vector factorial vanishes. In more exact terms, we can define a set that enumerates these singular points for a bubble $i$ in a diagram, 
\begin{subequations}
    \begin{align}
    \mathbf{s}_i^{(\mu)} = \left \{s \big| \mu_i^1 + \mu_i^2 + \cdots + \mu_i^s = 0 \right \} \\
    \mathbf{s}_i^{(\nu)} = \left \{s^{\prime} \big| \nu_i^1 + \nu_i^2 + \cdots + \nu_i^{s^{\prime}} = 0 \right \}
    \end{align}
\end{subequations}
Defined for a particular bubble correction with a vector of left-modes $\vec{\mu}_i$ and a vector of right-modes $\vec{\nu}_i$. For example, for a bubble left-modes $\mu_i = (0, 1, 2, -3)$ we would have $s_i = \{1, 4\}$, since the first factor is $0$ and vanishes trivially, and the fourth factor  $0 + 1 + 2 - 3 = 0$ sums up to zero and vanishes as well. 

Let us define the contribution due to a single diagram as $\epsilon_d$. In order obtain the finite contribution near these singular terms, we expand the diagram near the poles by taking each frequency $\omega \rightarrow \omega + d\omega$ and expanding to the order of the pole.
\begin{equation}
\begin{split}
\epsilon_d
=
\epsilon_d^{(0)} \sum_{n_1 \ge 0, u_1 \ge 0, l_1 \ge 0} \cdots \sum_{n_{||d||} \ge 0, u_{||d||} \ge 0, l_{||d||} \ge 0}&
\left (
a_{n_i}(\vec{\mu}_i ,\vec{\nu}_i) + \frac{b_{u_i}(\vec{\mu}_i)\cdot c_{l_i}(\vec{\nu}_i) }{\prod_{k \in s_i^{(\mu)}, l \in s_i^{(\nu)}} k \cdot l }
\right )\\
&\quad
\times \delta \left ( \sum_i(n_i + u_i + l_i) - \sum_i(|s_i| + |s^{\prime}_{i}|) = 0 \right )
\end{split}
\end{equation}
where the Kronecker-delta $\delta$ denotes that the sum is only over solutions to the partition problem in the argument and the sum is over $3||d||$ indices, for example for $n_i$ we have $(n_1, n_2, ..., n_{||d||})$.

$\epsilon_d^{(0)}$ is just the simple product of the correction factors excluding the resonant terms in the factorial. 
\begin{equation}
\epsilon_d^{(0)}
=
\left ( -1 \right )^{r + ||d|| - 1} \prod^{||d||}_{i=1} \frac{1}{\vec{\mu}_{i}!^{\ast} \cdot \vec{\nu}_{i}!^{\ast}} f \left ( \sum_j \left ( \mu_i^{(j)} + \nu_i^{(j)} \right ) \right )
\end{equation}
and we define the non-singular vector factorial as $\vec{v}_{i}!^{\ast} = \prod_{j \notin s_i^{(v)}} (v^1 + \cdots + v^j)$.

Additionally, we define the auxiliary functions,
\begin{subequations}
    \begin{align}
    a_{n_i}(\vec{\mu}_i, \vec{\nu}_i) &= \sum_{k=0}^{\lfloor\frac{n_i}{2} \rfloor} \frac{c(n_i, k)}{n_i!} \tau^{2(n_i - k)} \left (\sum_i \mu_i + \sum_i \nu_i \right )^{n_i - 2 k} (||\mu_i||_1 + ||\nu_1||_1)^{n_i} \\
    b_{u_i}(\vec{\mu}_i) &= \sum_{m}  \prod_{j \notin s^{(\mu)}_i} \left ( \frac{-j}{\mu_i^1 + \cdots + \mu_i^j} \right )^{m_j} \delta \left ( \sum_{j \notin s_i} m_j - u_i\right ) \\
    c_{l_i}(\vec{\nu}_i) &= \sum_{m_j}  \prod_{j \notin s_i} \left ( \frac{-j}{\nu_i^1 + \cdots + \nu_i^j} \right )^{m_j} \delta \left ( \sum_{j \notin s^{(\nu)}_i} m_j - l_i\right ) \\
    \end{align}
\end{subequations}
where $c(n,k)$ are the expansion coefficient, defined recursively through,
\begin{subequations}
    \begin{align}
    c(0,0) &= 1 \\
    c(n, -1) &\equiv 0 \text{ for  any } n \\
    c(n,k) &\equiv 0 \text{ for } n < 2k
    \end{align}
\end{subequations}
In that form, the diagram corrections are always finite and well defined, and can be explicitly calculated without relying on explicit limits, which are difficult to calculate numerically.

\section{The TCG effective model -- examples}
\subsection{Rabi-model contraction coefficients and the effective master equation}
\label{app:rabi-model_contraction_coefficients}
\begin{figure}[!h]
    \centering
    \includegraphics[width=0.5\textwidth]{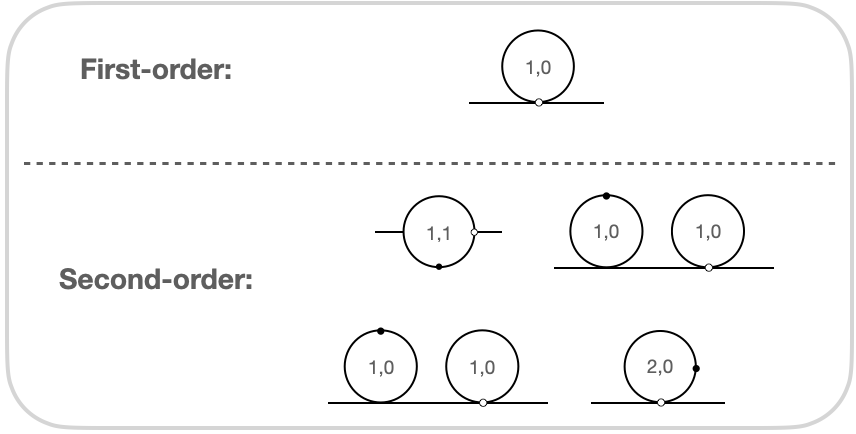}
    \caption{The different diagrams contributing to the TCG superoperators. The top diagram shows the first-order (single-mode) contributions, whereas the bottom diagrams show the second-order (two-mode) contributions.}
    \label{fig:tcg-rabi}
\end{figure}

These coefficients are determined by a set of diagrams. In this simple example, we have five diagrams contributing to the process, as illustrated in Fig.  \ref{fig:tcg-rabi}. For the first-order contractions we have a single diagram with only a single mode.

The diagram contributions yield the following coefficients,
\begin{equation}
    C_{1,0}(\omega) = e^{-\frac{\omega^2\tau^2}{2}}.
\end{equation}
Similarly, for the second-order contractions we have
\begin{equation}
C_{2,0}(\omega, \omega^{\prime})
=
-
C_{1,1}(\omega, \omega^{\prime})
=
\frac{e^{-\frac{(\omega + \omega^{\prime})^{2} \tau^{2}}{2}}}{\omega^{\prime}}
-
\frac{e^{-\frac{\omega^{2}+\omega^{\prime 2}}{2}\tau^{2}}}{\omega^{\prime}}.
\end{equation}
Adding up contributions from all the diagrams, we end up with the second-order effective Hamiltonian
\begin{equation}
\label{eq: rabi_H}
\begin{split}
\hat{H}_{\rm TCG}^{(2)}
=&
\hat{H}_{\textrm{r}}^{(2)}
+
\hat{H}_{\textrm{cr}}^{(2)}
\end{split}
\end{equation}
where the ``co-rotating'' terms $\hat{H}_{\textrm{r}}^{(2)}$ are not exponentially suppressed in the $\omega_{a} \sim \omega_{c} \gg \frac{1}{\tau}$ limit:
\begin{equation}
\label{eq: rabi_rotating_H}
\begin{split}
\hat{H}_{\textrm{r}}^{(2)}
=&
\frac{g}{2}
e^{-\frac{(\omega_{a}-\omega_{c})^{2}\tau^{2}}{2}}
\Big(
e^{-i(\omega_{a}-\omega_{c})t} \hat{a} \hat{\sigma}_{+}
+
e^{+i(\omega_{a}-\omega_{c})t} \hat{a}^{\dagger} \hat{\sigma}_{-}
\Big)
+
\frac{g^{2}}{8}
\Big(
\frac{1 - e^{-(\omega_{a} - \omega_{c})^{2}\tau^{2}}}{\omega_{a} - \omega_{c}}
+
\frac{1}{\omega_{a} + \omega_{c}}
\Big)
\hat{\sigma}_{z}\\
&
+
\frac{g^{2}}{4}
\Big(
\frac{1 - e^{-(\omega_{a} - \omega_{c})^{2}\tau^{2}}}{\omega_{a} - \omega_{c}}
+
\frac{1}{\omega_{a} + \omega_{c}}
\Big)
\hat{a}^{\dagger}\hat{a} \hat{\sigma}_{z}
\end{split}
\end{equation}
while the ``counter-rotating'' terms $\hat{H}_{\textrm{cr}}^{(2)}$ will be exponentially suppressed in the same limit:
\begin{equation}
\begin{split}
\hat{H}_{\textrm{cr}}^{(2)}
=&
\frac{g}{2}
e^{-\frac{(\omega_{a}+\omega_{c})^{2}\tau^{2}}{2}}
\Big(
e^{i(\omega_{a}+\omega_{c})t} \hat{a}^{\dagger} \hat{\sigma}_{+}
+
e^{-i(\omega_{a}+\omega_{c})t} \hat{a} \hat{\sigma}_{-}
\Big)\\
&
+
\frac{g^{2}}{4}
\Big(
e^{-2\omega_{c}^{2}\tau^{2}}
-
e^{-(\omega_{a}^{2} + \omega_{c}^{2})\tau^{2}}
\Big)
\frac{\omega_{a}}{\omega_{a}^{2} - \omega_{c}^{2}}
\big[
e^{-2i \omega_{c} t} \hat{a}^{2} \hat{\sigma}_{z}
+
e^{2i \omega_{c} t} \hat{a}^{\dagger 2} \hat{\sigma}_{z}
\big].
\end{split}
\end{equation}
In addition to the effective Hamiltonian, TCG also gives rise to effective pseudo-dissipators. In particular, at the second order, we can write $\hat{D}^{(k)}$ as
\begin{equation}
\begin{split}
\hat{D}^{(k)}
=
\hat{D}^{(k)}_{\textrm{r}}
+
\hat{D}^{(k)}_{\textrm{cr}}
\end{split}
\end{equation}
where the rotating and counter-rotating pseudo-dissipators can be written as
\begin{equation}
\begin{split}
\hat{D}^{(k)}_{\textrm{r}}
=
\frac{g^{2}}{2}\cdot
\frac{
e^{-(\omega_{a}-\omega_{c})^{2}\tau^{2}}
-
e^{-2(\omega_{a}-\omega_{c})^{2}\tau^{2}}
}{\omega_{a} - \omega_{c}} e^{-2i(\omega_{a} - \omega_{c})t}
\mathcal{D}\big[ \hat{a}^{\dagger} \hat{\sigma}_{-}, \hat{a}^{\dagger} \hat{\sigma}_{-} \big]
+
h.c.
\end{split}
\end{equation}
and
\begin{equation}
\begin{split}
\hat{D}^{(k)}_{\textrm{cr}}
=&
\frac{g^{2}}{2}
\Big(
e^{-2\omega_{c}^{2}\tau^{2}}
-
e^{-(\omega_{a}^{2} + \omega_{c}^{2})\tau^{2}}
\Big)
\frac{\omega_{c}}{\omega_{a}^{2} - \omega_{c}^{2}}
e^{-2i\omega_{c} t}
\Big(
\mathcal{D}\big[ \hat{a} \hat{\sigma}_{+}, \hat{a} \hat{\sigma}_{-} \big]
+
\mathcal{D}\big[ \hat{a} \hat{\sigma}_{-}, \hat{a} \hat{\sigma}_{+} \big]
\Big)\\
&
+
\frac{g^{2}}{2}
\Big(
e^{-2\omega_{a}^{2}\tau^{2}}
-
e^{-(\omega_{c}^{2} + \omega_{a}^{2})\tau^{2}}
\Big)
\frac{\omega_{a}}{\omega_{c}^{2} - \omega_{a}^{2}}
e^{-2i\omega_{a} t}
\Big(
\mathcal{D}\big[ \hat{a} \hat{\sigma}_{-}, \hat{a}^{\dagger} \hat{\sigma}_{-} \big]
+
\mathcal{D}\big[ \hat{a}^{\dagger} \hat{\sigma}_{-}, \hat{a} \hat{\sigma}_{-} \big]
\Big)\\
&
+
\frac{g^{2}}{2}
\Big(
e^{-(\omega_{a}+\omega_{c})^{2}\tau^{2}}
-
e^{-2(\omega_{a}+\omega_{c})^{2}\tau^{2}}
\Big)
\frac{1}{\omega_{a}+\omega_{c}}
e^{-2i(\omega_{a} + \omega_{c})t}
\mathcal{D}\big[ \hat{a} \hat{\sigma}_{-}, \hat{a} \hat{\sigma}_{-} \big]
+
h.c.
\end{split}
\end{equation}
respectively.

\subsection{The driven Kerr-parametron}
\label{app: parametron}
Assuming the numerical values of the parameters presented in Eq.(\ref{eq: parametron parameter values}), we present all TCG superoperators with coefficients greater than $0.08 \textrm{MHz}$ for a coarse-graining time scale of $\tau = 0.125 \textrm{ns}$.
The resulting TCG effective Hamiltonian and dissipators can be written as
\begin{equation}
\begin{split}
\hat{H}_{\textrm{TCG}}(t)
=&
g_{11} \hat{a}^{\dagger} \hat{a}
+
g_{22} \hat{a}^{\dagger 2} \hat{a}^{2}
+
g_{33} \hat{a}^{\dagger 3} \hat{a}^{3}
+
\big(
g_{20} \hat{a}^{\dagger 2}
+
g_{31} \hat{a}^{\dagger 3} \hat{a}
+
h.c.
\big)
\end{split}
\end{equation}
and
\begin{equation}
\begin{split}
\hat{D}_{\textrm{TCG}}(t)
=&
\Gamma_{2,0;0,2} \mathcal{D}[\hat{a}^{\dagger 2}, \hat{a}^{2}]
+
\Gamma_{0,2;2,0} \mathcal{D}[\hat{a}^{2}, \hat{a}^{\dagger 2}]
+
\Big(
\Gamma_{2,0;2,0} \mathcal{D}[\hat{a}^{\dagger 2}, \hat{a}^{\dagger 2}]
+
\Gamma_{2,0;2,2} \mathcal{D}[\hat{a}^{\dagger 2}, \hat{a}^{\dagger 2} \hat{a}^{2}]\\
&
+
\Gamma_{0,2;2,2} \mathcal{D}[\hat{a}^{2}, \hat{a}^{\dagger 2} \hat{a}^{2}]
+
\Gamma_{1,1;2,0} \mathcal{D}[\hat{a}^{\dagger} \hat{a}, \hat{a}^{\dagger 2}]
+
\Gamma_{1,1;0,2} \mathcal{D}[\hat{a}^{\dagger} \hat{a}, \hat{a}^{2}]
+
h.c.
\Big)
\end{split}
\end{equation}
respectively, with
\begin{equation}
\begin{split}
g_{11}
=&
\Delta(t)
-
\frac{2 \big( \beta(t)^{2} + \beta(\tau)^{2} \big)}{\omega_{p}}
-
\frac{4\chi^{2}}{\omega_{p}}\\
&
+
\Delta_{0} \beta(\tau)^{2} \Big( 20 \tau^{2} + \frac{11}{2 \omega_{p}^{2}} \Big)
-
\Delta(t) \beta(\tau)^{2} \Big( 26 \tau^{2} + \frac{13}{\omega_{p}^{2}} \Big)
-
\Big( \frac{\tau^{2} \Delta_{0}}{T} \Big)^{2} \Delta(t)
+
\frac{3}{2} \frac{\beta_{0} \Delta_{0}}{T^{2}\omega_{p}^{4}} \beta(t)\\
&
+
2 \Big( \frac{\beta(t)}{\omega_{p}} \Big)^{2} \Delta(t)
-
\frac{51}{2} \Big(\frac{\beta_{0}}{T\omega_{p}^{2}}\Big)^{2} \chi
-
\Big(
6 \big( \frac{\beta(t)}{\omega_{p}} \big)^{2}
-
4 \big( \frac{\beta(\tau)}{\omega_{p}} \big)^{2}
\Big) \chi
+
8 \Big( \frac{\chi}{\omega_{p}} \Big)^{2} \Delta(t)
-
\frac{151 \chi^{3}}{6 \omega_{p}^{2}}
\end{split}
\end{equation}
\begin{equation}
\begin{split}
g_{22}
=&
-
\frac{\chi}{2}
-
\frac{17}{4} \frac{\chi^{2}}{\omega_{p}}
+
\Delta_{0} \beta(\tau)^{2} \Big( 16\tau^{2} + \frac{26}{\omega_{p}^{2}} \Big)
-
\Delta(t) \beta(\tau)^{2} \Big( 24\tau^{2} + \frac{52}{\omega_{p}^{2}} \Big)
-
\Big( \frac{\tau^{2} \Delta_{0}}{T} \Big)^{2} \big( 3 \Delta(t) - 2\chi \big)
\\
&
+
7 \beta(\tau)^{2} \tau^{2} \chi
-
9 \Big( \frac{\beta_{0}}{T \omega_{p}^{2}} \Big)^{2} \chi
-
\Big( 4 \big(\frac{\beta(t)}{\omega_{p}}\big)^{2} - \frac{137}{4} \big(\frac{\beta(\tau)}{\omega_{p}}\big)^{2} \Big)\chi
+
\frac{17}{2} \Big(\frac{\chi}{\omega_{p}}\Big)^{2} \Delta(t)
-
\frac{199}{4} \frac{\chi^{3}}{\omega_{p}^{2}}
\end{split}
\end{equation}
\begin{equation}
\begin{split}
g_{33}
=&
-
\frac{17}{18} \frac{\chi^{2}}{\omega_{p}}
+
\Delta_{0} \beta(\tau)^{2} \Big( 4\tau^{2} + \frac{17}{2 \omega_{p}^{2}} \Big)
-
\Delta(t) \beta(\tau)^{2} \Big( 6\tau^{2} + \frac{17}{\omega_{p}^{2}} \Big)
-
\Big( \frac{\tau^{2} \Delta_{0}}{T} \Big)^{2} \Big( \Delta(t) - \frac{5}{2}\chi \Big)\\
&
+
\beta(\tau)^{2} \Big( 6\tau^{2} + \frac{191}{6\omega_{p}^{2}} \Big) \chi
+
\frac{5}{2} \Big( \frac{\beta_{0}}{T \omega_{p}^{2}} \Big)^{2} \chi
+
\frac{17}{9} \Big( \frac{\chi}{\omega_{p}} \Big)^{2} \Delta(t)
-
\frac{233\chi^{3}}{9\omega_{p}^{2}}
\end{split}
\end{equation}
\begin{equation}
\begin{split}
g_{20}
=&
\beta(t)
+
\frac{5\chi}{2\omega_{p}} \beta(t)
-
\frac{17}{4} \frac{\chi^{2}}{\omega_{p}}
+
i \Big(
\beta(\tau) \tau \big( \Delta_{0} - \frac{\chi}{2} \big)
-
\frac{9 \beta_{0} \chi}{8 T \omega_{p}^{2}}
\Big)\\
&
+
\Big( 16 \Delta_{0} - 17 \beta(t) \Big) \beta(\tau)^{2} \tau^{2}
-
24 \Delta(t) \beta(\tau)^{2} \tau^{2}
-
\big( 7 \beta(t) - 4 \beta_{0} \big) \Big( \frac{\tau^{2} \Delta_{0}}{T} \Big)^{2}
-
3 \Delta(t) \Big( \frac{\tau^{2} \Delta_{0}}{T} \Big)^{2}\\
&
+
\frac{15}{2} \beta(\tau^{2}) \tau^{2} \chi
-
2 \frac{
(3\beta_{0}-\Delta_{0}) \Delta_{0} \chi \tau^{4}
}{T^{2}}
+
\Big( \frac{11\beta_{0}}{2 T \omega_{p}^{2}} \Big)^{2} \beta(t)
-
\frac{53}{6} \Big( \frac{\beta_{0}}{T \omega_{p}^{2}} \Big)^{2} \chi
-
\frac{3 \beta_{0} \Delta_{0} \chi}{16 \big(T \omega_{p}^{2}\big)^{2}}\\
&
-
\frac{7 \beta(t)^{3}}{\omega_{p}^{2}}
-
\Big( \frac{\beta_{0} \tau}{T \omega_{p}} \Big)^{2}
\big(
74 \beta(t)
+
52 \Delta(t)
-
26 \Delta_{0}
\big)
+
\frac{197}{12} \Big(\frac{\chi}{\omega_{p}}\Big)^{2} \beta(t)
+
\frac{17}{2} \Big(\frac{\chi}{\omega_{p}}\Big)^{2} \Delta(t)
-
\frac{199\chi^{3}}{4\omega_{p}^{2}}\\
&
+
\frac{
\frac{559}{16} \beta(\tau)^{2}
-
\frac{83}{48} \beta(\tau) \frac{\tau}{T} \Delta_{0}
-
4 \beta(t)^{2}
-
\frac{11}{4} \beta(t) \Delta(t)
}{\omega_{p}^{2}} \chi\\
&
+
i \Big[
\Big(
2 \beta(t)^{2} \tau^{2}
+
\frac{7
\big(
\beta(t)^{2} + \beta(\tau)^{2}
\big)
}{2 \omega_{p}^{2}}
\Big) \frac{\beta_{0}}{T \omega_{p}}
-
\frac{
\chi \big( \frac{\beta_{0}}{6} - \frac{15}{8}\Delta_{0} \big)
}{T \omega_{p}^{3}} \beta(t)
+
\frac{\beta_{0} \chi}{T \omega_{p}^{3}} \Big( \frac{\Delta_{0}}{4} - \frac{167 \chi}{24} \Big)\\
&\qquad
+
\frac{5 \beta_{0} \Delta_{0} \chi \tau^{2}}{2 T \omega_{p}}
-
\frac{5 \beta_{0} \tau^{2} \chi^{2}}{T \omega_{p}}
\Big]
\end{split}
\end{equation}
\begin{equation}
\begin{split}
g_{31}
=&
\frac{5\chi}{3\omega_{p}} \beta(t)
-
i \Big[
\tau \beta(\tau) \chi
+
\frac{3\beta_{0}\chi}{4T\omega_{p}^{2}}
\Big]\\
&
-
3\tau^{2} \Big(
4\beta(\tau)^{2}
+
3\Big(\frac{\tau \Delta_{0}}{T}\Big)^{2}
\Big) \beta(t)
-
9 \Big( \frac{\beta_{0}}{T \omega_{p}^{2}} \Big)^{2} \beta(t)
-
52 \Big( \frac{\beta(\tau)}{\omega_{p}} \Big)^{2} \beta(t)
+
\frac{\beta_{0} \Delta_{0} \tau^{4}}{T^{2}} \big( 6\Delta_{0} - 5\chi \big)\\
&
-
\frac{11 \chi}{6 \omega_{p}^{2}} \Delta(t) \beta(t)
-
\frac{\beta_{0}\Delta_{0}\chi}{8T^{2}\omega_{p}^{2}} \Big( 27\tau^{2} + \frac{1}{\omega_{p}^{2}} \Big)
+
\frac{119}{6} \Big(\frac{\chi}{\omega_{p}}\Big)^{2} \beta(t)\\
&
+
i \Big[
\frac{5\Delta_{0}\chi}{4T\omega_{p}^{3}} \beta(t)
-
\frac{9 \beta_{0}}{T \omega_{p}^{3}} \Big( \beta(t)^{2} + \beta(\tau)^{2} \Big)
+
\frac{\beta_{0} \Delta_{0} \chi}{6 T \omega_{p}^{3}}
-
\frac{50 \beta_{0} \chi^{2}}{3T \omega_{p}^{3}}
+
\frac{5 \beta_{0} \Delta_{0} \chi \tau^{2}}{3T \omega_{p}}
-
\frac{49 \beta_{0} \tau^{2} \chi^{2}}{3T \omega_{p}}
\Big]
\end{split}
\end{equation}
\begin{equation}
\begin{split}
&
\Gamma_{2,0;0,2}
=
\Gamma_{0,2;2,0}
=
\frac{\beta_{0}}{T} \Big( 2\tau^{2} + \frac{1}{2\omega_{p}^{2}} \Big) \beta(t)
;\quad
\Gamma_{2,0;2,0}
=
\frac{2\tau^{2} \beta_{0}}{T} \beta(t)
;\quad
\Gamma_{2,0;2,2}
=
\Gamma_{0,2;2,2}
=
-
\frac{\tau \chi}{2} \beta(\tau)\\
&
\Gamma_{1,1;2,0}
=
-
\frac{45 \beta_{0}}{2T\omega_{p}^{3}} \Big( \beta(t)^{2} + \beta(\tau)^{2} \Big)
-
\frac{2\tau^{2}\beta_{0}}{T \omega_{p}} \Big( 3\beta(t)^{2} + 2\beta(\tau)^{2} \Big)
-
\frac{5}{2} \Big( 2\beta(t) - \beta_{0} \Big) \frac{\tau^{2}\Delta_{0}\chi}{T \omega_{p}}\\
&\qquad\quad\;\;\;
-
\frac{2\Delta_{0}\chi}{T \omega_{p}^{3}} \beta(t)
-
\frac{19\beta_{0}\chi^{2}}{3T\omega_{p}^{3}}
-
\frac{4\tau\chi^{2}}{\omega_{p}} \beta(\tau)
+
i \Big[
3 \tau^{2} \Big( \Big( \frac{\tau\Delta_{0}}{T} \Big)^{2} - 4 \beta(\tau)^{2} \Big) \beta(t)
-
4 \Big( \frac{\tau^{2}\Delta_{0}}{T} \Big)^{2} \beta_{0}\\
&\qquad\quad\;\;\;\,
+
2 \tau \beta(\tau) \Big( \frac{\tau^{2}\Delta_{0}}{T} \Big) \chi
+
\frac{183}{4} \Big(\frac{\beta_{0}}{T \omega_{p}^{2}}\Big)^{2} \beta(t)
-
17 \frac{\beta(t)^{3}}{\omega_{p}^{2}}
-
39 \Big(\frac{\beta(\tau)}{\omega_{p}}\Big)^{2} \beta(t)
-
\frac{2\beta(t) \Delta(t)}{\omega_{p}^{2}} \chi\\
&\qquad\quad\;\;\;\,
+
\frac{13\tau^{2}\beta_{0}\Delta_{0}\chi}{4T^{2}\omega_{p}^{2}}
-
\Big(\frac{\chi}{\omega_{p}}\Big)^{2} \beta(t)
\Big]\\
&
\Gamma_{1,1;0,2}
=
-
\frac{33 \beta_{0}}{2T \omega_{p}^{3}}
\Big( \beta(t)^{2} + \beta(\tau)^{2} \Big)
-
\frac{2\tau^{2}\beta_{0}}{T \omega_{p}} \Big( 3\beta(t)^{2} + 2\beta(\tau)^{2} \Big)
-
\frac{5}{2} \Big( 2\beta(t) - \beta_{0} \Big) \frac{\tau^{2}\Delta_{0}\chi}{T \omega_{p}}\\
&\qquad\quad\;\;\;\,
-
\frac{2\Delta_{0}\chi}{T \omega_{p}^{3}} \beta(t)
-
\frac{25\beta_{0}\chi^{2}}{6T\omega_{p}^{3}}
-
\frac{4\tau\chi^{2}}{\omega_{p}} \beta(\tau)
+
i \Big[
3 \tau^{2} \Big( \Big( \frac{\tau\Delta_{0}}{T} \Big)^{2} - 4 \beta(\tau)^{2} \Big) \beta(t)\\
&\qquad\quad\;\;\;\,
-
\frac{159}{4} \Big( \frac{\beta_{0}}{T \omega_{p}^{2}} \Big)^{2} \beta(t)
+
17 \frac{\beta(t)^{3}}{\omega_{p}^{2}}
+
63 \Big(\frac{\beta(\tau)}{\omega_{p}}\Big)^{2} \beta(t)
+
\frac{2\beta(t)\Delta(t)}{\omega_{p}^{2}} \chi
-
\frac{\tau^{2}\beta_{0}\Delta_{0}\chi}{T^{2}\omega_{p}^{2}}\\
&\qquad\quad\;\;\;\,
+
\Big(\frac{\chi}{\omega_{p}}\Big)^{2} \beta(t)
\Big]
\end{split}
\end{equation}
Although we assumed particular numerical values of the parameters for truncating the TCG master equation, the formulas above can be used as good approximations as long as the coarse-graining time scale is in the range $\frac{1}{\abs{\beta(t)}}, \frac{1}{\abs{\Delta(t)}}, \frac{1}{\abs{\chi}} \gg \tau \gg \frac{1}{\omega_{p}}$. Here the first inequality ensures the validity of the TCG perturbative expansion regardless of the frequencies of the terms involved, whereas the second inequality allows us to ignore the highly oscillatory terms in the TCG master equation.

\subsection{The driven Duffing oscillator}
\label{subsec: The driven Duffing}
The fourth-order effective TCG Hamiltonian can be written as
\begin{equation}
\begin{split}
\hat{H}_{\textrm{TCG}}^{(4)}
\equiv
K^{(4)}_{1} \hat{a}^{\dagger} \hat{a}
+
K^{(4)}_{2} \hat{a}^{\dagger 2} \hat{a}^{2}
+
K^{(4)}_{3} \hat{a}^{\dagger 3} \hat{a}^{3}
+
K^{(4)}_{4} \hat{a}^{\dagger 4} \hat{a}^{4}
+
K^{(4)}_{5} \hat{a}^{\dagger 5} \hat{a}^{5}
\end{split}
\end{equation}
where the approximate expressions for the coefficients $K^{(4)}_{n}$ are given in subsection \ref{subsec: duffing}, and the exact analytical formulas for the coefficients can be written as
\begin{fleqn}
\begin{equation}
\begin{split}
&
K^{(4)}_{1}\\
=&
\frac{g_{4}^{2}}{\omega}
\big(
-
\frac{288}{5}
+
\frac{240448}{385} \abs{\Pi}^{2}
+
\frac{29232}{55} \abs{\Pi}^{4}
\big)
+
\frac{g_{4}^{2}}{\omega^{2}} \delta
\big(
\frac{288}{25}
+
\frac{403234112}{444675} \abs{\Pi}^{2}
+
\frac{2011248}{3025} \abs{\Pi}^{4}
\big)\\
&
+
\frac{g_{4}^{3}}{\omega^{2}}
\big(
\frac{14328}{25}
+
\frac{1228651264}{88935} \abs{\Pi}^{2}
+
\frac{6411937792}{148225} \abs{\Pi}^{4}
+
\frac{5904212871552}{270438025} \abs{\Pi}^{6}
\big)\\
&
+
\frac{g_{4}^{4}}{\omega^{3}}
\big(
-
\frac{39168}{5}
+
\frac{8297258809465088}{33612242125} \abs{\Pi}^{2}
+
\frac{215864405353424}{102719925} \abs{\Pi}^{4}\\
&\qquad\quad
+
\frac{4629035610176328876032}{1525442189145875} \abs{\Pi}^{6}
+
\frac{16321516698840554304}{15121001291825} \abs{\Pi}^{8}
\big)\\
&
+
\frac{g_{4}^{3} \delta}{\omega^{3}}
\big(
-
\frac{28656}{125}
+
\frac{3609939830272}{102719925} \abs{\Pi}^{2}
+
\frac{18278157853696}{171199875} \abs{\Pi}^{4}
+
\frac{228204470088870144}{4447353321125} \abs{\Pi}^{6}
\big)\\
&
+
\frac{g_{4}^{2} \delta^{2}}{\omega^{3}}
\big(
-
\frac{288}{125}
+
\frac{439080132928}{513599625} \abs{\Pi}^{2}
+
\frac{107346672}{166375} \abs{\Pi}^{4}
\big)
\end{split}
\end{equation}

\end{fleqn}
\begin{fleqn}

\begin{equation}
\begin{split}
K^{(4)}_{2}
=&
\frac{g_{4}^{2}}{\omega}
\big(
-
\frac{306}{5}
+
\frac{120224}{385} \abs{\Pi}^{2}
\big)
+
\frac{g_{4}^{2}}{\omega^{2}} \delta
\big(
\frac{306}{25}
+
\frac{201617056}{444675} \abs{\Pi}^{2}
\big)\\
&
+
\frac{g_{4}^{3}}{\omega^{2}}
\big(
\frac{25164}{25}
+
\frac{531220656}{29645} \abs{\Pi}^{2}
+
\frac{3205968896}{148225} \abs{\Pi}^{4}
\big)\\
&
+
\frac{g_{4}^{4}}{\omega^{3}}
\big(
-
\frac{513234}{25}
+
\frac{22048747044860544}{33612242125} \abs{\Pi}^{2}
+
\frac{2825906273812}{1037575} \abs{\Pi}^{4}\\
&\qquad\quad
+
\frac{2314517805088164438016}{1525442189145875} \abs{\Pi}^{6}
\big)\\
&
+
\frac{g_{4}^{3}\delta}{\omega^{3}}
\big(
-
\frac{50328}{125}
+
\frac{529778123296}{11413325} \abs{\Pi}^{2}
+
\frac{9139078926848}{171199875} \abs{\Pi}^{4}
\big)\\
&
+
\frac{g_{4}^{2}\delta^{2}}{\omega^{3}}
\big(
-
\frac{306}{125}
+
\frac{219540066464}{513599625} \abs{\Pi}^{2}
\big)
\end{split}
\end{equation}

\end{fleqn}
\begin{fleqn}

\begin{equation}
\begin{split}
K^{(4)}_{3}
=&
-
\frac{g_{4}^{2}}{\omega} \cdot \frac{68}{5}
+
\frac{g_{4}^{2}}{\omega^{2}} \delta \cdot \frac{68}{25}
+
\frac{g_{4}^{3}}{\omega^{2}}
\big(
480
+
\frac{354147104}{88935} \abs{\Pi}^{2}
\big)\\
&
+
\frac{g_{4}^{4}}{\omega^{3}}
\big(
-
\frac{399092}{25}
+
\frac{286401882242048}{806693811} \abs{\Pi}^{2}
+
\frac{5651812547624}{9338175} \abs{\Pi}^{4}
\big)\\
&
+
\frac{g_{4}^{3}\delta}{\omega^{3}}
\big(
-192
+
\frac{1059556246592}{102719925} \abs{\Pi}^{2}
\big)
-
\frac{g_{4}^{2}\delta^{2}}{\omega^{3}} \frac{68}{125}
\end{split}
\end{equation}

\end{fleqn}
\begin{fleqn}

\begin{equation}
\begin{split}
K^{(4)}_{4}
=&
\frac{g_{4}^{3}}{\omega^{2}} \cdot 60
+
\frac{g_{4}^{4}}{\omega^{3}} \big(
-
\frac{21378}{5}
+
\frac{35800235280256}{806693811} \abs{\Pi}^{2}
\big)
-
\frac{g_{4}^{3}\delta}{\omega^{3}} \cdot 24 
\end{split}
\end{equation}

\end{fleqn}
\begin{fleqn}

\begin{equation}
\begin{split}
K^{(4)}_{5}
=&
-
\frac{g_{4}^{4}}{\omega^{3}} \cdot \frac{42756}{125}
\end{split}
\end{equation}
\end{fleqn}

The TCG dissipators in $D_{\textrm{TCG}}(t)$ start to appear at the third-order in the perturbative expansion. Keeping the lowest-order contributions, we have
\begin{equation}
\begin{split}
&
\hat{D}_{\textrm{TCG}}^{(3)} \equiv \Gamma^{(3)}_{2,1 ; 0,1} \mathcal{D}[\hat{a}^{\dagger 2} \hat{a}, \hat{a}]
+
\Gamma^{(3)}_{0,1 ; 2,1} \mathcal{D}[\hat{a}, \hat{a}^{\dagger 2} \hat{a}] \\
&+
\Gamma^{(3)}_{2,2 ; 1,1} \mathcal{D}[\hat{a}^{\dagger 2}\hat{a}^{2}, \hat{a}^{\dagger}\hat{a}]
+
\Gamma^{(3)}_{2,3 ; 1,0} \mathcal{D}[\hat{a}^{\dagger 2}\hat{a}^{3}, \hat{a}^{\dagger}]\\
&
+
\Gamma^{(3)}_{1,0 ; 2,3} \mathcal{D}[\hat{a}^{\dagger}, \hat{a}^{\dagger 2}\hat{a}^{3}]
+
\Gamma^{(3)}_{1,3 ; 2,0} \mathcal{D}[\hat{a}^{\dagger}\hat{a}^{3}, \hat{a}^{\dagger 2}] \\
&+
\Gamma^{(3)}_{2,0 ; 1,3} \mathcal{D}[\hat{a}^{\dagger 2}, \hat{a}^{\dagger}\hat{a}^{3}]
+
\Gamma^{(3)}_{3,3 ; 1,1} \mathcal{D}[\hat{a}^{\dagger 3}\hat{a}^{3}, \hat{a}^{\dagger}\hat{a}]\\
&
+
\Gamma^{(3)}_{4,2 ; 0,2} \mathcal{D}[\hat{a}^{\dagger 4}\hat{a}^{2}, \hat{a}^{2}]
+
\Gamma^{(3)}_{0,2 ; 4,2} \mathcal{D}[\hat{a}^{2}, \hat{a}^{\dagger 4}\hat{a}^{2}] \\
&+
\Gamma^{(3)}_{4,1 ; 0,3} \mathcal{D}[\hat{a}^{\dagger 4}\hat{a}, \hat{a}^{3}]
+
\Gamma^{(3)}_{0,3 ; 4,1} \mathcal{D}[\hat{a}^{3}, \hat{a}^{\dagger 4}\hat{a}]\\
&+
\Gamma^{(3)}_{3,4 ; 1,0} \mathcal{D}[\hat{a}^{\dagger 3}\hat{a}^{4}, \hat{a}^{\dagger}]
+
\Gamma^{(3)}_{1,0 ; 3,4} \mathcal{D}[\hat{a}^{\dagger}, \hat{a}^{\dagger 3}\hat{a}^{4}] \\
&+
\Gamma^{(3)}_{3,2 ; 1,2} \mathcal{D}[\hat{a}^{\dagger 3}\hat{a}^{2}, \hat{a}^{\dagger}\hat{a}^{2}]
+
\Gamma^{(3)}_{1,2 ; 3,2} \mathcal{D}[\hat{a}^{\dagger}\hat{a}^{2}, \hat{a}^{\dagger 3}\hat{a}^{2}]\\
&
+
\Gamma^{(3)}_{3,3 ; 2,2} \mathcal{D}[\hat{a}^{\dagger 3}\hat{a}^{3}, \hat{a}^{\dagger 2} \hat{a}^{2}]
+
\Gamma^{(3)}_{4,4 ; 1,1} \mathcal{D}[\hat{a}^{\dagger 4}\hat{a}^{4}, \hat{a}^{\dagger} \hat{a}] \\
&+
\Gamma^{(3)}_{4,2 ; 1,3} \mathcal{D}[\hat{a}^{\dagger 4}\hat{a}^{2}, \hat{a}^{\dagger} \hat{a}^{3}]
+
\Gamma^{(3)}_{1,3 ; 4,2} \mathcal{D}[\hat{a}^{\dagger} \hat{a}^{3}, \hat{a}^{\dagger 4}\hat{a}^{2}]\\
&+
\Gamma^{(3)}_{5,1 ; 0,4} \mathcal{D}[\hat{a}^{\dagger 5}\hat{a}, \hat{a}^{4}]
+
\Gamma^{(3)}_{0,4 ; 5,1} \mathcal{D}[\hat{a}^{4}, \hat{a}^{\dagger 5}\hat{a}] \\
&+ \Gamma^{(3)}_{3,4 ; 2,1} \mathcal{D}[\hat{a}^{\dagger 3}\hat{a}^{4}, \hat{a}^{\dagger 2}\hat{a}]
+
\Gamma^{(3)}_{2,1 ; 3,4} \mathcal{D}[\hat{a}^{\dagger 2}\hat{a}, \hat{a}^{\dagger 3}\hat{a}^{4}]\\
&+
\Gamma^{(3)}_{5,2 ; 0,3} \mathcal{D}[\hat{a}^{\dagger 5}\hat{a}^{2}, \hat{a}^{3}]
+
\Gamma^{(3)}_{0,3 ; 5,2} \mathcal{D}[\hat{a}^{3}, \hat{a}^{\dagger 5}\hat{a}^{2}] \\
&+
\Gamma^{(3)}_{3,5 ; 2,0} \mathcal{D}[\hat{a}^{\dagger 3}\hat{a}^{5}, \hat{a}^{\dagger 2}]
+
\Gamma^{(3)}_{2,0 ; 3,5} \mathcal{D}[\hat{a}^{\dagger 2}, \hat{a}^{\dagger 3}\hat{a}^{5}]\\
&
+
\Gamma^{(3)}_{4,4 ; 2,2} \mathcal{D}[\hat{a}^{\dagger 4}\hat{a}^{4}, \hat{a}^{\dagger 2}\hat{a}^{2}]
+
\Gamma^{(3)}_{6,2 ; 0,4} \mathcal{D}[\hat{a}^{\dagger 6}\hat{a}^{2}, \hat{a}^{4}] \\
&+
\Gamma^{(3)}_{0,4 ; 6,2} \mathcal{D}[\hat{a}^{4}, \hat{a}^{\dagger 6}\hat{a}^{2}]
+
\Gamma^{(3)}_{3,5 ; 3,1} \mathcal{D}[\hat{a}^{\dagger 3}\hat{a}^{5}, \hat{a}^{\dagger 3}\hat{a}^{1}]\\
&
+
\Gamma^{(3)}_{3,1 ; 3,5} \mathcal{D}[\hat{a}^{\dagger 3}\hat{a}^{1}, \hat{a}^{\dagger 3}\hat{a}^{5}]
+
h.c.
\end{split}
\end{equation}
The analytical expressions for the coefficients $\Gamma^{(3)}_{\alpha,\beta;\gamma,\rho}$ are given below.
\begin{fleqn}
\begin{equation}
\begin{split}
\Gamma^{(3)}_{2,1 ; 0,1}
=&
-i
\frac{g_{4}^{3}}{\omega^{2}}
\big(
\frac{421632}{121} \abs{\Pi}^{2}
+
\frac{1475712}{121} \abs{\Pi}^{4}
+
\frac{131946344448}{10817521} \abs{\Pi}^{6}
+
\frac{37726754304}{10817521} \abs{\Pi}^{8}
\big)\\
&
-
i \frac{g_{4}^{2}\delta}{\omega^{2}}
\big(
\frac{35136}{121} \abs{\Pi}^{2}
+
\frac{52704}{121} \abs{\Pi}^{4}
+
\frac{1571948096}{10817521} \abs{\Pi}^{6}
\big)
\end{split}
\end{equation}

\end{fleqn}
\begin{fleqn}

\begin{equation}
\begin{split}
\Gamma^{(3)}_{0,1 ; 2,1}
=&
i
\frac{g_{4}^{3}}{\omega^{2}}
\big(
\frac{843264}{121} \abs{\Pi}^{2}
+
\frac{2108160}{121} \abs{\Pi}^{4}
+
\frac{150809721600}{10817521} \abs{\Pi}^{6}
+
\frac{37726754304}{10817521} \abs{\Pi}^{8}
\big)\\
&
+
i
\frac{g_{4}^{2}\delta}{\omega^{2}}
\big(
\frac{35136}{121} \abs{\Pi}^{2}
+
\frac{52704}{121} \abs{\Pi}^{4}
+
\frac{1571948096}{10817521} \abs{\Pi}^{6}
\big)
\end{split}
\end{equation}

\end{fleqn}
\begin{fleqn}

\begin{equation}
\label{Gamma2211}
\begin{split}
\Gamma^{(3)}_{2,2 ; 1,1}
=&
-i
g_{4}^{3} \big( 1 + \abs{\Pi}^{2} \big)
\big(
5184 \tau^{2}
+
17280 \tau^{2} \abs{\Pi}^{2}
+
( \frac{48}{\omega^{2}} + 13824 \tau^{2} ) \abs{\Pi}^{4}
\big)\\
&
-
i g_{4}^{2} \delta
\big(
936 \tau^{2}
+
2592 \tau^{2} \abs{\Pi}^{2}
+
(1728 \tau^{2} + \frac{2}{\omega^{2}}) \abs{\Pi}^{4}
\big)
-
i g_{4} \delta^{2} \tau^{2} (54 + 72 \abs{\Pi}^{2})
-
i \delta^{3} \tau^{2}
\end{split}
\end{equation}

\end{fleqn}
\begin{fleqn}

\begin{equation}
\begin{split}
\Gamma^{(3)}_{2,3 ; 1,0}
=
-i
\frac{g_{4}^{3}}{\omega^{2}}
\big(
\frac{737856}{121} \abs{\Pi}^{2}
+
\frac{1264896}{121} \abs{\Pi}^{4}
+
\frac{47126011008}{10817521} \abs{\Pi}^{6}
\big)
-
i
\frac{\delta_{4}^{2}\delta}{\omega^{2}}
\frac{17568}{121} ( \abs{\Pi}^{2} + \abs{\Pi}^{4} )
\end{split}
\end{equation}

\end{fleqn}
\begin{fleqn}

\begin{equation}
\begin{split}
\Gamma^{(3)}_{1,0 ; 2,3}
=
i
\frac{g_{4}^{3}}{\omega^{2}}
\big(
\frac{527040}{121} \abs{\Pi}^{2}
+
\frac{1054080}{121} \abs{\Pi}^{4}
+
\frac{47126011008}{10817521} \abs{\Pi}^{6}
\big)
+
i \frac{g_{4}^{2} \delta}{\omega^{2}} \frac{17568}{121} ( \abs{\Pi}^{2} + \abs{\Pi}^{4} )
\end{split}
\end{equation}

\end{fleqn}
\begin{fleqn}

\begin{equation}
\begin{split}
\Gamma^{(3)}_{1,3 ; 2,0}
=&
-
i
\frac{g_{4}^{3}}{\omega^{2}}
\big(
\frac{108}{5}
+
\frac{2088}{25} \abs{\Pi}^{2}
+
\frac{1284408}{3025} \abs{\Pi}^{4}
+
\frac{763344}{3025} \abs{\Pi}^{6}
\big)\\
&
-
i
\frac{g_{4}^{2}\delta}{\omega^{2}}
\big(
\frac{3}{5}
+
\frac{48}{25} \abs{\Pi}^{2}
+
\frac{31806}{3025} \abs{\Pi}^{4}
\big)
\end{split}
\end{equation}

\end{fleqn}
\begin{fleqn}

\begin{equation}
\begin{split}
\Gamma^{(3)}_{2,0 ; 1,3}
=
i
\frac{g_{4}^{3}}{\omega^{2}}
\big(
\frac{36}{5}
+
\frac{936}{25} \abs{\Pi}^{2}
+
\frac{521064}{3025} \abs{\Pi}^{4}
+
\frac{763344}{3025} \abs{\Pi}^{6}
\big)
+
i
\frac{g_{4}^{2}\delta}{\omega^{2}}
\big(
\frac{3}{5}
+
\frac{48}{25} \abs{\Pi}^{2}
+
\frac{31806}{3025} \abs{\Pi}^{4}
\big)
\end{split}
\end{equation}

\end{fleqn}
\begin{fleqn}

\begin{equation}
\label{Gamma3311}
\begin{split}
\Gamma^{(3)}_{3,3 ; 1,1}
=&
-
i
g_{4}^{3}
\big(
3456 \tau^{2}
+
10368 \tau^{2} \abs{\Pi}^{2}
+
( \frac{12}{\omega^{2}} + 6912 \tau^{2} ) \abs{\Pi}^{4}
\big)
-
i
g_{4}^{2} \delta
\big( 432\tau^{2} + 576\tau^{2} \abs{\Pi}^{2} \big)\\
&
-
i 12 g_{4} \delta^{2} \tau^{2}
\end{split}
\end{equation}

\end{fleqn}
\begin{fleqn}

\begin{equation}
\begin{split}
\Gamma^{(3)}_{4,2 ; 0,2}
=
-
i
\frac{g_{4}^{3}}{\omega^{2}}
\big(
\frac{198}{25}
+
\frac{648}{25} \abs{\Pi}^{2}
+
\frac{225684}{3025} \abs{\Pi}^{4}
\big)
-
i \frac{g_{4}^{2}\delta}{\omega^{2}}
\frac{6}{25} \big( 1 + 2 \abs{\Pi}^{2} \big)
\end{split}
\end{equation}

\end{fleqn}

\begin{fleqn}

\begin{equation}
\begin{split}
\Gamma^{(3)}_{0,2 ; 4,2}
=
i
\frac{g_{4}^{3}}{\omega^{2}}
\big(
\frac{342}{25}
+
\frac{936}{25} \abs{\Pi}^{2}
+
\frac{225684}{3025} \abs{\Pi}^{4}
\big)
+
i \frac{g_{4}^{2}\delta}{\omega^{2}}
\frac{6}{25} \big( 1 + 2 \abs{\Pi}^{2} \big)
\end{split}
\end{equation}

\end{fleqn}
\begin{fleqn}

\begin{equation}
\begin{split}
\Gamma^{(3)}_{4,1 ; 0,3}
=
-
i
\frac{g_{4}^{3}}{\omega^{2}}
\frac{3712}{1323}
\big(
\abs{\Pi}^{2}
+
2 \abs{\Pi}^{4}
\big)
-
i \frac{g_{4}^{2}\delta}{\omega^{2}} \frac{928}{3969} \abs{\Pi}^{2}
\end{split}
\end{equation}

\begin{equation}
\begin{split}
\Gamma^{(3)}_{0,3 ; 4,1}
=
i
\frac{g_{4}^{3}}{\omega^{2}}
\frac{7424}{1323}
\big(
2 \abs{\Pi}^{2}
+
\abs{\Pi}^{4}
\big)
+
i
\frac{g_{4}^{2}\delta}{\omega^{2}} \frac{928}{3969} \abs{\Pi}^{2}
\end{split}
\end{equation}

\begin{equation}
\begin{split}
\Gamma^{(3)}_{3,4 ; 1,0}
=
-
i
\frac{g_{4}^{3}}{\omega^{2}}
\frac{105408}{121}
\big(
\abs{\Pi}^{2}
+
\abs{\Pi}^{4}
\big)
;\qquad\qquad
\Gamma^{(3)}_{1,0 ; 3,4}
=
i
\frac{g_{4}^{3}}{\omega^{2}}
\frac{105408}{121}
\big(
\abs{\Pi}^{2}
+
\abs{\Pi}^{4}
\big)
\end{split}
\end{equation}

\end{fleqn}
\begin{fleqn}

\begin{equation}
\begin{split}
\Gamma^{(3)}_{3,2 ; 1,2}
=
-
i
\frac{g_{4}^{3}}{\omega^{2}}
\frac{527040}{121}
\big(
\abs{\Pi}^{2}
+
\abs{\Pi}^{4}
\big)
-
i
\frac{g_{4}^{2}\delta}{\omega^{2}}
\frac{17568}{121} \abs{\Pi}^{2}
\end{split}
\end{equation}

\begin{equation}
\begin{split}
\Gamma^{(3)}_{1,2 ; 3,2}
=
i
\frac{g_{4}^{3}}{\omega^{2}}
\frac{105408}{121}
\big(
7 \abs{\Pi}^{2}
+
5 \abs{\Pi}^{4}
\big)
+
i
\frac{g_{4}^{2}\delta}{\omega^{2}}
\frac{17568}{121} \abs{\Pi}^{2}
\end{split}
\end{equation}

\end{fleqn}
\begin{fleqn}

\begin{equation}
\label{Gamma3322and4411}
\begin{split}
\Gamma^{(3)}_{3,3 ; 2,2}
=
-
i
1728 g_{4}^{3} \tau^{2}
\big(
1
+
\abs{\Pi}^{2}
\big)
-
i 72 g_{4}^{2} \delta \tau^{2}
;\quad
\Gamma^{(3)}_{4,4 ; 1,1}
=
-
i
432 g_{4}^{3} \tau^{2}
\big(
1
+
2 \abs{\Pi}^{2}
\big)
-
i 36 g_{4}^{2} \delta \tau^{2}
\end{split}
\end{equation}

\end{fleqn}
\begin{fleqn}

\begin{equation}
\begin{split}
\Gamma^{(3)}_{4,2 ; 1,3}
=
-
i
\frac{g_{4}^{3}}{\omega^{2}}
\frac{12}{25}
\big(
11
+
14 \abs{\Pi}^{2}
\big)
-
i \frac{4}{25} \frac{g_{4}^{2} \delta}{\omega^{2}}
;\quad\;\;
\Gamma^{(3)}_{1,3 ; 4,2}
=
i
\frac{g_{4}^{3}}{\omega^{2}}
\frac{12}{25}
\big(
19
+
14 \abs{\Pi}^{2}
\big)
+
i \frac{g_{4}^{2}\delta}{\omega^{2}}
\frac{4}{25}
\end{split}
\end{equation}

\end{fleqn}
\begin{fleqn}

\begin{equation}
\begin{split}
\Gamma^{(3)}_{5,1 ; 0,4}
=
-
i
\frac{g_{4}^{3}}{\omega^{2}}
\frac{3}{100}
\big(
1
+
2 \abs{\Pi}^{2}
\big)
-
i \frac{g_{4}^{2}\delta}{\omega^{2}} \frac{1}{400}
;\quad\;\;
\Gamma^{(3)}_{0,4 ; 5,1}
=
i
\frac{g_{4}^{3}}{\omega^{2}}
\frac{3}{100}
\big(
5
+
2 \abs{\Pi}^{2}
\big)
+
i
\frac{g_{4}^{2}\delta}{\omega^{2}}
\frac{1}{400}
\end{split}
\end{equation}

\end{fleqn}
\begin{fleqn}

\begin{equation}
\begin{split}
\Gamma^{(3)}_{3,4 ; 2,1}
=
-
\Gamma^{(3)}_{2,1 ; 3,4}
=
-
i
\frac{g_{4}^{3}}{\omega^{2}}
\frac{105408}{121}
\abs{\Pi}^{2}
;\qquad\quad\;
\Gamma^{(3)}_{5,2 ; 0,3}
=
-
\Gamma^{(3)}_{0,3 ; 5,2}
=
-
i
\frac{g_{4}^{3}}{\omega^{2}}
\frac{1856}{1323}
\abs{\Pi}^{2}
\end{split}
\end{equation}

\end{fleqn}
\begin{fleqn}

\begin{equation}
\begin{split}
\Gamma^{(3)}_{3,5 ; 2,0}
=
-
\Gamma^{(3)}_{2,0 ; 3,5}
=
-
i
\frac{g_{4}^{3}}{\omega^{2}}
\frac{36}{25}
\big(
1
+
2\abs{\Pi}^{2}
\big)
;\qquad\,
\Gamma^{(3)}_{4,4 ; 2,2}
=
-
i
216 g_{4}^{3} \tau^{2}
\end{split}
\end{equation}

\end{fleqn}
\begin{fleqn}

\begin{equation}
\begin{split}
\Gamma^{(3)}_{6,2 ; 0,4}
=
-
\Gamma^{(3)}_{0,4 ; 6,2}
=
-
i
\frac{g_{4}^{3}}{\omega^{2}} \frac{3}{200}
;\qquad\qquad\qquad\;\;
\Gamma^{(3)}_{3,5 ; 3,1}
=
-
\Gamma^{(3)}_{3,1 ; 3,5}
=
-
i
\frac{g_{4}^{3}}{\omega^{2}} \frac{24}{25}
\end{split}
\end{equation}
\end{fleqn}
In particular, we notice that the coefficients of the pseudo-dissipators with excitation-conserving operators on both sides of the density matrix (e.g. $\Gamma^{(3)}_{2,2 ; 1,1}$, $\Gamma^{(3)}_{3,3 ; 1,1}$, $\Gamma^{(3)}_{3,3 ; 2,2}$, $\Gamma^{(3)}_{4,4 ; 1,1}$, and $\Gamma^{(3)}_{4,4 ; 2,2}$) are explicitly dependent on the coarse-graining time scale $\tau$, whereas all the other coefficients are independent of $\tau$. And for the same reason explained in the last paragraph of \ref{subsec: parametron}, we expect the analytical expression of the TCG master equation above to be accurate as long as one works in the regime where $\frac{1}{g_{4}} \gg \tau \gg \frac{1}{\omega}$.


\end{document}